\documentclass[aps,pra,twocolumn,showpacs,amsmath,amssymb,nofootinbib,longbibliography,superscriptaddress
]{revtex4-1}

\usepackage[pdftex]{graphicx}
\usepackage[usenames,dvipsnames]{xcolor}
\usepackage{epstopdf}
\usepackage[hidelinks]{hyperref}

\usepackage[normalem]{ulem}

\graphicspath{{FIGs/}}

\usepackage[utf8]{inputenc}
\usepackage[T1]{fontenc}

\usepackage{beramono}
\usepackage{listings}

\usepackage{lmodern}

\usepackage{amssymb,amsmath,amsthm}

\makeatletter
\DeclareFontEncoding{LS1}{}{}
\DeclareFontEncoding{LS2}{}{\noaccents@}
\DeclareFontSubstitution{LS1}{stix}{m}{n}
\DeclareFontSubstitution{LS2}{stix}{m}{n}
\makeatother

\DeclareMathAlphabet\mathscr{LS1}{stixscr}{m}{n}
\SetMathAlphabet\mathscr{bold}{LS1}{stixscr}{b}{n}

\makeatletter
\lst@InstallKeywords k{attributes}{attributestyle}\slshape{attributestyle}{}ld
\lst@InstallKeywords k{attributes2}{attributestyle2}\slshape{attributestyle2}{}ld
\lst@InstallKeywords k{attributes3}{attributestyle3}\slshape{attributestyle3}{}ld
\makeatother

\definecolor{macros}{rgb}{0.64,0.55,0.08}
\definecolor{types}{rgb}{0,0.55,0.55}
\definecolor{functions}{rgb}{1,0.27,0}

\lstdefinelanguage{Julia}%
  {morekeywords={abstract,break,case,catch,const,continue,do,else,elseif,%
      end,export,false,for,function,immutable,import,importall,if,in,nothing,%
      macro,module,otherwise,quote,return,switch,true,try,type,typealias,%
      using,while,mutable},%
   sensitive=true,%
   alsoother={$},%
   morecomment=[l]\#,%
   morecomment=[n]{\#=}{=\#},%
   morestring=[s]{"}{"},%
   morestring=[m]{'}{'},%
   moreattributes={@inline,@inbounds,@simd,Threads.@threads},
   attributestyle = \bfseries\color{macros}, 
moreattributes2={AbstractArray,Array,genColType,Union,Number,Integer,Tuple,Bool,Colon,struct,Real,Float64,include},
   attributestyle2 = \bfseries\color{types}, 
moreattributes3={eigen,contract,ccontract,contractc,ccontractc,TensType,MPS,MPO,tens,Env,,intType,svd,reshape,unreshape,denstens,qarray,qr,lq,permutedims,move,spinOps,makeMPO,convert2MPS,expect,largeMPO,largeMPS,largeEnv,fermionOps,spinOps,tJOps,environment,makeEnv,applyMPO,largematrixproductstate,largematrixproductoperator,largeenvironment,correlation,correlationmatrix,polar,matrixproductoperator,matrixproductstate,loadMPS,loadMPO,loadLenv,loadRenv,transfermatrix,dmrg,Lupdate,Rupdate,boundaryMove,div!,krylov,twosite_update,twositeOps,simpledmrg,fullH,randMPS,applyOps!,applyOps,fullpsi,sub!,div!,add!,mult!,norm},   
   attributestyle3 = \bfseries\color{functions},
}[keywords,comments,strings]%

\graphicspath{{FIGs/}}

\begin{document}

\title{Build your own tensor network library: \\DMRjulia I. Basic library for the density matrix renormalization group}
\author{Thomas E.~Baker}
\affiliation{Department of Physics, University of York, Heslington,York YO10 5DD, United Kingdom}
\affiliation{Institut quantique \& Département de physique, Université de Sherbrooke, Sherbrooke, Québec J1K 2R1 Canada}
\author{Martin P.~Thompson}
\affiliation{Institut quantique \& Département de physique, Université de Sherbrooke, Sherbrooke, Québec J1K 2R1 Canada}
\date{\today}

\begin{abstract}

An introduction to the density matrix renormalization group is contained here, including coding examples. The focus of this code is on basic operations involved in tensor network computations, and this forms the foundation of the DMRjulia library.  Algorithmic complexity, measurements from the matrix product state, convergence to the ground state, and other relevant features are also discussed. The present document covers the implementation of operations for dense tensors into the Julia language.  The code can be used as an educational tool to understand how tensor network computations are done in the context of entanglement renormalization or as a template for other codes in low level languages.   A comprehensive Supplemental Material is meant to be a "Numerical Recipes" style introduction to the core functions and a simple implementation of them.  The code is fast enough to be used in research and can be used to make new algorithms.

\end{abstract}

\maketitle

\tableofcontents


\section{Introduction}

Solving models in quantum physics takes on a great importance when considering the vast array of systems that these models can imitate.  A broad class of efficient algorithms to do this have appeared that make the solutions possible, especially in the case of a lattice problem.  These include tensor network methods which use the loaclity of a system to obtain efficient ground state results.  

Tensor networks have several advantages \cite{schollwock2005density,schollwock2011density}.  First, there is a site-by-site decomposition of the problem to avoid the exponentially large system size encountered in exact diagonalization. The resulting algorithms can be extended to many hundreds or thousands of sites, far beyond the memory limits of present day computers with exact diagonalization. The second advantage of tensor network methods is the modeling of local interactions, which guides the form that a tensor network takes \cite{hastings2004locality}.  In comparison with quantum Monte Carlo methods, tensor network methods have no sign problem.

One tensor network algorithm known as the density matrix renormalization group (DMRG) is approaching its 30 year anniversary \cite{white1992density}. While the original derivation of the algorithm can be considered as an extension of the numerical renormalization group (NRG) \cite{krishna1980renormalization} to more than one site, the modern formulation of the the algorithm has been successfully framed in terms of a matrix product state (MPS) \cite{affleck1988valence,ostlund1995thermodynamic}. This formulation is a very clear representation of the problem and has many connections with quantum information.  By connecting the DMRG algorithm to the MPS framework, the method can be identified in a broader class of algorithms known as tensor network methods.  The key feature of DMRG and lattice solvers in general from more general tensor network methods is that they renormalize the problem with entanglement by discarding states that do not contribute much to the ground state. This means that not all operations applied on the tensor network are unitary and that a specific correlation structure is assumed for the problems solved with these methods.  

The steps in the DMRG algorithm are very simple if the core concepts of a tensor network are understood. In the traditional DMRG algorithm, two adjacent sites are contracted together, an algorithm (power method, Lanczos, Davidson, random perturbation, etc.) is applied to update them, the singular value decomposition is applied to decompose the tensors, and a special site known as the center of orthogonality is moved to the next site.  The resulting algorithm is very quickly convergent for locally entangled models.

The level of understanding in the previous paragraph only contains a mechanical understanding of the algorithm.  Often, more questions should be answered so that the algorithm is well understood.  An even larger question is also unanswered by only understanding the algorithm: how do the concepts of information theory and renormalization group influence the method?  Why is it called the density matrix renormalization group?

Even aside from these high level questions related to theory, there is a need for explanations of this method that allow for a user to apply DMRG efficiently, understand the relevant components for a computation, and even make new algorithms. As tensor network computations are used more and more in physics, the need for more groups to manage their own code and program their own algoithms will increase. Even with a suitable understanding of the algorithms of a tensor network, the implementation in a code can be slow. With a fast code, an inexperienced user can cause some inefficiencies in the code.  The need for easy to use code that is transparently written  can be a great asset for someone working with a limited amount of time or just starting.

The library that accompanies the discussion here implements the DMRG algorithm in the Julia language with a learnable number of functions and lines, while still implementing a wide variety of algorithms in tensor networks (hosted at the link in Ref.~\onlinecite{dmrjulia}).  The resulting code is nearly 6,000 lines at its core (including comments) while being competitively fast for large systems.  The true advantage of this code is the ability to make informed user independently through documentation like this and also to be small enough to be manipulated, focusing on basic operations as building blocks.  This makes manipulating the code easier and therefore more accessible so the best algorithms can be applied in as many situations as possible.

This paper expands on the discussion from Ref.~\onlinecite{bakerCJP21,*baker2019m}. If the reader is not familiar with the material in that paper or if the reader wishes a more general overview of tensor networks, then it is encouraged to start with that paper (available in English and French \cite{bakerCJP21,*baker2019m}) or other introductory papers on DMRG \cite{bakerCJP21,*baker2019m,schollwock2005density,hallberg2006new,schollwock2011density,orus2014practical,bridgeman2017hand}. This introduction assumes only a basic knowledge of quantum physics which will be left for other existing works on the subject \cite{townsend2000modern,desai2010quantum,shankar2012principles}. All functions developed will correspond to the conventions used in Julia so that the use of these functions is as visibly similar to using plain Julia code as if any other computation was performed.

This paper will first begin with a surface level introduction to the core theoretical concepts in DMRG in Sec.~\ref{whyDMRG}: density matrices and the renormalization group.  Once introduced in a way that is geared towards understanding the steps in DMRG, the basic operations and how to program them are covered in Sec.~\ref{tensorsAndOps} and the Supplemental Material \cite{tensor_recipes}. In Sec.~\ref{networks} focuses on the storage and manipulation of a network of tensors representing the wavefunction or operators.  Sec.~\ref{DMRG} reviews the steps of the DMRG algorithm with the tools developed in the previous steps.  The last section, Sec.~\ref{useDMRG}, contains many useful concepts for converging to the ground state in several cases of interest and how to take measurements on the MPS.\footnote{There are two notable omissions from the discussion here that is included in other summaries of the same topic.  The wavefunction ansatz from Affleck-Kennedy-Lieb-Tasaki, known as the AKLT state \cite{affleck1988valence}, is not covered.  This is because DMRG takes a known matrix product operator (MPO) which can be found from a given Hamiltonian and then solves for the ground state wavefunction.  While the AKLT state is useful as an original and exact example of the MPS, it is not strictly necessary to understand the use of the MPS in DMRG.  For further details, the interested readers is referred to Ref.~\onlinecite{schollwock2005density}.}

\section{Why is it the called the density matrix renormalization group?} \label{whyDMRG}

A very brief history and review of relevant concepts as it pertains to DMRG is presented here. This section can be skipped and returned to later if only implementation details of DMRG are required.

The theoretical background of DMRG requires both density matrices and renormalization group techniques to be introduced.   For density matrices, there is a deep and fundamental connection with information theory, and this will be explained in the next section. 

It is sometimes incorrectly commented that DMRG is not truly related to the renormalization group.  A review of the history \cite{fraser2021twin} shows that the ideas for DMRG stem from Kenneth Wilson who extended concepts originally introduced by St\"uckelberg and Peterman \cite{stueckelberg1953normalization} and Gell-Mann and Low \cite{gell1954quantum}. Wilson won the Nobel Prize for his application of the renormalization group technique particularly to particle physics and critical systems, and the technique of renormalization has been used in a variety of contexts and applies equally to condensed matter physics as well as particle physics \cite{wilson1975renormalization}.  It is also the inspiration for several algorithms, including the numerical renormalization group (NRG) \cite{krishna1980renormalization}.  Both of these concepts will be introduced before moving on to computations with tensor networks.

\subsection{Density matrices}\label{densitymatrix}

Density matrices are one of the most useful quantities from statistical physics, although they are often presented as a curiousity in many introductory texts.  They become a core quantity in more advanced quantum solvers and deserve some extended exposition here.  In DMRG, the coefficients of the density matrix are utilized directly.

Choosing a basis for the density matrix, the operator can be written as
\begin{equation}
\hat\rho=\sum_{ij}\rho_{ij}|i\rangle\langle j|
\end{equation}
where $i$ and $j$ index the basis functions of the matrix.  The reason that the density matrix plays a minor role in many texts is that its deep connection with entropy, which is often presented as a purely abstract quantity without much physical interpretation. However, understanding entropy is crucial to the understanding of DMRG here, so some time is spent developing a connection with the physical world and this quantity. 

The main idea behind tensor networks is to break the problem up into smaller parts. In order to connect the subsets back with the full problem, the information passed between the subset and the rest of the system must be known. One way to do this is with the density matrix.  The justification for this is motivated by the following.

\subsubsection{Connection to information theory and entanglement}

To understand entropy, return to the foundations of information theory which were very important in the 1940s for code breaking purposes.  Quantifying mathematically how much information is passed from one person to another became critically important for espionage around this time, although its use has grown to error correction in computers and encryption since, among others \cite{gardner2005phaselock}.

A key statement by Shannon in the 1950s formalized the mathematical quantity that describes the information passed from one source to a receiver.  The full situation is that a message can be passed from one source to a receiver.  For example, the word
\begin{equation}\nonumber
\mathrm{F\quad A\quad C\quad E}
\end{equation}
might be transmitted. But if there is some noise in the letters individually, then the message received might be 
\begin{equation}\nonumber
\mathrm{F\quad A\quad T\quad E}
\end{equation}
and this would represent a problem when passing messages because it can not be determined with certainty that the correct message was read. Importantly, this may greatly affect the meaning of the message!

The broad solution to ensuring that the correct message got through is to repeatedly transmit the message and to compare the messages via an agreed upon strategy. To keep the example simple, once the message is passed and received, a systematic set of questions can be asked to the source by the receiver to determine if the correct message was obtained.  The source can be asked whether the first letter "is A, is B, is C..." and on until the letter is found.  Then the process can be repeated for the next letter.  Eventually, the message will be verified completely, and the mathematical question to ask is: can another algorithm be used more efficiently, and how can the average uncertainty ($S$) in the letters be quantified with a mathematical expression?

\subsubsection{Basic postulates in information theory}

In order to quantify the uncertainty of a transmitted message, four reasonable axioms can be established. Following the presentation in Ref.~\onlinecite{ash1965information}:
\begin{enumerate}
\item Monotonically increasing

If the number of objects increases, then so should the uncertainty in finding the correct element.  For example, finding a single person in a small town should be easier than finding someone in a large city because there are fewer cases to search through.  In the previous example of sending messages, searching a smaller alphabet should have lower uncertainty.

The generic expression of the average uncertainty, $S$, in a message can be written as
\begin{equation}
S\left(\rho_1,\rho_2,\rho_3,\ldots,\rho_N\right)
\end{equation}
where the elements are assigned a probability, $\rho$, of being correct. For the example above, each letter in a word would have a probability of 1/26 for a 26-letter alphabet.

\item Additivity

Searching both a small town and a large city for a single person should appear as the search of the both populations added together. Simillarly, a smaller alphabet should be easier to search than a larger one. The resulting relationship between elements of sets $A$ and $B$ can be cast as
\begin{equation}
S(AB) = S(A) + S(B)
\end{equation}
which resembles the addition of two logarithms.

\item Grouping axiom

In contrast to the additivity axiom, one can reasonably ask how two sets of objects can be separated and what effect this has on the function $S$. This also implies that the probabilities must be resummed. Considering the first $w$ values of the first set.  The probability to pick the $i$th element should be $\rho_i/\rho_A$ where $\rho_A=\sum_{g=1}^w\rho_g$, placing the new sum of the probabilities in the subset in the denominator.

The function $S$ under this grouping of elements should be decomposed into three functions.   The first is to identify the uncertainty in picking either group $A$ or group $B$
\begin{align}
S\left(\rho_1,\ldots,\rho_N\right)=&S\left(\rho_1,\ldots,\rho_w,\rho_{w+1},\ldots,\rho_N\right)\\
&+\rho_AS\left(\frac{\rho_1}{\rho_A},\ldots,\frac{\rho_w}{\rho_A}\right)\nonumber\\
&+\rho_BS\left(\frac{\rho_{w+1}}{\rho_B},\ldots,\frac{\rho_N}{\rho_B}\right)\nonumber
\end{align}
where $\rho_B=\sum_{g=w+1}^N\rho_g$. This describes how the probabilities can be partitioned into sets.

\item Continuous

The continuous nature of the function results from the fact that we should expect that the function $S$ is perturbative with a small change in $\rho_i$.

\end{enumerate}
The unique solution to these four axioms is  known to be (see Ref.~\onlinecite{ash1965information} for a proof)
\begin{equation}\label{shannonentropy}
S=-C\sum_i\rho_i\log_2\rho_i\quad[\mathrm{Shannon}]
\end{equation}
for an arbitrary constant $C$ which essentially allows for a change of base of the logarithm.

Take note of the implication here.  Passing a message from source to receiver has an inherent uncertainty that can be characterized by an entropy function.  To complete the analogy of passing a code from source to receiver, the entropy represents the average number of questions that must be asked to verify a message for some algorithm that verifies the message \cite{ash1965information}.  Some algorithms will be more efficient and require fewer questions to be asked (lower $S$).

\subsubsection{Entropy for quantum systems}

Shannon's entropy function applies to the classical case \cite{shannon1948mathematical}.  Meanwhile, the extension to the quantum case was performed by von Neuman and has the same functional form \cite{von1955mathematical}, but the elements of the density matrix now define the $\rho_i$ elements,
\begin{equation}\label{vNentropy}
S=-\mathrm{Tr}\left(\hat\rho\ln\hat\rho\right)\quad[\mathrm{von\;Neumann}]
\end{equation}
where the trace of an operator is defined as $\mathrm{Tr}(\mathcal{A})=\sum_n\langle\psi_n|\mathcal{A}|\psi_n\rangle$, the sum of the diagonal entries which is invariant under a unitary transformation of $\mathcal{A}$. 

The quantum case reinterprets the classical message passing as the information passed between two parts of a system.  Consider the simple example of two spins in the state $|\psi\rangle=|\uparrow\downarrow\rangle$ with the first site in the up state and the second site in the down state ($=|\uparrow\rangle\otimes|\downarrow\rangle$). The resulting density matrix has one element (see discussion in Ref.~\onlinecite{bakerCJP21,*baker2019m}) of amplitude 1, and this signifies that the state is a classical state. There is no entanglement by direct evaluation of Eq.~\eqref{vNentropy} with $\rho_1=1$.

Taking another example of a singlet pair of spins, $|\psi\rangle=(|\uparrow\downarrow\rangle-|\downarrow\uparrow\rangle)/\sqrt2$, the resulting density matrix has two elements of amplitude $1/\sqrt2$ and this is a maximally entangled state ($S$ is at its maximum value) \cite{bakerCJP21,baker2019m}. This state is inherently quantum mechanical and displays a superposition of states.  The uncertainty of which spin is in which state is not known.  By drawing the partition between the two spins (which will be explained formally next and in Ref.~\onlinecite{bakerCJP21,*baker2019m}), $S$ provides a measure of how uncertain the spin states are.  The information being passed is characterized by the entanglement and therefore the quantum aspects of a system are fully characterized by the entanglement between states.

There are frequently examples on entangled states where one spin of the singlet is received by one observer ("Alice") and the other spin by another observer ("Bob").  Since neither spin can be determined beforehand (but upon measurement, the other observer knows both of the spins that were measured), this lends itself to the idea that keeping track of the information between the two observers is the more useful quantity to record than the individual spin states that Alice and Bob measure.

The entropy depends on how the system is partitioned. When it is commented that the quantum case displays entanglement, this is specifically stating that the partition between sites in a lattice has some non-trivial entropy. 

Earlier, it was remarked that the entanglement was regarded as a curiosity or a clever--but specific--quantity that was useful for some problems in many statistical physics books and historically.  This can be attributed to the need to develop introductory statistical physics courses from the historical perspective, starting from classical considerations where no superpositions exist.  However, this connection between information theory, uncertainties, superpositions, and entropies motivates a new way of looking at the quantum problem because they are inherent to those quantum states.  The hint here is that getting the entanglement correct in a given basis will give the full ground state.  That will be the motivation for using the density matrix in DMRG computations, as opposed to another quantity such as the Hamiltonian.

\subsubsection{Connection to other entropies}

The entropy in Eq.~\eqref{shannonentropy} and the quantum analogue with elements of the density matrix in Eq.~\eqref{vNentropy} are somewhat different but related to the original equation for entropy owed to Boltzmann and Gibbs.  The equation is 
\begin{equation}
S=k_B\ln W\quad[\mathrm{Boltzmann}]
\end{equation}
where $W$ is the number of microstates of a given system.  To recover this form from von Neumann's entropy, note that the probability of being in the $i$th state is $\rho_i=1/\sqrt[W]W$ when all states are equally probabalistic, giving the form above. The original derivation was to identify that the relation between $W$ and $S$ must behave like a logarithm \cite{reif2009fundamentals}, and the Boltzman coefficient $k_B$ was only determined long after the relationship was deduced. In these early introductions of entropy, the idea was to essentially provide a convenient way to describe the Carnot cycle, which led to an abstract interpretation of the quantity.

There are other definitions of the entropy, such as the Renyi entropy, which do not assume all the axioms above.  For example, by ignoring axiom 3, the Renyi entropy can be suitably defined
\begin{equation}
S=\frac1{1-\alpha}\log\mathrm{Tr}\left(\sum_i\rho_i^\alpha\right)\quad\mathrm{[Renyi]}
\end{equation}
where the limit as $\alpha\rightarrow1$ recovers the von Neumann entropy.  These alternative measures are useful in other contexts, but it is sufficient to simply be aware of them.

While it may seem trivial to characterize information in this way, this is quite a powerful connection between information and the physically relevant quantity of the density matrix.  With the recent rise of topological physics and the characterization of these states through entanglement \cite{simonknots,kitaev2006topological,jiang2012identifying}, it has become undeniable that information plays a major role in the description of quantum states.  It turns out to be crucial detail in DMRG that this connection exists and has real significance for expressing quantum wavefunctions. Identifying exactly what quantity in a physics system identifies the information of the full system is very valuable.  

\subsubsection{Subsystems and partial traces}\label{partialtraces}

Since the entropy has been established as both passing information, a question can be asked: if a system is partitioned into a subset of systems, how can one of those subsets convery information about the entire system?  The answer is crucial to realizing DMRG.

The key element to defining a part of a system is the partial trace. To understand a partial trace, consider the density matrix for two spin-half sites
\begin{equation}
\hat \rho=\left(\begin{array}{cccc}
\rho_{11} & \rho_{12} & \rho_{13} & \rho_{14} \\ 
\rho_{21} & \rho_{22} & \rho_{23} & \rho_{24} \\ 
\rho_{31} & \rho_{32} & \rho_{33} & \rho_{34} \\ 
\rho_{41} & \rho_{42} & \rho_{43} & \rho_{44} 
\end{array}\right)
\end{equation}
in a basis of $\{|\uparrow\uparrow\rangle,|\uparrow\downarrow\rangle,|\downarrow\uparrow\rangle,|\downarrow\downarrow\rangle\}$.  If the reduced density matrix for the second spin is sought, then the first spin can be traced out.  So, the subscript 1 here represents all spin states on the first site,
\begin{align}
\mathrm{Tr}_1\hat \rho=
%
%
%
\left(\begin{array}{cc}
\rho_{11}+\rho_{33} & \rho_{12}+\rho_{34}\\
\rho_{21}+\rho_{43} & \rho_{22}+\rho_{44}\\
\end{array}\right)
\end{align}
and a similar expression could be written if the first spin were to be traced out.  The resulting basis only contains spin states on the second site, and this is one partition of the original system. In the continuum limit, this is related to integrating out variables \cite{schwartz2014quantum}.

The only lesson from this discussion is that the trace of a partial trace is equal to the full trace of the original system,
\begin{equation}
\mathrm{Tr}\left(\hat \rho\right)=\mathrm{Tr}\left(\mathrm{Tr}_1\hat\rho\right),
\end{equation}
so the trace of the density matrix should be equal whether tracing a density matrix representing one partition of the system or the other.

The above discussion on partial traces can be applied on the wavefunction to partition a system while keeping track of the information passed between partitions. The full density matrix of a given $N$-site system is written as
\begin{equation}\label{exrho}
\hat\rho=|\psi_{\sigma_1\sigma_2\sigma_3\ldots\sigma_N}\rangle\langle\psi_{\sigma_1\sigma_2\sigma_3\ldots\sigma_N}|
\end{equation}
where some of the lattice sites (a set of sites $B$) can be traced out,
\begin{equation}\label{traceA}
\mathrm{Tr}_B\hat\rho=|\psi_{\sigma_1\sigma_2}\rangle\langle\psi_{\sigma_1\sigma_2}|=\hat\rho_A
\end{equation}
leaving the first two sites.

If the complementary set of sites had been traced out in Eq.~\eqref{exrho} (trace over a set $A$), then the resulting density matrix would have been
\begin{equation}
\mathrm{Tr}_A\hat\rho=|\psi_{\sigma_3\ldots\sigma_N}\rangle\langle\psi_{\sigma_3\ldots\sigma_N}|=\hat\rho_B.
\end{equation}
It can be said that the lattice was "cut" or "partitioned" along the bond between sites 2 and 3 \cite{bakerCJP21,*baker2019m}. The resulting reduced density matrices describing two sides of the system are then related representations of the same system.

The main question is how to connect the the description of the reduced system $A$ and $B$. To do that, both density matrices $\hat\rho_A$ and $\hat\rho_B$ can be decomposed with an eigenvalue decomposition as
\begin{equation}
\hat\rho_A=\hat U\hat \Omega_L\hat U^\dagger\quad\mathrm{and}\quad\hat\rho_B=\hat V\hat \Omega_R\hat V^\dagger
\end{equation}
where $\hat U$ contains the basis of states (denoted as combinations of spins $\sigma_1$ and $\sigma_2$ for $A$) for the left partition of the system (for more discussion and diagrams see Ref.~\onlinecite{bakerCJP21,*baker2019m}).  The right partition of the system, $B$, has basis functions in $\hat V$.

The weights of the basis functions are contained in the matrices $\hat\Omega_L$ and $\hat\Omega_R$. Note that a cyclic property of the trace implies that
\begin{equation}
\mathrm{Tr}(\hat\rho_A)=\mathrm{Tr}\left(\hat\Omega_L\hat U^\dagger\hat U\right)=\mathrm{Tr}\left(\hat\Omega_L\right)
\end{equation}
and similar for $\hat\rho_B$.  At this point, we can impose that the trace of the partial trace for either system must be equal.  This could be argued on physical grounds.  The information as determined for the $A$ partition from the $B$ partition must be equal to the information between the $B$ partition from the $A$ partition. So,
\begin{equation}\label{informationsymmetry}
\hat\Omega_L=\hat\Omega_R=\hat \Omega
\end{equation}
can be enforced for these decompositions.

\subsubsection{Partitioning the wavefunction}\label{whySVD}

In terms of the components of the reduced densities matrices derived so far, an expression for the wavefunction $|\psi_{\sigma_1\sigma_2\sigma_3\ldots\sigma_N}\rangle$ can be derived from the partitioned wavefunctions $|\psi_{\sigma_1\sigma_2}\rangle$ and $|\psi_{\sigma_3\ldots\sigma_N}\rangle$.  The basis for each partitioned wavefunction is contained in $\hat U$ and $\hat V$, respectively. The relationship is not as simple as the tensor product, $\otimes$, for systems with entanglement. The correct form combining all of these elements can be guessed to be (regardless of the chosen partition)
\begin{equation}\label{svd_decompose}
|\psi_{(\sigma_1\sigma_2)(\sigma_3\ldots\sigma_N)}\rangle=\hat U\hat D\hat V^\dagger
\end{equation}
where $\hat\Omega=\hat D^2$.  In this example, $\hat U$ has the indices $U^{(\sigma_1\sigma_2)}_{a}$ where $a$ is the index connecting the tensors, implying $D^a_a$ is the form for the middle tensor. Similarly, $\hat V^\dagger$ has indices $(V^\dagger)^{a}_{(\sigma_3\ldots\sigma_N)}$. Note that raised and lowered indices here have no meaning formally but are used to signify indices on the "left" and "right" of the tensors.  

If the wavefunction is decomposed into this form, then the entirety of the density matrix will be discovered.  As has already been seen, the density matrix contains all information once a chosen partition is selected for a given system.  The linear algebra operation that has the form of Eq.~\eqref{svd_decompose} is the singular value decomposition (SVD).  In practice, it is not difficult to obtain all of the elements used so far to describe reduced density matrices.  The positive definite elements of the diagonal matrix $\hat D$ are known as the singular values.  The square of the singular values are the eigenvalues of the density matrix.

Having now developed the density matrix and how it is useful in decomposing a quantum problem, an understanding of a broader concept of how to change a problem to an equivalent problem is contained in the study of the renormalization group.

\subsection{Renormalization group}

Mandelbrot (known for fractals) wrote in an article \cite{mandelbrot1967long} that began the discussion with the observation that finding the length of the coastline of Britain was completely dependent on how it was measured. The answer obtained by viewing the coastline from miles above the earth would be different than if the coastline was walked or zoomed in.   

The difference is one of scale.  If using the coarser perspective from higher up, the finer features (every bend and twist) of the coastline will not be measured.  Neither answer is wrong. Measuring the coastline at both scales answers two different questions, and the connection between those computations is a question of the relevant scale which is captured by the correlation length of a system.  Understanding the correlation length between those two computations will introduce the core ideas behind the renormalization group which, in this example, is the connection between the coastline measured at the fine and coarse scale.

\subsubsection{Correlation lengths}

The correlation length is one of the most fundamental properties of a physical system \cite{cardy1996scaling,francesco2012conformal}. This quantity can describe how far away from a perturbation the disturbance would be felt. Take for example the operator $\hat c^\dagger_{i\sigma}\hat c_{j\sigma}$ for fermions of spin $\sigma$ can measure the amplitude of adding a particle at site $i$ and destroying a particle at site $j$.  The correlation function is an expectation value of the operator, $\langle\Psi|\hat c^\dagger_{i\sigma}\hat c_{j\sigma}|\Psi\rangle$.  There are many other types of correlation functions in physics, and this is just one example of a correlation function in quantum physics.

There is a simple analogy that can be introduced to give a conceptual picture of a second-order phase transition and how the correlation length behaves in this case. The idea of a correlation length can be explained with a simple example of the night and day cycle. Consider the transition from day to night (and the reverse) as a phase transition.  Day and night take the place of phases with some defined correlation functions.  The natural correlation length to define here is the length of a shadow of a stick stuck into the ground.

While the sun is out, the stick's shadow has a particular length.  When the sun is highest in the sky (noon), the shadow's length is almost zero.  When the sun sets, the shadow's length will diverge, becoming infinite at the moment when day transitions to night.  When the moon rises, the shadow's length will once again grow.

The description of the last paragraph closely mirrors the behavior of a correlation length for a second order phase transition.  At the phase transition, the correlation length diverges, meaning that the phase (a mixture of day and night at dusk and dawn) is extending to all regions of the system.  This is a feature of critical points.  

Understanding correlation lengths is useful in order to see what the length of an interaction is. The idea that the same object can be described with a theory with a larger correlation length or with a smaller correlation length is similar to the idea that a basis chosen for a problem should not affect the end result. However, choosing a theory with the appropriately sized correlation length can be advantageous for computing quantities.

To connect with the example of measuring Britain's coastline, the observer far above the earth is measuring the coastline with larger units, one that does not capture all of the zigs and zags of the coastline.  The observer who walks the coastline is using a smaller unit of measurement and is able to capture more of the features of the coastline.  Note that an ant, who uses a very small length-scale to measure the same problem would take a very long time to obtain the result.  So, clearly choosing the appropriately sized correlation length is a very useful concept when making theories to measure phenomena. Note that unlike the example with day and night transitions, Mandelbrot's example includes no phase transition.

The perspective here applies equally to classical as well as quantum systems at zero temperature and finite temperature. Quantum systems exhibit critical phenomena and a major topic of interest is understanding how exactly to describe these critical points and systems. 

When studying phase transitions numerically, it is important to remember that the finite size of the lattice will impose some cutoff. This means that these numerical cutoffs will prevent the correlation length from ever truly going to infinity.  This will mean that a quantity that should be infinite ({\it i.e.}, the specific heat) will have some round off, meaning that a peak in the specific heat will not generally be computed in the true thermodynamic limit, as would be analytically.

While the focus here was on second order phase transitions, there are also first order phase transitions which are defined to have a discontinuity in the first derivative of the partition function \cite{reif2009fundamentals}.  In this case, the correlation length does not diverge across a phase transition. The guideline is that the phase transitions are named after the first derivative of the partition function where a discontinuity appears \cite{ehrenfest1933phasenumwandlungen,jaeger1998ehrenfest}.

Understanding how to go from one problem to another is the core concept behind the renormalization group ({\it i.e.}, rewrite the theory for a given time of day in the above day and night analogy, or different perspective for the coastline example).  There are two historically recognized ways in which the renormalization group was realized and they are both relevant to DMRG, although both of these techniques are closely related.  One way in which renormalization group techniques can be used is to reduce the size of a system, keeping the relevant degrees of freedom as a low-pass filter would.  The second way in which renormalization is known and will be useful here is to impose cutoffs on the theory being studied to best understand where that theory applies and what it can not describe.

\subsubsection{Kadanoff's renormalization group: Coarse graining to eliminate trivial degrees of freedom}

In general, the idea of breaking a problem into smaller parts has a long history. One of the notable early cases where this strategy was applied was on a Bethe lattice where spins from a two-dimensional Ising model were unraveled into a tree-like network called a Cayley graph \cite{bethe1935statistical}.  In order to obtain values from this network, the individual tensors needed to be contracted together.

A major step forward for computing systems with small effective models was the original use of renormalization group by Kadanoff in the 1960s \cite{kadanoff1966scaling}. The spin-blocking technique grouped together spins on a lattice. This is a coarse graining operation.  It is well known that solving a large Hamiltonian can contain too many degrees of freedom to be properly represented on a modern computer, so the strategy was to rewrite the Hamiltonian (for example the Heisenberg spin model with spins $\mathbf{S}=(S^x,S^y,S^z)$ indexed by $i$ and $j$ with a coupling constant $J$)
\begin{equation}\label{heisenberg}
\mathcal{H}=-J\sum^N_{\langle ij\rangle}\mathbf{S}_i\cdot\mathbf{S}_j\rightarrow\mathcal{H}_{RG}=-J'\sum^M_{\langle ij\rangle}\mathbf{S}_i\cdot\mathbf{S}_j
\end{equation}
into a renormalization Hamiltonian $\mathcal{H}_{RG}$ where $M<N$ (and in the example of a two-dimensional model being $N/4^d=M$ for some number of coarse graining steps $d$).  The summation is over fewer spin sites, representing an advantage.  However, if the reduced model is to correspond to the full physical system, an adjusted parameter $J'$ must be taken into account in order to ensure that the energy (or generally some other property) remains the same between the two models.  The computational time to solve the renormalized Hamiltonian will be much smaller.

This procedure of reducing the degrees of freedom in a model and modifying the parameters, while maintaining the relevant physical quantities, is referred to as a renormalization of that quantity. The true procedure in Kadanoff's case will not be necessary here, so the interested reader is referred to Ref.~\onlinecite{kadanoff1966scaling}.

\subsubsection{Wilson's renormalization group: Where theories apply and cutoffs}

Renormalization group techniques take on a specific meaning in the history and context of high energy physics. Wilson's renormaization group was a good analysis tool to analyze critical points in these systems.  Wilson's application of the renormalization techniques essentially establishes an energy range over which a field theory is accurate, similar to a radius of convergence. At the time, field theorists were derided for imposing artificial cutoffs to make naturally divergent integrals convergent. Yet, there was no explanation for why a seemingly arbitrary constant should be imposed.  The mathematical origins of this aspect stemmed from regularization techniques \cite{peskin1995quantum} that reside in the field of mathematics known as asymptotic analysis \cite{de1981asymptotic}. Wilson's theory formalized these cutoffs and give the theory a solid interpretation.  They take on the meaning of extra physics beyond the model being studied.

For the purposes here, the cutoffs in particle physics are not useful to explain more detail; however, renormalization cutoffs appear in a variety of contexts.  Very broadly, any coefficient that is used in a physical system that is derived from some other theory could be referred to as an RG cutoff parameter. One commonly known coefficient is the coefficient of friction used in classical physics, and derived from experimental considerations, to avoid a full computation of the electromagnetic interactions that would produce such a quantity.  A more relevant quantity for an example of an RG cutoff is the Debye frequency \cite{reif2009fundamentals}.  In a condensed matter system, the Debye cutoff corresponds to the lattice spacing in a model, or the highest possible energy supported in a lattice model.  Integrating beyond this cutoff would require knowledge of physics that is not contained in the model studied, so it can be safely neglected.

If one wishes to attribute the condensed matter implementations of renormalization group to Kadanoff and the particle physics implementations to Wilson (although this is only how the literature often represents the situation, the line is not so clear between them), then Kadanoff is using the equivalent of a low-pass filter \cite{horowitz1980art} to keep only the relevant degrees of freedom (low energy states). This shrinks the size of the system while only introducing a small amount of error contained in the high energy degrees of freedom for the eventual answer. Meanwhile, Wilson's best known use of renormalization group is establishing a radius of convergence for a given theory.  

Note that the numerical renormalization group (NRG) was invented by Wilson and used prominently in condensed matter applications to solve impurity models to illuminate Kondo physics \cite{wilson1975renormalization}, illustrating the less than clear line between Wilson and Kadanoff's contributions here. Much more should be said about the renormalization group methods, but we will instead refer to other works \cite{fraser2021twin,zinn2007phase,pathria2011statistical,bakerPRB18}.

The DMRG algorithm uses both types of renormalization group presented here. Kadanoff's ideas are found in the rewriting of the problem into an equivalent, few-site problem.  Meanwhile, the selection of short-range entangled states imposes an effective cutoff, just as is attributed to Wilson here.  The key innovation that the DMRG algorithm uses is to impose this cutoffs for density matrices, not in real space or momentum space.

\subsection{Why does the density matrix renormalization group work?}\label{whyDMRGworks}

Having now derived the density matrix decomposition from the wavefunction in the above, all necessary ingredients to perform the density matrix renormalization group are available.  The relevant objects in the system (wavefunction, Hamiltonian, etc.) can all be decomposed into a site-by-site representation with the SVD. Since this decomposition is exact, in principle the entire system could be renormalized down to one site without losing any information, so long as the interaction with the rest of the lattice is retained.  However, this program of activity will provide for a means to approximate the tensors to avoid a large number of degrees of freedom kept.  This will keep the algorithm fast and give a controlled accuracy.

One key advantages of the density matrix has is that it provides access to expectation values for operators in quantum physics. Note that when taking the trace of the density matrix and Hamiltonian,
\begin{equation}\label{expectVal}
\mathrm{Tr}\left(\hat\rho{\mathcal{H}}\right)=E
\end{equation}
the energy, $E$, results which is the expectation value of the Hamiltonian.  If the density matrix were diagonalized as
\begin{equation}
\hat\rho=\sum_k\rho_k|\Phi_k\rangle\langle\Phi_k|
\end{equation}
then it can be noticed that the eigenvalues $\rho_k$ can be ordered from greatest to least.  Clearly, the best approximation of the full density matrix is to keep the largest $\rho_k$ values.  This retains the best accuracy on expectation values and matrices in general.

A concept that is natural in quantum chemistry is that of the natural orbitals, $\Phi_k$.  They are the orbitals that result from a diagonalization of $\hat\rho$, and they constitute a rapidly convergent basis set for quantum systems \cite{lowdin1956natural,lowdin1955quantum}. If the natural orbitals were chosen through some method, then this would be an excellent way to approximate the density matrix.  The issue is that the full wavefunction must be found to obtain the full natural orbitals, so clearly some approximation must taken place to obtain these quantities.  This makes computing with the density matrix somewhat unnatural, but it was justified in Sec.~\ref{whySVD} that the density matrix's elements can be accessed via the SVD, so this will provide a route to obtaining the quantities necessary variationally.  Using the density matrix also has the advantage that purely real space partitioning can produce misleading results.  For example, if the Hamiltonian had been chosen as the core object to partition, then it would be found that partitioning a system such as the particle in a box would produce the first excited states as was found in early implementations of this \cite{schollwock2005density,noack1993real}.

In summary, using the density matrix is a strong candidate to find the solutions for quantum systems because it is known to only require a few states to accurately describe a system.

It is also worth noting here that the entire, large system can be renormalized to a few sites, just was attempted in Kadanoff's renormalization group.  The fact that the density matrix is a good candidate to search for the most relevant degrees of freedom is linked to the ability for a tensor network to be represented with only a few sites.  This is one of the reasons that DMRG will be able to solve very large systems.

\subsubsection{Area law of entanglement and locality}\label{arealaw}

The study of entropy in the context of black holes led to a very novel concept, known as the area law \cite{eisert2010colloquium,kuwahara2020area}. This concept appears in tensor networks to justify why only a few states need to be kept yet still accurately describe the ground state wavefunction.  

In order for black holes to satisfy the second law of thermodynamics \cite{reif2009fundamentals}, some radiation should be emitted. In computing the entropy of the black hole under these conditions, Hawking found that the entropy of the black hole was proportional to the area of the black hole \cite{hawking1976particle}.

This connection between a black hole's entropy and its area caught the imagination of the community and in particular those working on conformal field theory. There is a deep connection between theories that admit conformal transformations and quantum theories of gravity.  This should not take up too much of the discussion here, but it is sufficient to state that information in a quantum system is bound by some upper limit and that physical systems with local correlations have a definite, useful relationship.  Knowing this fact has a direct application in tensor networks. It will turn out that the amount of information that can be passed from one site to another can inform the graph that a tensor network will take. 

It turns out that physical Hamiltonians are inherently local and states at the edge of the eigenvalue spectrum follow the area law. A proof by Hastings demonstrates that there are only two types of correlations: gapped and gapless \cite{hastings2004locality}. If a gap in the eigenvalues is present, then the correlations will be exponential in nature. If there is no gap between eigenvalues, then there is a power law decay for the correlation functions \cite{bakerCJP21,*baker2019m}. This effectively provides a proof of Kohn's near-sightedness principle \cite{kohn1996density,prodan2005nearsightedness}.

There are notable counter-examples to the discussion above \cite{movassagh2016supercritical,*movassagh2014power} but the principle holds for a many physical Hamiltonians. There are other conventional cases where the area law does not apply. For example, when investigating states in the bulk where the states scale as a volume law instead \cite{abanin2019colloquium}.  But the largest singular values represent the shortest-range entangled states and these states typically denote shorter range interactions which should be kept since they contribute more to ground states.

This can be stated in another way.  Of the exponentially many states in the entire Hilbert space, ground states for physical, gapped Hamiltonians tend to be found amongst exactly the few states that are described well by the MPS.   This provides a strong reasoning to tensor network to describe ground state efficiently, provided the correlations are local.  A quantum problem can be decomposed in a certain way to exploit the use of short-range interactions.

\subsubsection{How many states to keep?}

In practice, the coefficients $\rho_k$ decrease very quickly for systems with a gap in the lowest energies.  Even when the system has no gap, and the decay is much slower, it is still useful to think about how many states should be kept to retain an accurate density matrix.  The idea of keeping the most relevant degrees of freedom here is where Kadanoff's idea of renormalization enters the problem.

The number of states kept in the density matrix simply be referred to as "states" and the variable used to denote the number of states is typically assigned as $m$.  The geometry of the tensor network directly affects the size of the bond dimension, and by taking into account the properties of the the correlations in the connectivity of the graph representing the tensor network, the best representation can be sought \cite{bakerCJP21,*baker2019m}.  For here, only the MPS structure will be developed in this document.

It turns out that the number of states that must be kept to maintain an accurate representation of the full ground-state is not that much.  For even a 100-site spin-half system, 50 states gives an accurate energy in DMRG out to several digits. For this system and gapped systems in general, the singular values decay exponentially in value.  Likewise, for gapless systems, the singular values decay as a power law.  This means the method can be run on a laptop, although larger computations will take more dedicated computing services.

Crucially, the most important singular values are the largest and this implies that the short-range entangled components will be captured by the largest singular values.  By truncating some of the singular values, the effect is to limit the ultimate range of the entanglement captured by the final ground state.  This constitutes the effective range of entanglement that is captured by DMRG.

The maximum bond dimension that is typically sought in DMRG calculations is on the order of 30,000, but this is generally reserved for the hardest models.  Even very challenging two-dimensional models can be suitably handled with a few thousand many-body states kept.  There will be automatic methods to determine how to accurate a system is with a certain number of states kept.

\subsection{Who made the density matrix renormalization group?}

The fundamental contributors Kadanoff and Wilson were already mentioned, although many contributed to renormalization group concepts.  The story goes that Wilson gave a lecture and in the audience was Steven R. White \cite{white1999all} who took note of the some of the ideas for extending NRG to more than one site.

In a series of papers attempting to use real-space renormalization group methods, both Reinhardt Noack and White were actively working on developing these methods \cite{noack1993real}. At some point, allegedly while reading the Feynman Lectures \cite{feynman1965feynman}, it was realized that the density matrix would encode all the useful information in the system. White then published the DMRG algorithm \cite{white1992density,white1993density}. The original suggestion to use the density matrix in computatations was a guess, but it can be justified with the arguments above.

Shortly after the original DMRG article, Rommer and Ostlund published a paper connecting the DMRG algorithm to the matrix product state and AKLT wavefunctions \cite{ostlund1995thermodynamic,rommer1997class}.  This spawned a new way of annotating the method which helps keep track of the steps in a much more systematic manner.  This is the modern formulation that is used here, although the reader is encouraged to become familiar with the tensor notation to be able to read papers that are motivated from the other perspective.

The extension to the MPS formalism allowed for many proofs to be written using these developments. Many of these proofs were written by those working for Ignacio Cirac and Frank Verstraete. Another very notable contribution came from Guifre Vidal pertaning to the gauges of the MPS and the extension of the method to the multi-scale entanglement renormalization ansatz (MERA) \cite{vidal2008class}, although this will be covered in a subsequent paper. It should be noted that some proofs that are useful here by Matthew Hastings are exceptionally useful for understanding where the DMRG algorithm will work well and where more computational resources should be put to it \cite{hastings2004locality}.  All of the developments in this paragraph also have some connection to quantum computing or form useful ideas in that subfield.

The list given here of contributors to the theory is far from exhaustive, and the reader is encouraged to look at the fundamental literature to get ideas from all different angles even if they are not mentioned here.

Having explored the historical background of the method, the two major concepts of density matrices and renormalization group are surveyed before beginning to implement the core functions for the library.

\section{Tensors and operations}\label{tensorsAndOps}

In its simplest form, a tensor is a multidimensional array. Tensors are defined as a map from one vector space to another \cite{boas2006mathematical}.  While this definition is not wholly operationally useful, there are a few discussion points that can illuminate what these objects are.

This section will review the basic operations as outlined in Ref.~\onlinecite{bakerCJP21,*baker2019m}: permuting, reshaping, contracting, and decomposing tensors.  In order to implement these operations, a basic structure to manipulate them in a useful way must be defined, and this is done first.  Note that two operations (permuting and reshaping) manipulate a single tensor.  The other two operations (contraction and decomposition) involve more than one tensor.  All contractions can be composed of contractions between two tensors. Decompositions decompose a single tensor into 2 or more tensors.  For both contraction and decomposition, the problem is most efficiently treated by constructing a rank-2 tensor by permuting and reshaping the original tensor.


\subsection{Diagrammatic notation}

The mathematical notation for a tensor can be very cumbersome to treat analytically.  However, the field has chosen to use a graphical depiction of the tensors in order to readily transmit the information. The upper or lower indices, common in general relativity \cite{schutz2009first} and particle physics \cite{peskin1995quantum}, do not have a meaning typically in a tensor network computation.  The assignment of covariant and contravariant indices are not as useful here, especially once the diagrammatic notation is introduced. 

The diagrammatic notation is owed to Penrose who first introduced it in the context of quantum field theories \cite{penrose1971applications}.  In the diagrams, the tensor indices can be dropped in all cases.  It is very rare that a computation would require the preservation of this information.  Instead, it is always understood that indices have a partner that they are to be contracted onto.  If the index is to be changed, then it will be explicitly noted in an algorithm.

Each index of a tensor has a certain dimension.  The number of indices is known as the rank of the tensor.  A rank-3 tensor is denoted as
\begin{equation}
\includegraphics[width=0.35\columnwidth]{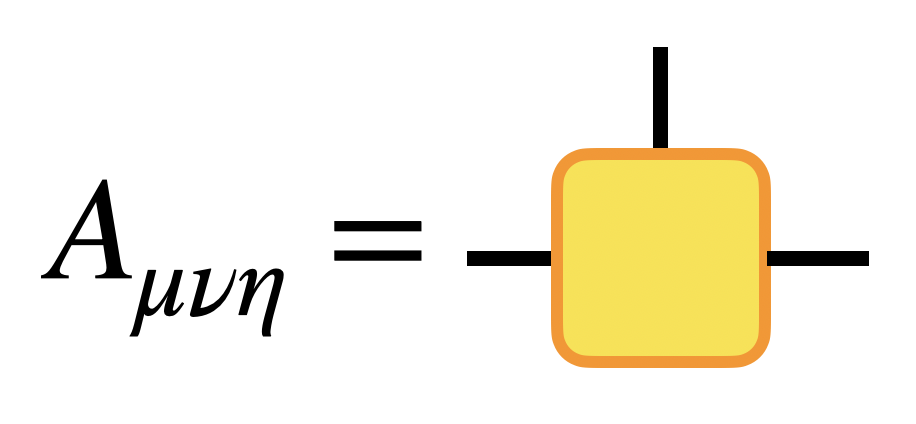}\nonumber
\end{equation}
where one simply counts the number of indices to recover the rank. A tensor of rank 2 is known as a matrix
\begin{equation}
\includegraphics[width=0.35\columnwidth]{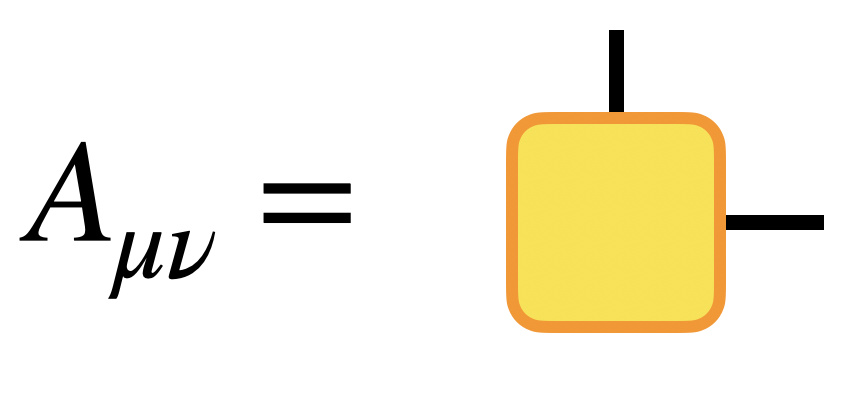}\nonumber
\end{equation}
A tensor of rank 1 is known as a vector
\begin{equation}
\includegraphics[width=0.35\columnwidth]{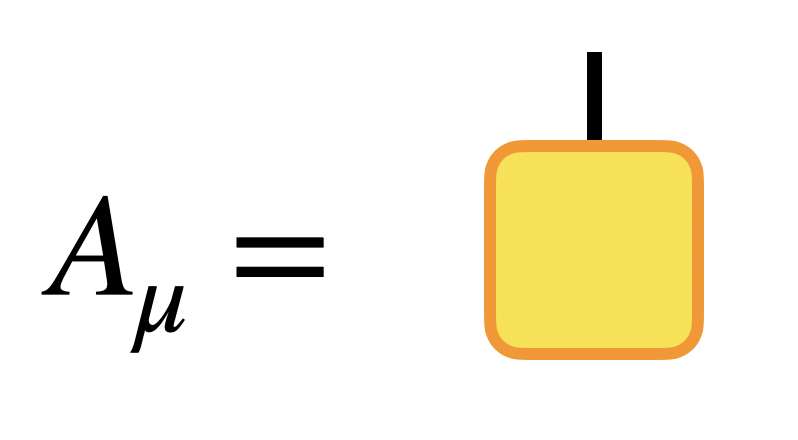}\nonumber
\end{equation}
The relative orientation of the indices does not matter for the tensors in general.  However, a convention will be imposed for the MPS tensors.  The vertical indices correspond to the physical degrees of freedom for a given model and the horizontal indices corresond to the link indices \cite{bakerCJP21,*baker2019m}.

\subsection{Defining tensors}

The main difference from Julia's internally defined tensor is that the rank of the tensor is not explicitly defined in the {\tt tens} object, the basic object for tensors in DMRjulia. This is because the rank of a tensor is an ever-changing quantity as can be seen with the basic operation of reshaping \cite{bakerCJP21,*baker2019m}.

There are a few functions from the Julia library that are coded for the {\tt denstens} class and are included in the Supplemental Material \cite{tensor_recipes}.  These correspond to the functions in Julia that are most relevant for the following.

\begin{lstlisting}[numbers=none]
mutable struct tens{W<:Number}
  size::Tuple
  T::Array{W,1}
end
\end{lstlisting}

There are only two fields that must be defined for a tensor.  The elements are stored in a field {\tt T} which is always a vector and the size of the tensor is stored in {\tt size}. The product of the {\tt size} field must be equal to the number of elements in the {\tt T} field.  So, for example, {\tt prod(A.size)==length(A.T)} should always be true.  It may appear unnatural to always store a tensor as a vector, but it turns out that the concept of the tensor's rank is very fluid in generic tensor network algorithm.  It is not worth keeping track of explicitly except to change the {\tt size} field appropriately.

There is one other reason to always store the tensor as a vector. The computer can more easily access a vector if the data is stored in sequence, making this a faster operation (as opposed to some implementations of sparse matrices, for example).  There can be other ways to store a tensor, but this method will be simple without sacrificing too much efficiency.

\subsection{Basic operation: permutation}

\begin{figure}[b]
\includegraphics[width=\columnwidth]{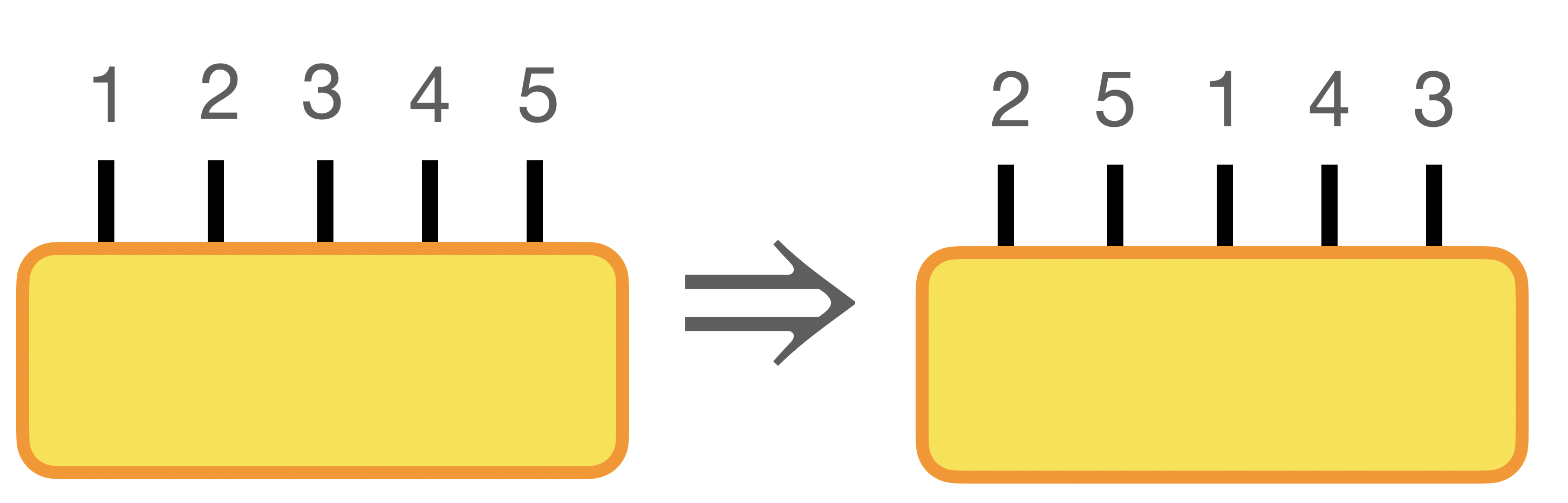}
\caption{A rank-5 vector is permuted. \label{permuteFIG}}
\end{figure}

Permuting a tensor is an operation that is used frequently in practice.  Several advanced libraries exist to implement this function and these can take advantage of properties of how the data is represented on a computer in several ways including memory management and other useful hurdles to overcome.  Very simply, the indices of the input tensor are permuted as in Fig.~\ref{permuteFIG}.

There are very efficient ways in which to program the functions.  The general idea is that the element of an array can be converted into the position of a tensor and then those position values can be swapped according to some reordering provided to the function. There are optimizations in terms of how the memory is laid out on  the computer, so it is recommended to use one of the existing library functions to use this function.

The syntax chosen here once again matches Julia's input functions for its own {\tt Array} class. The front-end is
\begin{lstlisting}[numbers=none]
pA = permutedims(A,[2,5,1,4,3])
#orm (in-place)
A = permutedims!(A,[2,5,1,4,3])
\end{lstlisting}
Note that permuting the tensor twice will result in the original tensor being returned, since the permutation is a representation of the permutation group.

There are in-place algorithms (denoted by {\tt !} at the end of the function name) for the permutations of tensors, but these tend to cost more time and are often not worth the trouble of coding or using. An in-place version is available in DMRjulia, but this calls the same function as the copy-based implementation.  As it is here, the cost to using a permuation is equivalent to the number of elements stored in the tensor, so this can be somewhat costly if used repeatedly and should be minimized, although this is not the largest concern when programming algorithms.

\subsection{Basic operations: reshape}

\begin{figure}[b]
\includegraphics[width=0.5\columnwidth]{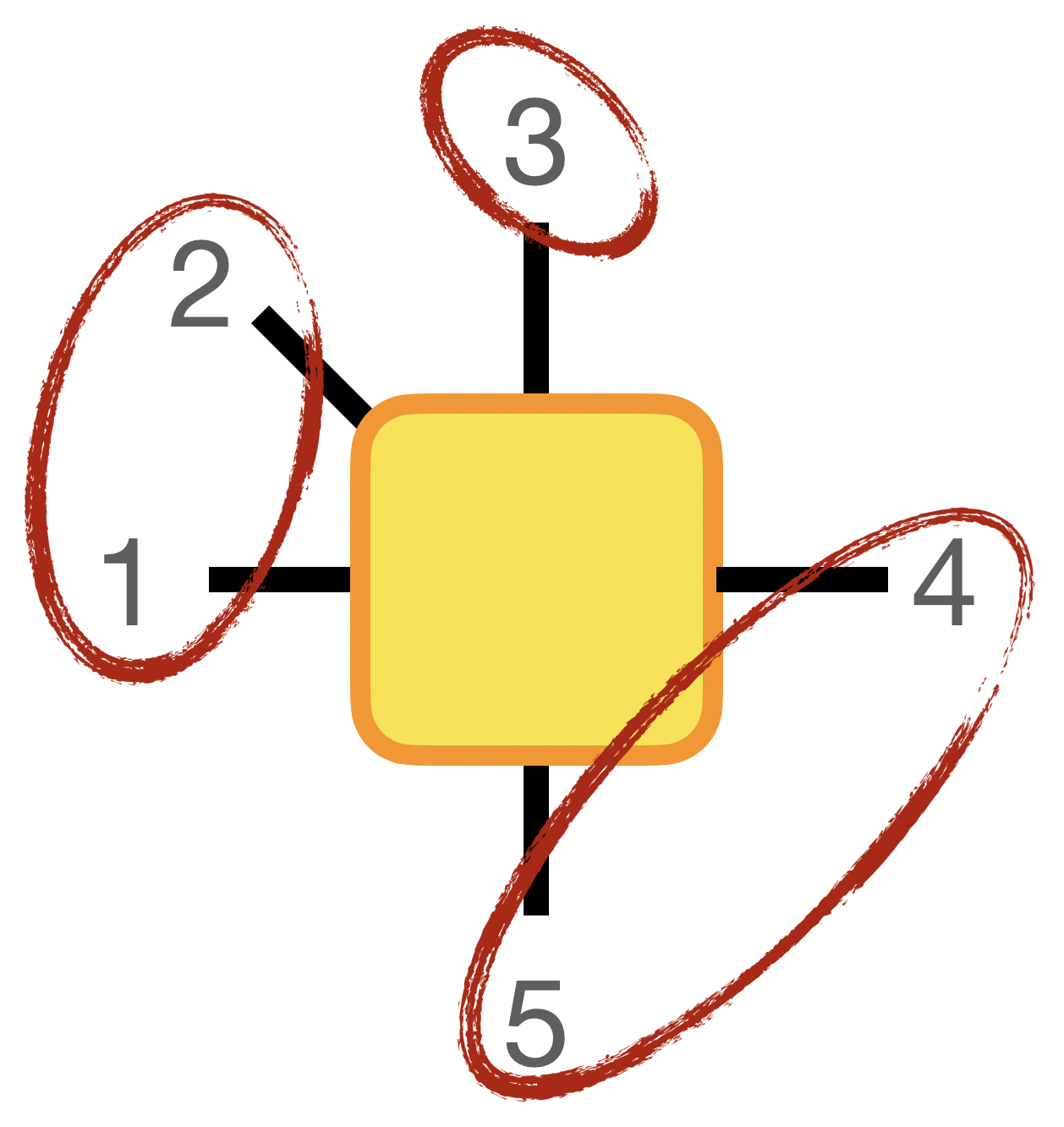}
\caption{An example of a reshape of a rank-5 tensor where there is a resulting rank-3 tensor. The numerals on each index correspond to the example in the text and are usually not shown on a diagram. \label{reshapeFIG}}
\end{figure}

As to how to reshape the tensors, it is sufficient to note that whole indices are reshaped together in a tensor network computation.  Splitting an index is not an operation that needs to be defined in general, so often the tensor will be reshaped (for the example of a rank-3 tensor) as $A^{\sigma_1}_{a_0,a_1}=A_{(\sigma_1a_0),a_1}$ where $\sigma_1$ and $a_0$ were joined together. Note that while the index was lowered in the notation on the page, the indices were not re-ordered. 

Note that any number of indices of size 1 can be reshaped onto a tensor with no issue.

With regards to the discussion in Sec.~\ref{partialtraces} (and extending the example in Ref.~\onlinecite{bakerCJP21,*baker2019m}), the reshaping operations can select the basis functions defined on the left and right sides of the cut of a particular model.  The physical significance of this is that the indices for a particular site are grouped together with this operation. This essentially defines the cut taken on a particular lattice when performing a decomposition, although the function has many other uses \cite{bakerCJP21,*baker2019m}.

Reshaping a tensor has such a small cost that this operation is essentially free.

The rank of a tensor is an ever-changing quantity in a tensor network computation, and it can be noticed that in the definition of the {\tt denstens} object of the last section that only the {\tt size} field changes when a tensor's rank is modified.  Note that when performing this operation that the order of the indices on a tensor remain unchanged, so this implies that the basic vector representing the elements of the tensor, {\tt T}, does not need to be changed when permuting the tensor.

The operation can be simply input into the function {\tt reshape} as
\begin{lstlisting}[numbers=none]
#A is a rank-5 tensor
A = tens(rand(10,20,30,10,40))
Lsize = size(A,1)*size(A,2)
Csize = size(A,3)
Rsize = size(A,4)*size(A,5)
rA = reshape(A,Lsize,Csize,Rsize)
# or (in-place)
A = reshape!(A,Lsize,Csize,Rsize)
\end{lstlisting}
The implementation here, as opposed to the implementation in Julia's basic array type includes no checks that the reshape is correct.  For debugging, one can always put the regular Julia {\tt Array} type into these functions (as they are overloaded on the basic functions defined in the library) to get the checks or by manually viewing the input and output tensors with the {\tt println} function.

Another input style for the reshape function is also available in the library, and this alternative can be much easier when using the reshape function.  Writing a line such as
\begin{lstlisting}[numbers=none]
rA = reshape(A,[[1,2],[3],[4,5]])
#or
rA = reshape!(A,[[1,2],[3],[4,5]])
\end{lstlisting}
and corresponds to the diagram in Fig.~\ref{reshapeFIG}. This performs the same reshape as the above block of code, but the sizes of the indices are computed explicitly by the function.  Note that the numbers in the vectors must appear sequentially or the original tensor will be permuted.

As it pertains to the implementation for dense tensors, another function {\tt unreshape} is defined in the library for ease of reading.  However, for the subsequent quantum number version, this function will have a different operation on the input tensor.  So, {\tt unreshape} performs the same as {\tt reshape} here.

\subsection{Basic operation: contraction}

\begin{figure}
\includegraphics[width=0.75\columnwidth]{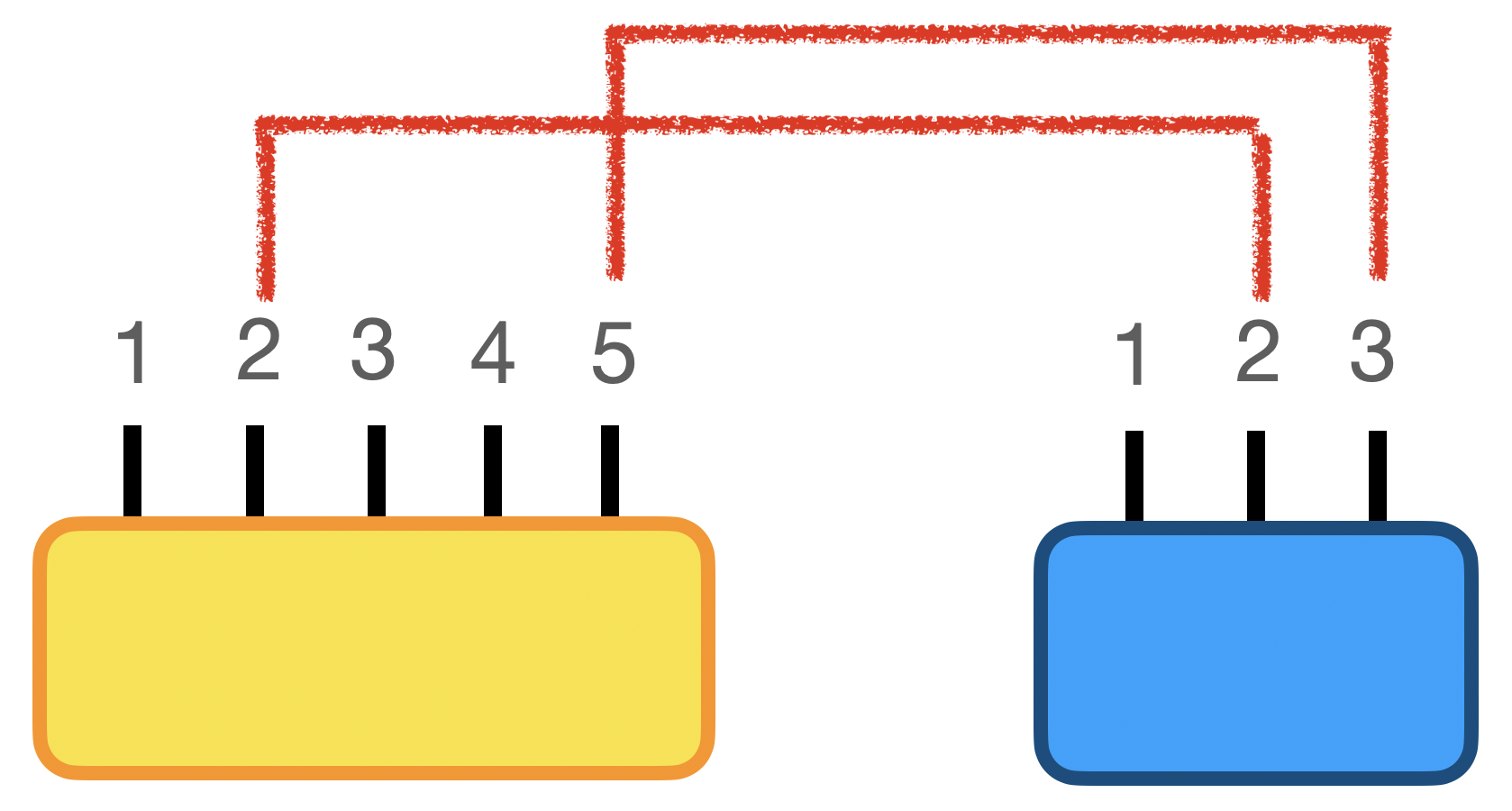}
\includegraphics[width=0.75\columnwidth]{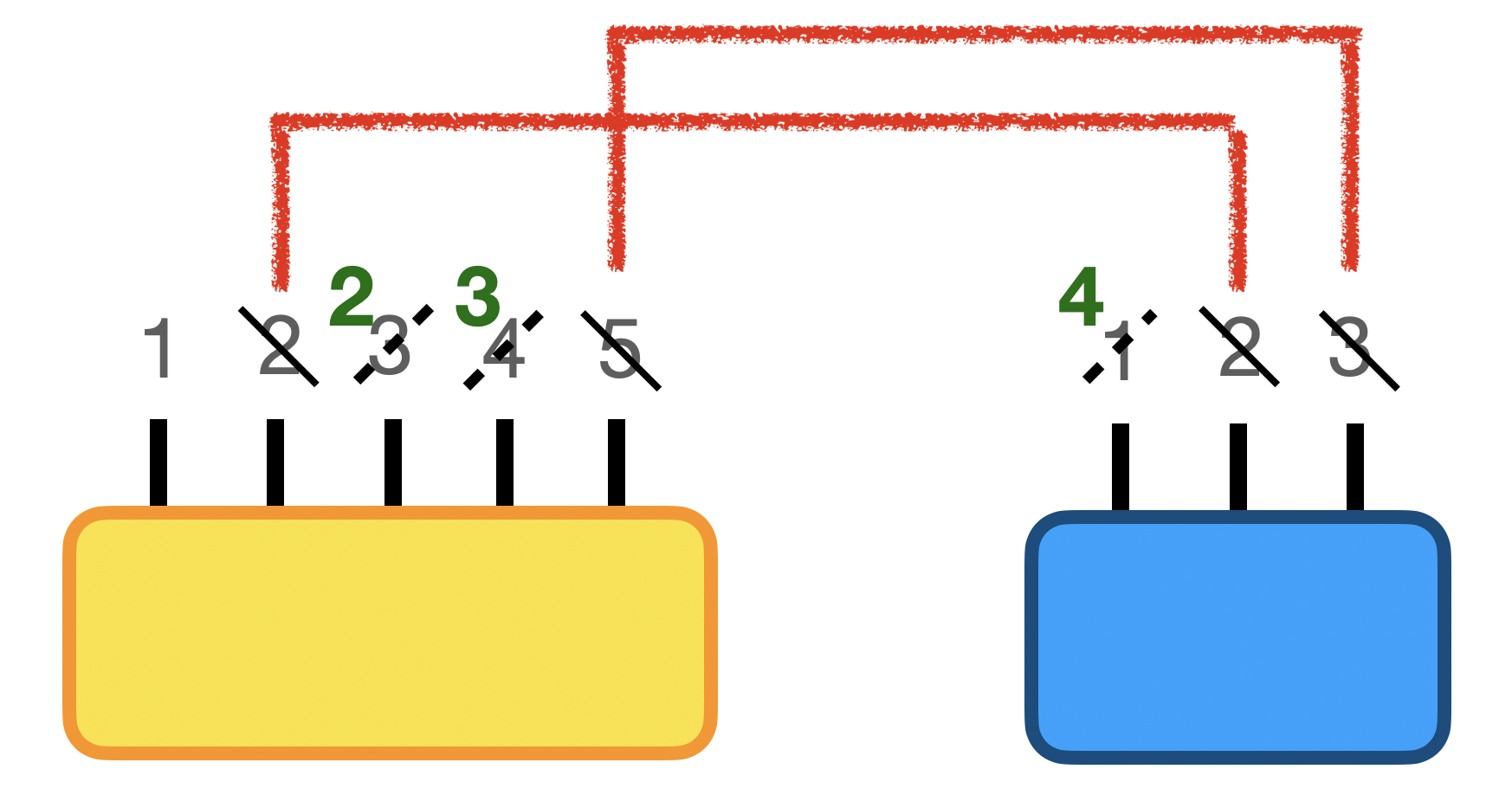}
\caption{Indices on a rank-5 tensor are contracted with indices on a rank-3 tensor in the first diagram.  The second diagram shows how the indices are relabeled (always from left to right). Number crossed out with a solid like are contracted out and not represented on the final tensor. Dashed lines are relabeled from the left to right tensor in order. It is recommended to draw a diagram similar to this for each contraction to master how the indices are relabeled for a given operation. The result of the contraction is a rank-4 tensor as shown in Eq.~\eqref{contractex} \cite{bakerCJP21,*baker2019m}.\label{contractionFIG}}
\end{figure}

A contraction over matching indices on two tensors is a central operation in any tensor network algorithm.  At the most basic level, this operation is required to recombine the tensors in the problem to the original problem, but it will play a fundamental role when constructing the reduced site representations of Schr\"odinger's equation.  A contraction displayed in Fig.~\ref{contractionFIG} can be represented with the summation notation as
\begin{equation}\label{contractex}
C_{\alpha\gamma\delta\zeta} =\sum_{\beta\gamma} A_{\alpha\beta\gamma\delta\eta}B_{\zeta\beta\eta}
\end{equation}
where matching indices are contracted over and have the same dimensions. The use of raised and lowered indices are not used here and typically the explicit summation of the indices is not written.

The question is then how to compute the contraction on a computer in the most efficient way.  The most naive way to accomplish this (but leads to a lot of code) is to manually code a series of for loops that contain all the required summations required.  This is very inefficient because many functions must be defined, the code will not be extendable to all possible situations, and this strategy is difficult to code in all cases to be as fast as another, more simple method.

The two, single-tensor basic operations of reshaping and permuting can be used to converted the tensors to matrices in all cases.  Then, very fast algorithms that have been refined over decades can be used to compute the resulting matrix multiplication in an efficient way. The existing libraries are known as LAPACK or BLAS, although there are many that can be used.   The matrix representation of a tensor that has a defined permutation and reshape (or simply two sets of indices) into a matrix will be referred to as the matrix-equivalent of a given tensor.  This will be how all of the contractions and decompositions in the next section are treated.

Because all tensors can be reshaped into a rank-2 tensor (matrix), all tensor contraction operations can be converted to a matrix-matrix multiplication.  The process of recovering the matrix form of a tensor with defined indices will be referred to as finding the matrix-equivalent of a given tensor.  It will often be necessary to also permute indices in order to obtain the correct ordering to find the matrix-equivalent.

Once the matrix-equivalent is formed, the question becomes how to accomplish the matrix multiplication efficiently.

Contraction of more than one tensor is not necessary, but it is sometimes useful to also add a leading coefficient and the option to add another tensor to the original. This resembles the "axpy" function present in some other programs for matrix multiplication \cite{press1992numerical} and has the form
\begin{equation}
C=\alpha A\cdot B+\beta Z
\end{equation}
where the dot here signifies tensor contraction over specified indices and $\alpha$ and $\beta$ are scalars.  Note that $Z$ must be input and shaped appropriately for the resulting contraction of $A$ and $B$.

\begin{lstlisting}[numbers=none]
C=contract(A,2,B,4,Z,alpha=2.,beta=9.)
\end{lstlisting}
which returns a tensor $C$.

There are four types of inputs for the contracted indices: tuples, arrays, rank-2 arrays (which are interpreted as $1\times m$ arrays or $m\times1$ arrays always), and integers.  Thus, the following all represent valid contraction inputs
\begin{lstlisting}[numbers=none]
contract(A,(1,2),B,(6,3)) #tuples
contract(A,[1,3],B,[2,4]) #rank-1 array
contract(A,[1;3],B,[2;4]) #rank-2 array
contract(A,2,B,2,alpha=2*im) #integers
\end{lstlisting}
or any combination of the types given above and provided that the contractions themselves are valid. As a general rule, the indices that are matched in the contraction function must match in size.  When kept track of on the page, any reshaped index (for example, two indices that are joined together), must be in order before being reshaped together for the contraction algorithm.  The function here will automatically perform this formatting. 

There are several additional aspects of the contractions that can be implemented for ease of use.

A discussion of how to compute the computational complexity is central to identifying how efficient an algorithm is, but the full discussion of how to do this is reserved for Appendix~\ref{complexity}.  For now, some useful functionality in the {\tt contract} function is discussed.

\subsubsection{Conjugation of input tensors}

When forming the matrix equivalent, there is a natural opportunity to conjugate the elements of a tensor, and this can be easily encoded into the contraction operation by extending the {\tt contract} to conjugate either the left, right, or both input tensors. This can save one copy of the input tensor by not conjugating beforehand.

\begin{lstlisting}[numbers=none]
ccontract(A,2,B,5) #conjugates A
contractc(A,2,B,5) #conjugates B
ccontractc(A,2,B,5) #conjugates A and B
\end{lstlisting}
Mostly this functionality is useful to shorten input code and as a short-hand.

Note that this form of the contraction has a particular interface with contracting the adjoint of an operator.  If the trivial computation $\hat U^\dagger\hat U$ were to be taken, then the following code would produce the identity matrix
\begin{lstlisting}[numbers=none]
#definition of U of rank 4
#the last index is the auxiliary index
#introduced on SVD
ccontract(U,[1,2,3],U,[1,2,3])
\end{lstlisting}
The inputs are the same despite one tensor being the adjoint! Unlike multiplying the two tensors on the page or in strict matrix multiplicy where the transpose of one of the tensors must be taken, the indices to contract are the same for both tensors. The only difference is that one is conjugated.  This can be kept in mind by remembering that indices have a specific character and must match when contracted.  The same indices on one tensor with the same indices on another.  Typically, these correspond to the same physical indices on the similar sites or matching auxiliary indices from the SVD.

\subsubsection{Reordering the output}

A vector can be added as the first field in any of the {\tt contract} functions and this will cause a permutation of the final tensor,
\begin{lstlisting}[numbers=none]
#permute output
contract([1,3,2,4],A,[1,3],B,[2,4])
\end{lstlisting}
This functionality is again useful for shortening code.

\subsubsection{Froebenius norm of a tensor}

The Frobenius norm of a tensor is simply the sum of all elements squared in the tensor and then the square root of the result.  This is equivalent to the $\mathcal{L}^2$ norm of a vector which the tensor can always be reshaped into.

Sometimes for checking, it is useful to compute the norm (scalar return value) and this can be accomplished with the {\tt norm} function or by 
\begin{lstlisting}[numbers=none]
contract(A) # =norm(A)
\end{lstlisting}
or any of the {\tt ccontract}, {\tt contractc}, or {\tt ccontractc} which will compute the contraction of $A$ with itself and either the left or right copy of $A$ is conjugated according to the function called as previous.

\subsubsection{Automatic contraction}

It is sometimes useful to simply contract all of the indices automatically.  This can be accomplished by assigning names to the indices and coding some algorithm to compute the best contraction.  This allows for an extension to more tensors without explicitly computing the indices to contract each time. 

The automatic contraction routine in the library, and how to best use it, will be discussed in a subsequent paper.  Many simple algorithms can be programmed using the tools presented here, although some more complicated models will benefit from this addition.

It is not recommended to learn how to make tensor network algorithms simply with this functionality.  It is best to master how to use the contraction algorithm with the diagrammatic procedure above and then switch to using an automatic contractor.  It is noticed that those using this style typically form a better picture of how to contract a tensor network than otherwise.

\subsection{Basic operation: decomposition}

\begin{figure}
\includegraphics[width=0.65\columnwidth]{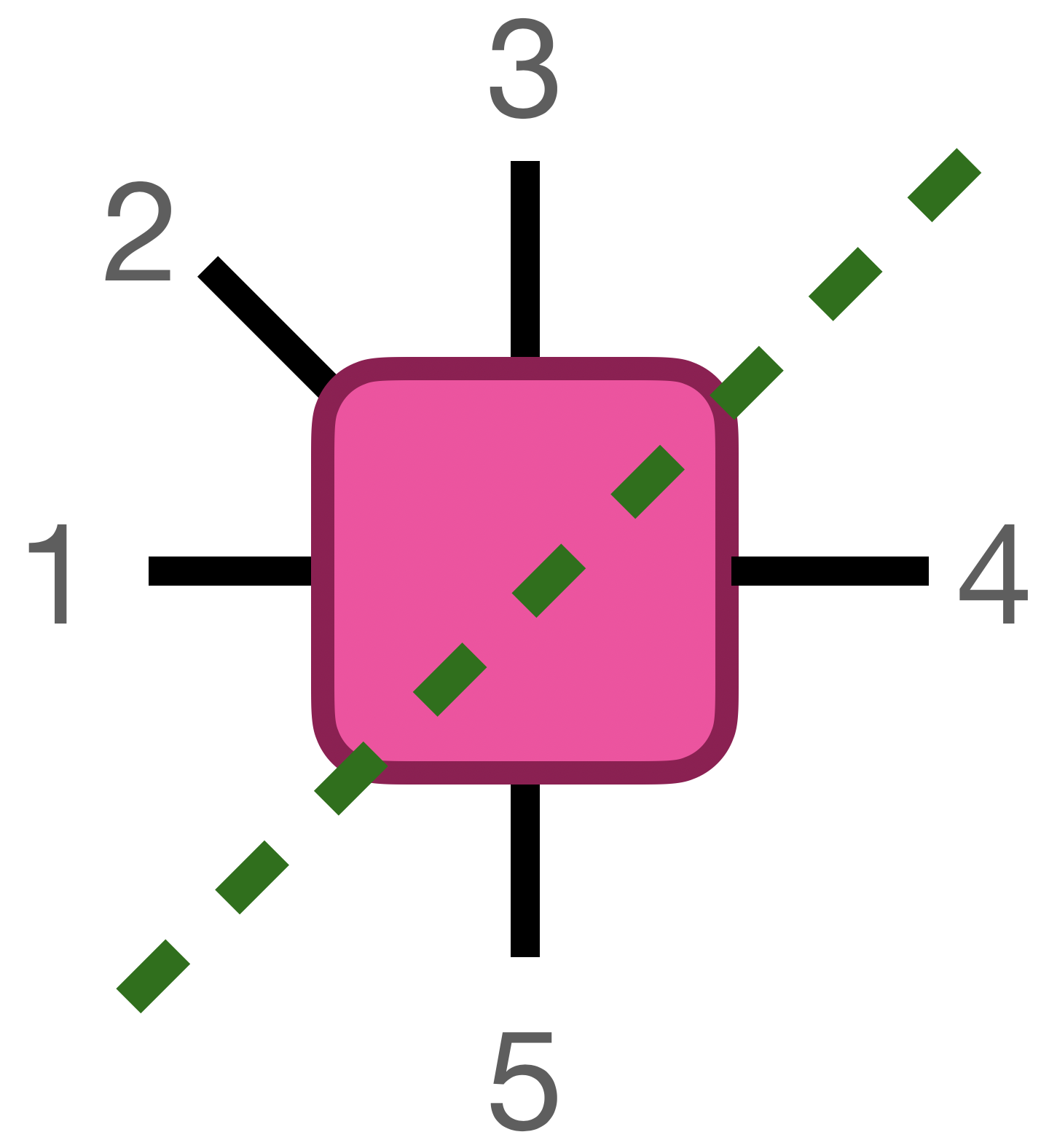}
\caption{A rank-5 tensor is partitioned into two groups of indices (those above and below the dashed line). The tensor is then decomposed into either two or three matrices depending on the decomposition used. \label{decomposeFIG}}
\end{figure}

As a competing operation to the contraction of tensors, there is also the possibility to split a tensor into two.  To do this, a matrix-equivalent is defined with two groups of indices as illustrated in Fig.~\ref{decomposeFIG}. The resulting tensor is then decomposed.  The example for the first, and most useful, decomposition is shown in Fig.~\ref{decomposeFIG} and was motivated in Sec.~\ref{whySVD}.

A common feature of these decomposition techniques is to group together two sets of indices to form a matrix equivalent.  Two groups of indices are picked, and the matrix-equivalent is decomposed (the resulting tensors can be then be unreshaped appropriately).

The result of the decompositions is to introduce a new index.  The size of this new index is often called the link index and is drawn horizontally on diagrams, although the strict use of the term bond dimension can apply to the size of other indices, so the meaning may change in some locations.  The most common abbreviation for the bond dimension is either $m$ or $\chi$.  Here, $m$ is used for the physical tensors and $\chi$ is used in Apx.~\ref{complexity} as the more common variable for complexity estimates.

\subsubsection{Singular value decomposition}

 A general rectangular matrix, $\hat M$, of dimensions $a\times b$ can be decomposed into the following form
\begin{equation}\label{svdform}
\hat M=\hat U\hat D\hat V^\dagger
\end{equation}
and this forms the most ubiquitously used decomposition in tensor networks. The decomposition is fully generic.  It will work for any grouping of indices of any size.  The resulting tensor introduces a new index into the system.  The physical significance of the elements of the diagonal $\hat D$ matrix is that they are the square roots of the positive-definite eigenvalues of the density matrix.  By decomposing a wavefunction $\psi$ with the SVD, the elements of the density matrix are automatically obtained as motivated in Sec.~\ref{whySVD}.

The tensors $\hat U$ and $\hat V$ will be referred to as unitaries without loss of generaity.  Technically, these are isometries (which are also called "row-unitary") since they are rectangular matrices \cite{bakerCJP21,*baker2019m}.  However, since they contract to the identity, they will be referred to here as unitaries to correspond to how they would be referred to in the eigenvalue decomposition of the density matrix. In full, this convention is chosen to match several pre-existing publications, but the distinction is very minor.

There are many definitions of the SVD \cite{press1992numerical}, all involving different sizes of the $\hat D$ matrix ($a\times m$, $m\times b$, etc.).  Appealing to the discussion of the information passed being symmetric from left to right as from right to left around Eq.~\eqref{informationsymmetry}, the size of the $\hat D$ matrix will be $m\times m$, causing $\hat U$ to be of size $a\times m$ and $\hat V$ to be $m\times b$.

There are a few ways in which a singular value decomposition can be computed.   One is to take the input matrix $\hat M$ and compute both (via eigenvalue decompositions) $\hat M\hat M^\dagger=\hat U\hat D^2\hat U^\dagger$ and $\hat M^\dagger \hat M=\hat V\hat D^2\hat V^\dagger$.  Computing the resulting square matrix $\hat M\hat M^\dagger$ or the adjoint is expensive.  

Another method to use is to write the super-matrix
\begin{equation}\label{superSVD}
\check M=\left(\begin{array}{cc}
\hat 0 & \hat M^\dagger\\
\hat M & \hat 0
\end{array}\right)
\end{equation}
with appropriately sized diagonal block matrices. Diagonalizing Eq.~\eqref{superSVD} gives two super-matrices $\check U$ and $\check D$.  The elements of the $\hat D$ matrix are contained in the upper left $m\times m$ block of the resulting eigenvalue super matrix,
\begin{equation}
\check D=\left(\begin{array}{cc}
\hat D& \cdots\\
\vdots&\ddots
\end{array}\right)
\end{equation}
where $m$ is chosen to be the minimum of the rows and columns of $\hat M$. The off-digonal blocks of the resulting $\check U$ matrix of the eigenvalue decomposition relate to $\hat U$ and $\hat V^\dagger$ as
\begin{equation}
\check U=\frac1{\sqrt2}\left(\begin{array}{cc}
\hat V & \cdots\\
\hat U & \ddots
\end{array}\right)
\end{equation}
where dots signify the other parts of the super-matrix.  The use of a pre-existing library is often key to making sure that this operation is efficient.  The BLAS and LAPACK libraries are two well known, long-standing and stable libraries that implement basic linear algebra operations. The second version of the SVD is preferred because it can be efficiently encoded into the existing libraries.  It is highly unlikely that a reasonable amount of research time will significantly improve on these implementations, so they should represent the maximum of efficiency and be trusted.

On some occasions, the default implementations of the SVD in some codes will require the implementation of these alternative methods of solving for the components of the SVD. This is largely a technical problem that can be encountered in some cases.  Also note that the SVD will calculate the precision of all singular values with relation to the value of the first singular value output.  One can re-enter the SVD into the algorithm recursively \cite{stoudenmire2013real}, selecting a subset of the matrix if more precision is needed.  Often, this will not be necessary.

There are two inputs that can be used to reshape the tensor.  One is to reshape the tensor manually, decompose the tensor, and then unreshape the resulting tensors.
\begin{lstlisting}[numbers=none]
sA = size(A) #A is a rank 4 tensor
Lsize = size(A,1)*size(A,2)
Rsize = size(A,3)*size(A,4)
rA = reshape(A,Lsize,Rsize)
rU,D,rV = svd(A)

newUsize = (sA[1],sA[2],size(rU,2))
U = unreshape!(rU,newUsize...)
newVsize = (size(rV,1),sA[3],sA[4])
V = unreshape!(rV,newVsize...)
\end{lstlisting}

Frequently when performing decompositions, it is more convenient to provide two vectors that contain the reshape.  So, the equivalent function is
\begin{lstlisting}[numbers=none]
U,D,V = svd(A,[[1,3],[2,4]])
\end{lstlisting}
will permute the indices according to the order {\tt [1,3,2,4]}, reshape the first two and last two indices together, perform the SVD, and then unreshape the resulting $\hat U$ and $\hat V^\dagger$ matrices into tensors. Even though the return variable is {\tt V} in the code, the actual return object is $\hat V^\dagger$ from the SVD.

\subsubsection{Truncated SVD  (Schmidt decomposition)}

With the exact same decomposed form as Eq.~\eqref{svdform}, the Schmidt decomposition can be used to truncate the SVD, leaving only the most relevant states in the density matrix.  In most cases, using the terms SVD and Schmidt decompositions are interchangeable, and after this brief mention, this operation will be called the SVD and occasionally the truncated SVD.

The truncated SVD then selects rows and columns of $\hat D$ corresponding to the largest values.  Then, the matching columns of $\hat U$ and rows of $V^\dagger$ are selected.  The remaining rows and columns are then discarded.

In order to call a truncated SVD, the command (bond dimension of 10 and cutoff of $10^{-6}$)
\begin{lstlisting}[numbers=none]
out = svd(A,cutoff=1E-6,m=10)
U,D,V,truncerr,mag = out
\end{lstlisting}
and optionally a vector such as {\tt [[1,3],[2,4]]} can be input as the second argument to permute and reshape (and then unreshape) the input tensor into a new form.

The two new outputs beyond the $\hat U$, $\hat D$, and $\hat V^\dagger$ matrices convey the truncation error and the magnitude (sum of elements) of the input matrix.  To understand both, the role of the truncation must be discussed.

\subsubsection{Truncation error}\label{truncerrdiscuss}

It is natural to ask how much of the norm of a wavefunction is left off of a given system when the SVD is truncated.  From a purely mathematical perspective, one can compute the $\mathcal{L}^2$ norm of a matrix and notice that truncating small elements of the $\hat D$ matrix mostly preserves the norm.  In the current problem, decomposing a matrix with the SVD is directly accessing the density matrix, a quantity that was argued to be an efficient way to organize relevant states in Sec.~\ref{whyDMRGworks}.

One can compute the full norm of the wavefunction, $\left\|\langle\Psi|\Psi\rangle\right\|^2$ which must always be one from the basic postulates of quantum physics \cite{von1955mathematical,baker2017methods,*cervera2017IPAM}.  However, the approximated wavefunction--after performing a truncation--will contain some error which can be evaluated as \cite{bakerCJP21,*baker2019m}
\begin{equation}\label{truncerr}
\||\psi\rangle-|\tilde\psi\rangle\|^2=1-\delta\equiv\epsilon
\end{equation}
where $\delta$ is the sum of all the kept  squared singular values (density matrix eigenvalues), $\delta=\sum_{i=1}^m\rho_i$.  For DMRG, it is not necessary to keep the accumulated truncation error, but for the evaluation of some quantities in other tensor network algorithms, it may be necessary.  The third output from {\tt svd}--the output variable named {\tt truncerr} above--is the truncation error.

Truncating the singular values imposes a cutoff on the effective range of the interactions. Keeping too few states may lead to a higher energy solution that has a minimal entanglement.  So, this operation implies that the solution found will be a minimally entangled solution, but not necessarily a bad representation fo the ground state.  By increasing the number of states kept, the correct solution can be reached.

The act of truncation constitutes the main difference between tensor network renormalization operations and the use of tensor networks in other contexts such as quantum computer where the operations are unitary throughout (no truncation) \cite{nielsen2010quantum}. This operation is similar to the principal component analysis in machine learning \cite{liPRB16}.  

\subsubsection{Cutoff}

Cutting off the many body states at a certain maximum is not always useful, especially when dealing with inhomogeneous systems when the bond dimension can vary significantly from bond to bond. Keeping many small values of the density matrix can cause large slow downs in the code, and so imposing a separate method to truncate the singular values is useful to ensure that the minimum descriptive bond size is kept.  For this, the {\tt cutoff} is defined as an input to the {\tt svd} function.  The {\tt cutoff} parameter imposes an upper limit on the summed value of the density matrix eigenvalues that are discarded.  So, defining $\eta$ for the cutoff gives $\eta = \sum_{k=m+1}^{w}\rho_k$ for $w$ density eigenvalues in total. 

\begin{figure}
\includegraphics[width=\columnwidth]{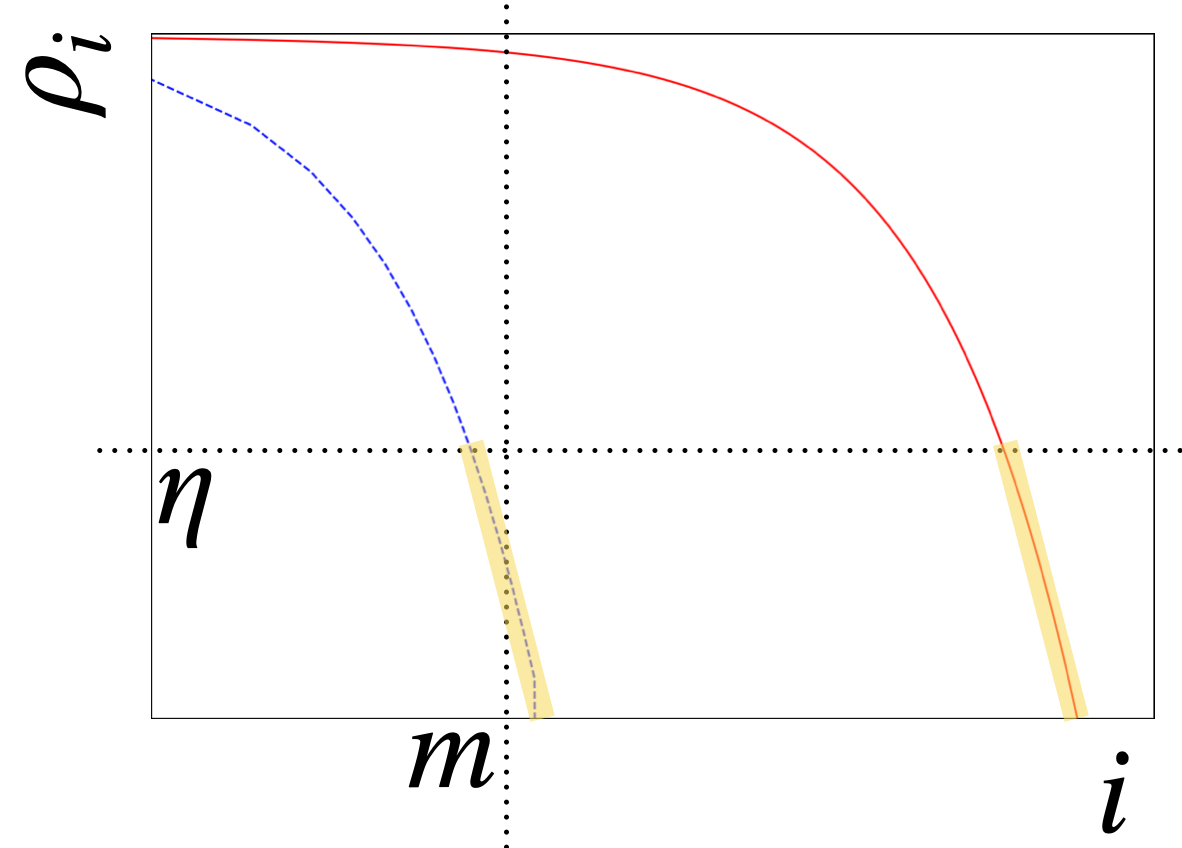}
\caption{A cartoon of the behavior of singular values for gapped (dashed) and gapless (solid line) systems. In practice, these are represented as a set of discrete points for each singular value in the SVD.  However, it is instructive to see the relationship on a log-log plot of an exponential and power law function.  The question is how much to truncate given a cutoff $\eta$ and maximum bond dimension $m$.  The horizontal line at value $\eta$ is the cutoff chosen for a given system.  The value $m$ is the maximum bond dimension chosen for a given system. The highlight on the values for the gapped (dashed line) system sum up to a value that is less than the cutoff $\eta$ despite leaving fewer values than the maximum bond dimension allowed, $m$. So, the bond can be truncated to be less than $m$. For the gapless system (solid line), the sum of the discarded values produces an error that is much larger than $m$, so the maximum bond dimension is used since it is lower. It can be noted that the power law-decaying curve tends to zero and has an exponential trend in deep into the spectrum because numerical noise and the truncation of the environment tensors in the full DMRG algorithm will impose an artificial cutoff, making the system look gapped, even though the solid line should continue on at constant slope for some distance. Actual systems will look very different and some gapless systems may even drop off dramatically, faster than some other gapped systems, depending on the coefficient involved.\label{cutoffFIG}}
\end{figure}

The concept of the cutoff is perhaps best understood diagrammatically in Fig.~\ref{cutoffFIG}.  The singular values are summed, starting from smallest to greatest, until the cutoff is reached.  Then, if the resulting inner bond dimension would be smaller than the maximum bond dimension specified ($m$), than the bond dimension determined from the cutoff is used.  At no point is the bond dimension ever allowed to be greater than $m$.

The full condition for the SVD will be that if the sum of a number of the lowest singular values squared is less than the cutoff, then they will be discarded. The default for the cutoff is 0 (no truncation) and 0 for the maximum bond dimension {\tt m} (no truncation). 

The {\tt mag} parameter was added in this implementation of the SVD because it was noticed that when computing the cutoff for tensors with norm not equal to 1, then it was best to choose a cutoff that scales with the size of the tensor.  It is not too computationally inefficient to compute this during the SVD (the loop time to do this even for large tensors is only a fraction of the rest of the algorithm).  Nevertheless, if it is known what the norm of the input tensor is beforehand, then the optional parameter {\tt mag} can be set to that value.

\begin{figure}
\includegraphics[width=\columnwidth]{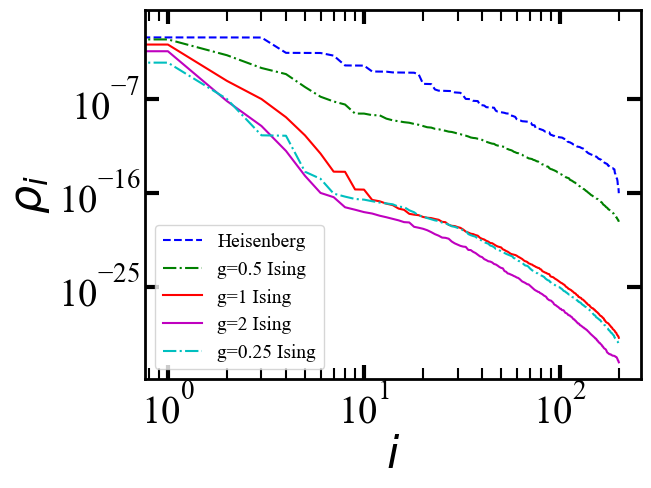}
\caption{Density matrix eigenvalues ($\rho_i$) for the center bond of a 100-site transverse field Ising model at criticality (gapless) and away from criticality (gapped) for several values $g$ from the Hamiltonian $\mathcal{H}=\sum_i\hat S^z_i\cdot\hat S^z_{i+1}+g\sum_i\hat S^x_i$. Note that $\hat S^z$ is defined with a factor of a half, justifying the critical point at $g=0.5$ instead of 1 if using Pauli matrices without the factor of 1/2. Unlike the cartoon version of the gapped curve in Fig.~\ref{cutoffFIG}, it can be seen that the singular values can drop off precipitously for the gapless case, despite having a power-law decay and dropping off slowly in comparison with the gapped Ising cases ($g=0.25, 1, 2$). For reference, a fourth curve for the singular values for a spin-half Heisenberg model are included.  All models are 100 sites long and the SVD is taken on the center bond.\label{isingcutoff}}
\end{figure}

\subsubsection{Eigenvalue decomposition}

The eigenvalue decomposition is one of the most ubiquitously used in quantum physics and takes the form
\begin{equation}
\hat M=\hat U\hat \Omega\hat U^\dagger
\end{equation}
which requires a square matrix (size $a\times a$) input.  In terms of a tensor, this implies that the same number of indices with the same dimensions appears in order on either the left or right group of indices on the matrix-equivalent of the tensor. It is possible to perform an eigenvalue decomposition on a tensor, provided that the matrix equivalent has equal rows and columns, making a square matrix equivalent of a tensor.  One example is the density matrix itself or one of the reduced density matrices.  At this point, the reduced density matrices can be obtained by tracing out (contracting pairs of indices with an identity matrix) to reduce the number of total indices on a tensor.

The eigenvalue decomposition can be truncated in the same way as the SVD was (including a cutoff parameter and maximum number of states kept). However, there is one crucial difference.  Instead of squaring the values of the diagonal matrix, they are simply summed here.  Recall that the singular values square to the eigenvalues of the density matrix.  In the eigenvalue decomposition, the diagonal elements are exactly the eigenvalues of the density matrix, without need for modification.

The input function for the library is the same but switching out {\tt svd} for {\tt eigen}.  Note that the {\tt LinearAlgebra} package in Julia is only imported so there is no name conflict with the DMRjulia library.

The input for the program is
\begin{lstlisting}[numbers=none]
D,U,truncerr,mag = eigen(A)
#or
out = eigen(A,[[1,2],[3,4]])
D,U,truncerr,mag = out
#or
D,U = eigen(A)
#or
D,U = eigen(A,[[1,2],[3,4]])
\end{lstlisting}
and the optional additional inputs {\tt m} and {\tt cutoff} also apply here.

If a generalized eigenvalue problem is to be solved ($\mathcal{H}\Psi=E\mathbf{B}\Psi$), then calling 
\begin{lstlisting}[numbers=none]
D,U = eigen(A,B,[[1,2],[3,4]])
\end{lstlisting}
or similar will allow for an overlap matrix {\tt B} to be input.

\subsubsection{QR and LQ decomposition}

In contrast to the previous two decomposition algorithms, the QR and LQ decompositions will not be programmed with a truncation procedure in this implementation.  In the specific case where the bond dimension of a tensor is not required to grow, either of the QR or the LQ decompositions can be used.  There may be useful mechanisms to impose a truncation of the matrices, but the SVD is sufficient here. Returning to the SVD, note that the $\hat D$ matrix can be contracted into one of the two unitary matrices
\begin{equation}\label{svd2qr}
\hat M=\hat U\left(\hat D\hat V^\dagger\right)\quad\mathrm{or}\quad\hat M=\left(\hat U\hat D\right)\hat V^\dagger
\end{equation}
to form two tensors on the result of the decomposition instead of 3. The QR decomposition essentially forms the $\hat D$ matrix congracted into one of the two tensors,
\begin{equation}\label{qrform}
\hat M=\hat Q\hat R\quad\mathrm{or}\quad\hat M=\hat L\hat Q
\end{equation}
where the $\hat Q$ matrix is the unitary. There are several forms of the QR and LQ decompositions, and the choice here will be to make the inner dimensions match.  Note that an identity $\hat W^\dagger\hat W$ can be inserted between the matrices to change the form without changing the overall character of the decomposition. So, the SVD decomposition as in Eq.~\eqref{svd2qr} may not be exactly the same as the return from the QR or LQ decomposition in Eq.~\eqref{qrform} from a library (notably, in the QR decomposition, the $\hat R$ matrix takes an upper diagonal form).   It is recommended to use a library to compute the QR or LQ forms as these routines have been heavily optimized.

In the library, the functions are called simply as {\tt qr} or {\tt lq} under the same inputs as {\tt svd} received
\begin{lstlisting}[numbers=none]
Q,R,truncerr,mag = qr(A)
#or
Q,R,truncerr,mag = qr(A,[[1,2],[3,4]])
#or
Q,R = qr(A)
#or
Q,R = qr(A,[[1,2],[3,4]])
\end{lstlisting}
but without the truncation parameters or similar.  A similar format for the LQ decomposition is also available
\begin{lstlisting}[numbers=none]
L,Q,truncerr,mag = lq(A)
#or
L,Q,truncerr,mag = lq(A,[[1,2],[3,4]])
#or
L,Q = lq(A)
#or
L,Q = lq(A,[[1,2],[3,4]])
\end{lstlisting}
The {\tt truncerr} and {\tt mag} parameters are returned 0 and 1 respectively and can be left off of the returns from the function.

\subsubsection{Polar decomposition}

Note that in some places the polar decomposition is employed.  This is simply an SVD that has one bond re-gauged.  There are two options, one is to generate two tensors of the form
\begin{equation}
\hat M=\left(\hat U \hat D\hat U^\dagger\right)\left(\hat U\hat V^\dagger\right)
\end{equation}
and the other is
\begin{equation}
\hat M=\left(\hat U \hat V^\dagger\right)\left(\hat V\hat D\hat V^\dagger\right)
\end{equation}
This can be viewed as a particular choice of an identity inserted into the SVD. The advantages of the polar decomposition is that it preserves the basis on the exterior indices of a given decomposition.

The input into the library can be called with {\tt polar} and is similar to the {\tt svd} which allows for truncation.
\begin{lstlisting}[numbers=none]
UV,VDV=polar(A,[[1,2],[3]])
#or
UDU,UV=polar(A,[[1,2],[3]],right=false)
\end{lstlisting}

\section{Networks: Matrix product states, operators, and environments}\label{networks}

Having defined the basic operations for manipulating a tensor, the broad concept of a wavefunction as a tensor network is developed here.  Instead of dealing with one or two tensors, a network will store enough tensors to describe a full quantum problem with one or more tensors per site.  This typically includes a wavefunction (here, an MPS), operators, and environment tensors. A fourth type is useful if running extremely large computations which stores the MPS, matrix product operator (MPO), and environments directly on the hard disk.  This is not useful for small systems, but it does allow for limited memory to be used efficiently without too much delay for reading and writing tensors to the disk.

Vertical lines are typically reserved for physical indices representing the local Fock space of a given model.  In short, the physical degrees of freedom for each site are represented in the vertical lines of each tensor while the horizontal lines represent the basis functions for every other site in the system.  For more discussion, see Ref.~\onlinecite{bakerCJP21,baker2019m}.

\subsection{Matrix product states}

In order to derive a MPS, it is best to start with a wavefunction represented as a tensor,
\begin{equation}
|\Psi\rangle=\sum_{\sigma_1\sigma_2\ldots\sigma_N}\psi_{\sigma_1\sigma_2\ldots\sigma_N}|\sigma_1\sigma_2\ldots\sigma_N\rangle
\end{equation}
where the coefficient $\psi_{\sigma_1\sigma_2\ldots\sigma_N}$ is most useful here. It will be taken from the results of many alternatives that using the density matrix is the right way to decompose the wavefunction.  So, the task becomes how to apply the SVD to separate the physical degrees of freedom in the wavefunction onto different sites.

Grouping the first index $\sigma_1$ by itself, the wavefunction becomes (after an appropriate reshape, with indices joined together with a parenthesis),
\begin{equation}
|\Psi\rangle=\sum_{\sigma_1\sigma_2\ldots\sigma_N}\psi^{\sigma_1}_{(\sigma_2\ldots\sigma_N)}|\sigma_1\sigma_2\ldots\sigma_N\rangle
\end{equation}
where indices on the left are lowered indices and indices on the right are raised. The first index is now isolated.  Applying a SVD produces
\begin{equation}
|\Psi\rangle=\sum_{\sigma_1\sigma_2\ldots\sigma_N}\sum_{a_1}A^{\sigma_1}_{a_1}\psi^{a_1}_{(\sigma_2\ldots\sigma_N)}|\sigma_1\sigma_2\ldots\sigma_N\rangle
\end{equation}
A new index has appeared, $a_1$. Continuing by grouping $\sigma_2$ with $a_1$ with an appropriate reshape and taking a SVD (or QR decomposition)
\begin{align}
|\Psi\rangle=&\sum_{\sigma_1\sigma_2\ldots\sigma_N}\sum_{a_1}A^{\sigma_1}_{a_1}\psi^{(a_1\sigma_2)}_{(\sigma_3\ldots\sigma_N)}|\sigma_1\sigma_2\ldots\sigma_N\rangle\\
=&\sum_{\sigma_1\sigma_2\ldots\sigma_N}\sum_{a_1a_2}A^{\sigma_1}_{a_1}A^{\sigma_2}_{a_1a_2}\psi^{a_2}_{(\sigma_3\ldots\sigma_N)}|\sigma_1\sigma_2\ldots\sigma_N\rangle
\end{align}
Continuing the program of reshaping and decompositions, the resulting wavefunction appears as
\begin{align}\label{leftMPS}
|\Psi\rangle=&\sum_{\sigma_1\sigma_2\ldots\sigma_N}\sum_{a_1a_2\ldots a_{N-1}}A^{\sigma_1}_{a_1}A^{\sigma_2}_{a_1a_2}A^{\sigma_3}_{a_2a_3}\ldots D^{\sigma_N}_{a_{N-1}}\\
&\hspace{3cm}\times|\sigma_1\sigma_2\ldots\sigma_N\rangle\nonumber
\end{align}
The last matrix is denoted as $\hat D$ since it represents the orthogonality center. Implied in the SVD step, the $\hat D$ matrix is simply contracted onto a given site.  Note a feature of the MPS.  Only one of the tensors will contract to something other than the identity.  All the $A$ tensors represent a left-normalized (relating to $\hat U$) tensor.  When contracted with its dual, each tensor will contract to the identity if they are properly represented ({\it e.g.} $\hat U$ with $\hat U^\dagger$ and not $\hat V^\dagger$) \cite{bakerCJP21,*baker2019m}.

A similar expression for the wavefunction can be obtained for right-normalized (relating to $\hat V$) tensors,
\begin{align}
|\Psi\rangle=&\sum_{\sigma_1\sigma_2\ldots\sigma_N}\sum_{a_1a_2\ldots a_{N-1}}D^{\sigma_1}_{a_1}B^{\sigma_2}_{a_1a_2}B^{\sigma_3}_{a_2a_3}\ldots B^{\sigma_N}_{a_{N-1}}\\
&\hspace{3cm}\times|\sigma_1\sigma_2\ldots\sigma_N\rangle\nonumber
\end{align}
by moving the SVD from right to left with appropriate reshapes.

\begin{figure}
\includegraphics[width=\columnwidth]{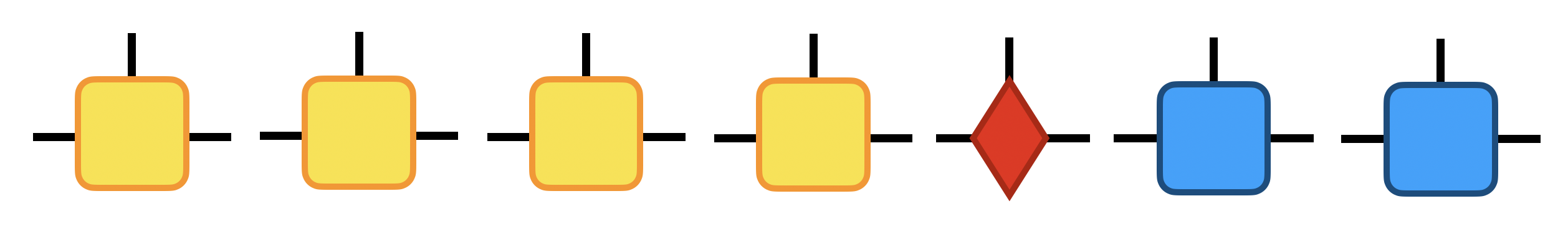}
\caption{A matrix product state (MPS) with rank-3 tensors is represented. Both left- and right-normalized tensors are present. The gauge center (diamond) can be moved to other sites by decomposing the site that it currently resides and repeating until the orthogonality center reaches another site. The first and last auxiliary index (horizontal) are size 1 for the finite MPS.\label{MPSgauge}}
\end{figure}

In DMRjulia, all MPS tensors will be rank-3, in contrast with the presentation analytically with rank-2 tensors. This is because it is easier to write a computer program with functions around a constant rank tensor than it is to make a variable rank tensor MPS.   Figure~\ref{MPSgauge} contains diagrams of the MPS tensors with the gauge.  So, an index of size 1 is reshaped onto the first and last tensors as presented above.

\subsubsection{Gauges}\label{gauge}

The $\hat D$ matrix does not need to be confined to the first or last site of the MPS. In fact, there is an entire family of MPS wavefunctions that are all equiavlent.    To justify this, note that on any bond, an identity operator of size $m\times m$ can be inserted of the form $\hat W^\dagger\hat W$ \cite{bakerCJP21,*baker2019m}.  These tensors can be absorbed into the horizontal indices, changing the tensors.  However, since the identity was inserted, no overall change to the wavefunction occurred.   

The $\hat D$ matrix can also be moved to a bond if necessary. Moving the $\hat D$ matrix is therefore known as a re-gauging of the MPS and produces an equivalent form for the MPS. This is known as re-gauging and will define a center of orthogonality for the MPS. In order to re-gauge the MPS so that the $\hat D$ matrix (orthogonality center) is represented on a different site, a SVD can be used to decompose the current orthogonality center.  Repeated application of this operation will move the $\hat D$ matrix along the MPS.

One movement of the orthogonality center to the left is contained in the following diagram
\includegraphics[width=\columnwidth]{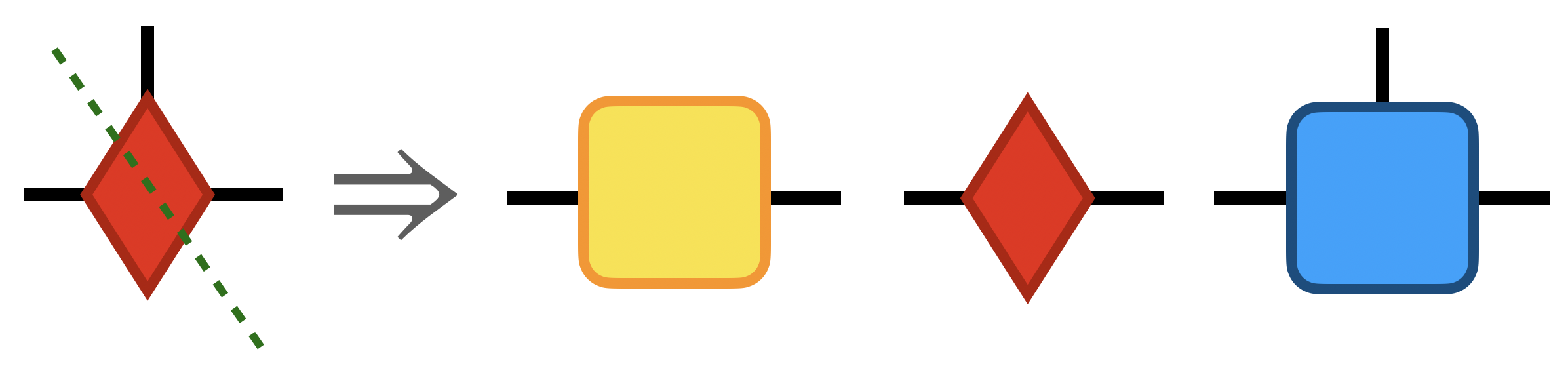}
where the first two tensors (both rank 2) are contracted into the left-normalized tensor to the right of the current site.  Alternatively, if a {\tt lq} decomposition is used, then only two tensors result and the $L$ matrix is multiplied into the next site.

Meanwhile, a move to the right is represented as
\includegraphics[width=\columnwidth]{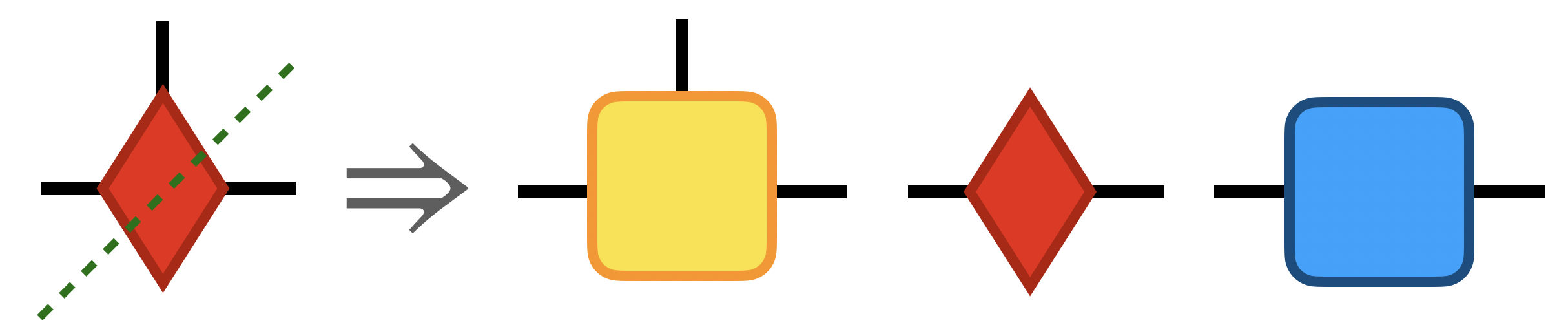}
where the second two tensors (both rank 2) are contracted into the right-normalized tensor to the right of the current site.  Alternatively, if a {\tt qr} decomposition is used, then only two tensors result and the $R$ matrix is multiplied into the next site.

The fact that only one tensor holds all of the information between the left and right partitions of the system is a remarkable feature, but this is simply a representation of the area law from Sec.~\ref{arealaw}.

\subsubsection{Dual of the MPS}

For any expectation value of the form $\langle\Psi|\mathcal{A}|\Psi\rangle$, the dual of the MPS tensors, $\langle\Psi|$, must also be defined. This is done by replacing $A\rightarrow A^\dagger$ as in Eq.~\eqref{leftMPS}, which appears in mathematical notation as to the tensors as
\begin{align}\label{dualMPS}
\langle\Psi|=&\sum_{\sigma_1'\sigma_2'\ldots\sigma_N'}\sum_{a_1'a_2'\ldots a_{N-1}'}A^{*\sigma_1'}_{a_1'}A^{*\sigma_2'}_{a_1'a_2'}A^{*\sigma_3'}_{a_2'a_3'}\ldots D^{*\sigma_N'}_{a_{N-1}'}\nonumber\\
&\hspace{3cm}\times\langle\sigma_1'\sigma_2'\ldots\sigma_N'|
\end{align}
where $*$ signifies a conjugation. Sometimes the tensor itself will be marked as $A^\dagger$ instead of using the star operator.  All indices are marked with an apostrophe to avoid combining them with the indices of the MPS tensors.  Graphically, the dual MPS tensors look the same, except that the vertical line signifying the physical index is pointed down on conjugation. When contracting onto the MPS, the apostrophe on the $\sigma_i$ indices is dropped so that those indices are contracted onto the MPS's physical indices.

\subsubsection{MPS in DMRjulia}

The generation of the MPS is one of the cases where the QR and LQ decompositions can be useful.  If the bond dimension will not be truncated, then the use of this function can slightly lessen the amount of time necessary to generate the MPS from an existing wavefunction. Appendix~\ref{ED} demonstrates how this is accomplished.

\begin{figure}[t]
\includegraphics[width=0.5\columnwidth]{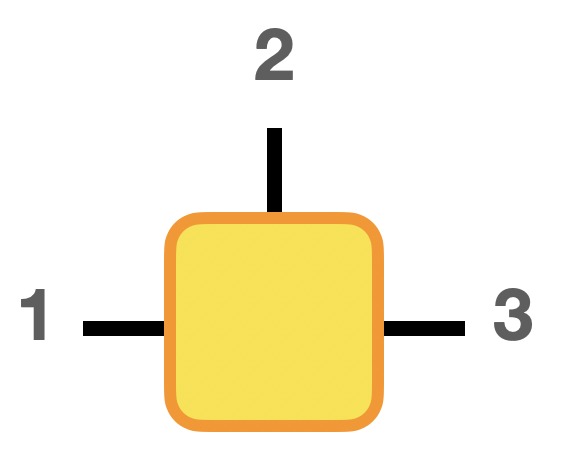}
\caption{The order for the indices of the MPS tensors (left-, right-, and orthgonality center are uniform) as chosen for DMRjulia.. \label{MPSorder}}
\end{figure}

The basic type that stores a MPS is listed below. In practice, the type can always be called through the parent type {\tt MPS}. So, the long name of this function type does not need to be memorized.
\begin{lstlisting}[numbers=none]
mutable struct matrixproductstate{W} 
  A::W
  oc::intType
end
\end{lstlisting}
and the indies are chosen as in Fig.~\ref{MPSorder}. Elements are called similar to the {\tt tens} object.  There is an array of tensors {\tt A} that can be called as {\tt psi.A} which give the elements of the MPS as derived in Eq.~\eqref{leftMPS}.  In practice, a classical MPS can be input into the program (a classical state has bond dimension 1 on every tensor). The DMRG algorithm will find the ground state in the MPS representation.

The orthothogonality center {\tt psi.oc} is useful to store here instead of doing some coarse search for the non-unitary tensor.

To move the orthogonality center, one simply calls
\begin{lstlisting}[numbers=none]
newoc = 3 #or another site
move!(psi,newoc)
\end{lstlisting}
This will modify the tensor's elements in-place without copying all the tensors into a new object.

If the MPS does not have the same Fock space on each physical index, the initial state can simply be initialized but with varying tensor sizes.  Then, the {\tt MPS} function works as before.

\subsection{Matrix product operators}

\begin{figure}
\includegraphics[width=\columnwidth]{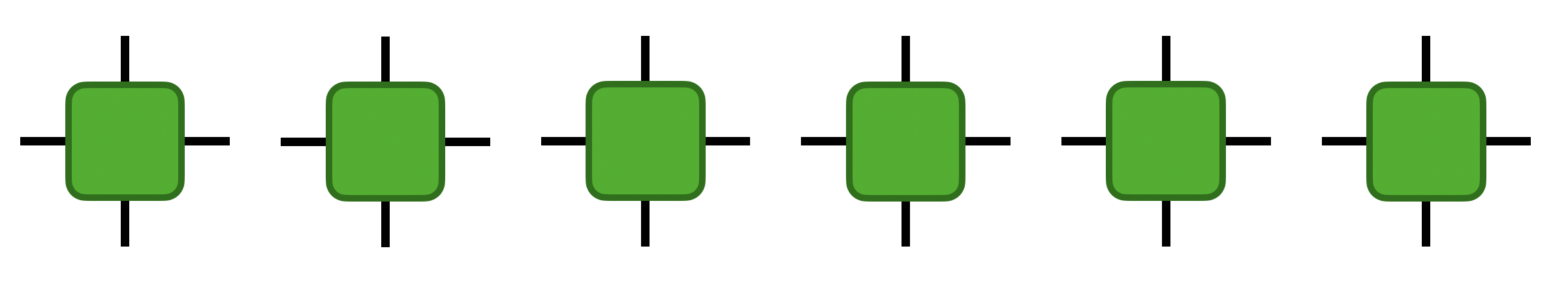}
\caption{The site-by-site representation of the matrix product operator (MPO) represents the Hamiltonian for the MPS wavefunction. \label{MPOtensors}}
\end{figure}

For the MPS wavefunction, it would be useful to have a form of the Hamiltonian that similarly is decomposed into a site-by-site representation. If a local form is not found for an operator, then all of the MPS tensors should be contracted onto the large operator, negating the advantage of the tensor network's ability to decompose a problem down a few sites.

The resulting matrix product operator (MPO) can be guessed to have the form
\begin{align}\label{MPOform}
\mathcal{H}=&\sum_{\sigma_1'\sigma_2'\ldots\sigma_N'}\sum_{\sigma_1\sigma_2\ldots\sigma_N}\sum_{b_1b_2\ldots b_{N-1}}H^{\sigma_1'\sigma_1}_{b_1}H^{\sigma_2'\sigma_2}_{b_1b_2}\ldots H^{\sigma_N'\sigma_N}_{b_{N-1}}\nonumber\\
&\hspace{2cm}\times|\sigma_1'\sigma_2'\ldots\sigma_N'\rangle\langle\sigma_1\sigma_2\ldots\sigma_N|
\end{align}
where there are now two physical indices ($\sigma_i$ and $\sigma_i'$) for each tensor so the operator can connect to both the MPS and the dual of the MPS when forming a quantity such as $\langle\psi|\mathcal{H}|\psi\rangle$. The auxiliary indices  $b_i$ appear for the same reason as the $a_i$ indices appeared in the MPS.

Graphically, the MPO is depicted as a series of rank-4 tensors in Fig.~\ref{MPOtensors}.  Two vertical lines now appear for the physical indices. The outer indices on the left and right edges of the system are of dimension 1 in the finite case, as they were for the MPS. These can be reshaped away or back onto the tensor if needed.  This is again for ease of programming to keep the rank of the tensor fixed and therefore have fixed functions operating on them.

There are systematic ways to construct the MPO, but it is better to understand the basic principles of how to write down the bulk MPO operator  and convert this to a tensor.  Some examples are given below.

\subsubsection{MPO example: Hubbard model}

The one-dimensional Hubbard model is \cite{hubbard1963electron}, 
\begin{equation}\label{hubbardham}
\mathcal{H}=\sum_{i\sigma}t\left(\hat c^\dagger_{i,\sigma}\hat c_{i+1,\sigma}+\hat c^\dagger_{i+1,\sigma}\hat c_{i,\sigma}\right)+\sum_i\Big(U\hat n_{i,\uparrow}\hat n_{i,\downarrow}+\mu\hat n_i\Big)
\end{equation}
where $\sigma=\uparrow,\downarrow$, $\hat n_i=\hat n_{i\uparrow}+\hat n_{i\downarrow}$, hopping parameter $t$, and a Coulombic repulsion $U$.  When represented on the page, the fermionic operators $\hat c^\dagger_{i\sigma}$ require an additional fermion sign operator as given in the Jordan-Wigner transformation from bosons to fermions \cite{fetter2012quantum}. Appendix~\ref{jordanwigner} explains the role of the new $\hat F$ operators and why they are necessary.  For now, the MPO for Eq.~\eqref{hubbardham} is (all dots are $\hat 0$)
\begin{align}\label{hubbardmpo}
H_{b_{i-1}b_i}^{\sigma_i\sigma_i'}&=\left(\begin{array}{cccccc}
\hat I & \cdot & \cdot & \cdot & \cdot & \cdot\\
-t\hat c_\uparrow^\dagger & \cdot & \cdot & \cdot & \cdot & \cdot\\
t\hat c_\uparrow & \cdot & \cdot & \cdot & \cdot & \cdot\\
-t\hat c_\downarrow^\dagger & \cdot & \cdot & \cdot & \cdot & \cdot\\
t\hat c_\downarrow & \cdot & \cdot & \cdot & \cdot & \cdot\\
\mathcal{V} & \hat c_\uparrow\hat F & \hat c_\uparrow^\dagger\hat F & \hat c_\downarrow\hat F & \hat c_\downarrow^\dagger \hat F& \hat I\\
\end{array}\right)\\
\mathcal{V}&=\mu\hat n + U\hat n_\uparrow\hat n_\downarrow\nonumber
\end{align}
where $\mathcal{V}$ is an onsite term with a chemical potential $\mu$.  The addition of the constants $t$ to the left column is by convention here as well and will be useful when defining MPOs with a quantum number symmetry.

This notation deserves some explanation in the context of the indices present on the MPO tensors in Eq.~\eqref{MPOform}. each entry in the super-matrix is a rank-2 tensor representing an operator. These two indices are represented by $\sigma_i$ and $\sigma_i'$.  The remaining two indices $b_i$ correspond to the position in Eq.~\eqref{hubbardmpo}.  Thus, the supermatrix can be reinterpreted as a rank-4 tensor.

Note that each of the inputs in Eq.~\eqref{hubbardmpo} do not have an index for the current site (although, one could be added to make position dependent variables in the operator).  This is because the bulk MPO operator is homogeneous in this case and the operators input into the MPO here are not the same multi-site operators as in exact diagonalization (see Apx.~\ref{ED}).

The MPO can also be represented in an upper right triangular form where the transpose of Eq.~\eqref{hubbardmpo} is represented, and this convention is taken in several works.  As a general guideline, the diagonal terms relate to long range interactions since it can be imagined that multiplying the MPO term will accumulate the full Hamiltonian in the lower left corner in this convention.

Performing an example with the dense operations for the Hubbard model will cause some of the fermions to disappear if the proper chemical potential is not chosen correctly, so a full example will be skipped here.   It is much better to run this calculation with conserved symmetries enforced as will be discussed in the next article in the series for this library.

\subsubsection{MPO example: Two-dimensional Heisenberg}\label{2DspinMPO}

\begin{figure}
\includegraphics[width=\columnwidth]{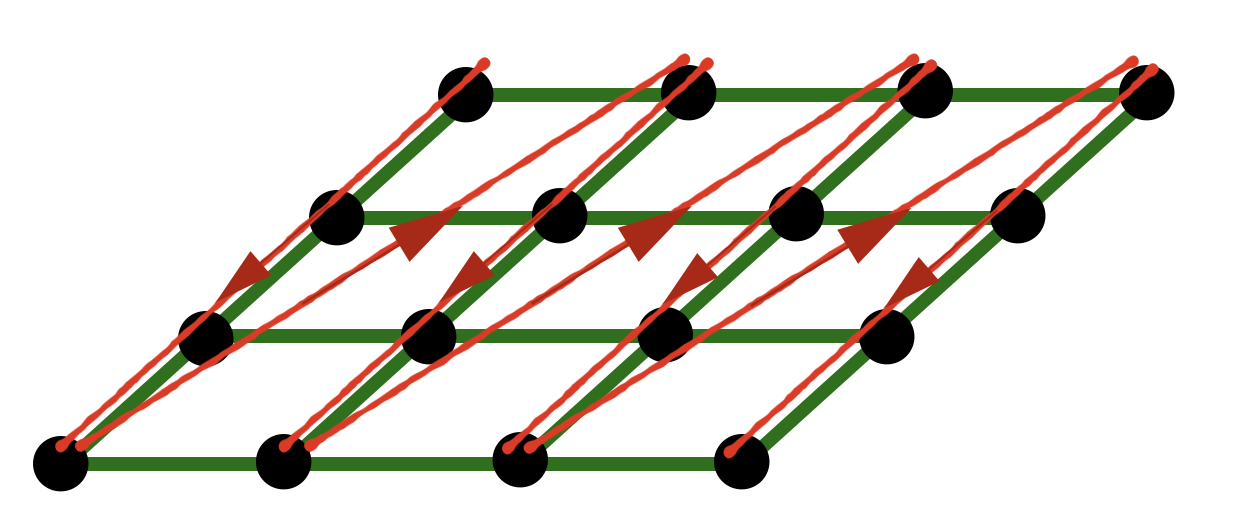}
\includegraphics[width=\columnwidth]{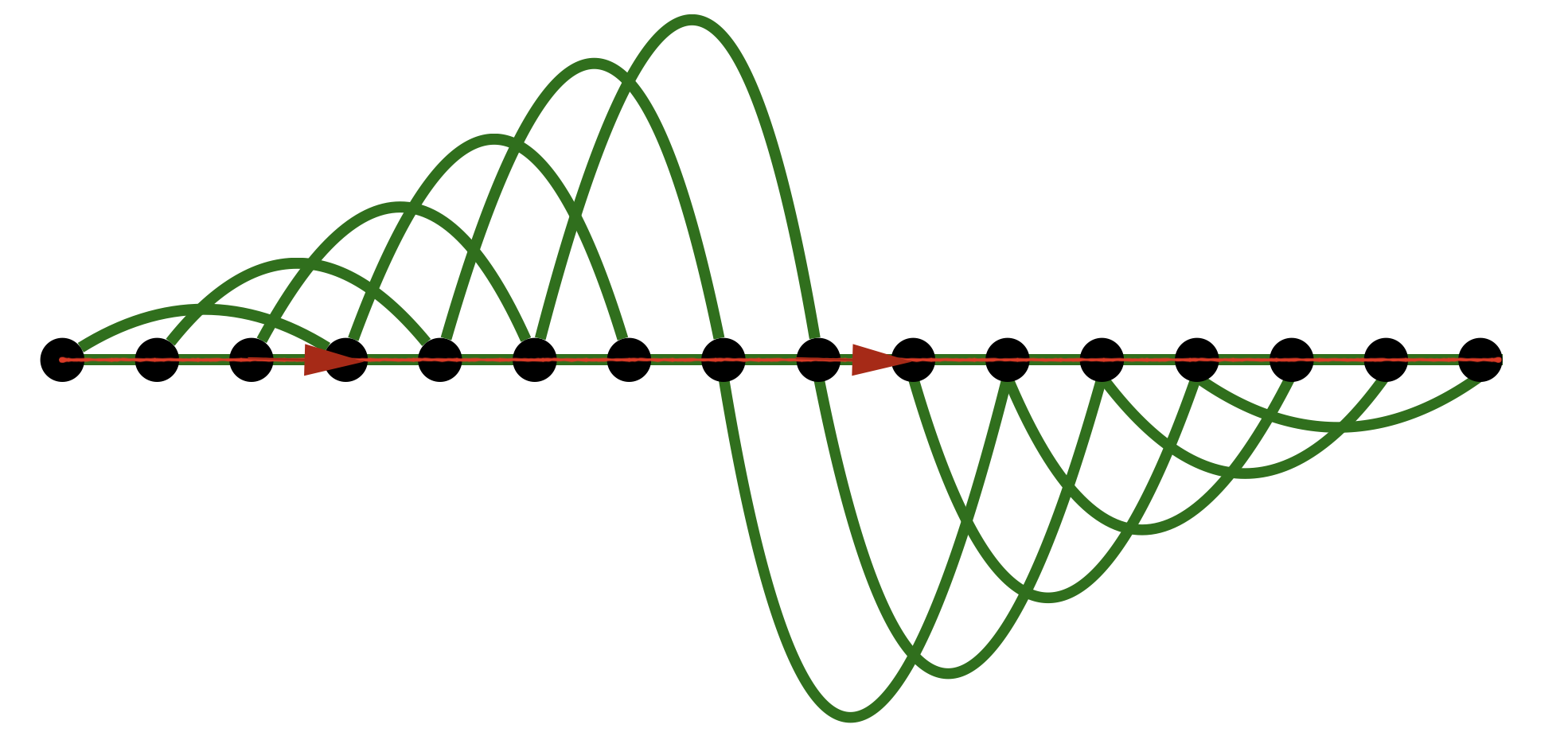}
\caption{A two-dimensional $4\times4$ lattice with a chosen path (arrows) to reveal a one-dimensional path. The lower figure is the unraveled one-dimensional path showing that longer ranged interactions appear and the MPO grows larger. \label{2Dsystem}}
\end{figure}

Another model that would be useful to list is the Heisenberg model as in Eq.~\eqref{heisenberg} on a two-dimensional lattice of size $4\times 4$ depicted in Fig.~\ref{2Dsystem}.

The path shown in the lower diagram in Fig.~\ref{2Dsystem} gives the MPO (all dots are $\hat 0$)
\begin{align}
&H_{b_{i-1}b_i}^{\sigma_i\sigma_i'}=\\
&\left(\begin{array}{cccccccccccccc}
\hat I & \cdot& \cdot & \cdot  & \cdot & \cdot & \cdot & \cdot & \cdot & \cdot & \cdot & \cdot & \cdot & \cdot\\
\frac12\hat S^+  & \cdot& \cdot & \cdot  & \cdot & \cdot & \cdot & \cdot & \cdot & \cdot & \cdot & \cdot & \cdot & \cdot\\
\cdot & \hat I & \cdot & \cdot  & \cdot & \cdot & \cdot & \cdot & \cdot & \cdot & \cdot & \cdot & \cdot & \cdot\\
\cdot & \cdot & \hat I & \cdot  & \cdot & \cdot & \cdot & \cdot & \cdot & \cdot & \cdot & \cdot & \cdot & \cdot\\
\cdot & \cdot & \cdot & \hat I & \cdot & \cdot & \cdot & \cdot & \cdot & \cdot & \cdot & \cdot & \cdot & \cdot\\
\frac12\hat S^- & \cdot& \cdot & \cdot  & \cdot & \cdot & \cdot & \cdot & \cdot & \cdot & \cdot & \cdot & \cdot & \cdot\\
\cdot & \cdot & \cdot & \cdot & \cdot & \hat I & \cdot & \cdot  & \cdot & \cdot & \cdot & \cdot & \cdot & \cdot\\
\cdot & \cdot & \cdot & \cdot & \cdot & \cdot & \hat I & \cdot  & \cdot & \cdot & \cdot & \cdot & \cdot & \cdot\\
\cdot & \cdot & \cdot & \cdot & \cdot & \cdot & \cdot & \hat I & \cdot & \cdot & \cdot & \cdot & \cdot & \cdot\\
\hat S^z & \cdot& \cdot & \cdot & \cdot  & \cdot & \cdot & \cdot & \cdot & \cdot & \cdot & \cdot & \cdot & \cdot\\
\cdot & \cdot & \cdot & \cdot & \cdot &  \cdot & \cdot & \cdot & \cdot & \hat I & \cdot & \cdot & \cdot & \cdot\\
\cdot & \cdot & \cdot & \cdot & \cdot &  \cdot & \cdot & \cdot & \cdot & \cdot & \hat I & \cdot & \cdot & \cdot\\
\cdot & \cdot & \cdot & \cdot & \cdot &  \cdot & \cdot & \cdot & \cdot & \cdot & \cdot & \hat I & \cdot & \cdot\\
\cdot & \hat S^- & \cdot & \cdot & \hat S^- &  \hat S^+ & \cdot & \cdot & \hat S^+ &  \hat S^z & \cdot & \cdot & \hat S^z & \hat I 
\end{array}\right)\nonumber
\end{align}
where the last site in a column is connected to the first site in the next column for ease of demonstration. First, note that the identity operators are not contained on the diagonal. This implies that the range of the operators have a limit. In this case, the identity operators will store the second operator on the lower, off-diagonal and eventually pass it to the first term in the column.

Note that if an onsite term, such as a Zeeman interaction $\sum_i\mathbf{B}\cdot\mathbf{S}_i$ for some magnetic field $\mathbf{B}$, can be added to the lower left corner.

It is recommended that when writing out the MPO in analytic form that an algebraic manipulation software ({\it e.g.} Mathematica, Maple, Maxima, etc.) is used to check that the terms accumulated in the lower left corner are indeed the Hamiltonian after all MPOs are contracted.  This is accomplished by simply multiplying appropriately defined matrices with the form given above.

From this form, it can be seen why MPS techniques will have a harder time with longer range interactions. The size of the MPO will make applying it to the MPS longer.  Additionally, more entangled pairs will need to be retained in the SVD at each step.

For the reason of longer-range interaction when running a two-dimensional model, it is often declared that tensor network methods will be harder to converge.  This is true, but sometimes this is taken too seriously.  The attitude that the reader should have when performing these computations is to try it and see how hard it is.  It is wise to remain vigilant of converging quantities properly, but it is recommended to at least try to apply these methods to all systems possible.

Several examples of MPO operators were presented in Ref.~\onlinecite{bakerCJP21,*baker2019m}, including the natural encoding of exponential functions. 

\subsubsection{MPOs in DMRjulia}

\begin{figure}
\includegraphics[width=0.5\columnwidth]{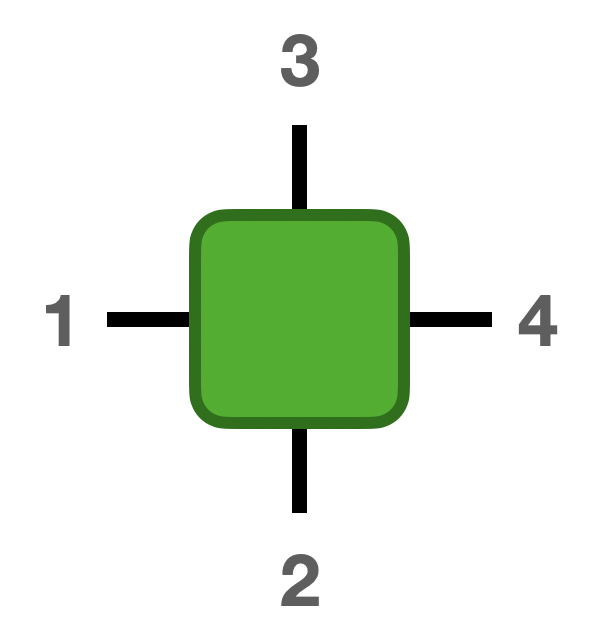}
\caption{The order for the indices of the MPO tensors as chosen for DMRjulia. \label{MPOorder}}
\end{figure}

The basic struct ({\it i.e.}, type) used to contain the MPO tensors has a single field
\begin{lstlisting}[numbers=none]
mutable struct matrixproductoperator{W}
  H::W
end
\end{lstlisting}
and the indices are chosen as in Fig.~\ref{MPOorder}.

The name of this container is long, but a short-hand {\tt MPO} can be used everywhere which will instruct the compiler that the object could be of this type.

DMRjulia allows for the input of the MPO tensor exactly as it appears in the previous section. The letter {\tt O} is reserved for a tensor with zeros in the library, $\hat 0$.  Meanwhile, {\tt Id} represents the identity matrix, $\hat I$.  In the examples in the Supplemental Material \cite{tensor_recipes}, this is prominently featured.

The MPO is initialized in the following form
\begin{lstlisting}[numbers=none]
out = fermionOps()
Cup,Cdn,Nup,Ndn,Ndens,F,O,Id = out
mu = -2.0
U = 4.0
t = 1.0
function H(i::Int64)
  V = mu*Ndens + U*Nup*Ndn
  return [Id  O O O O O;
    -t*Cup' O O O O O;
    conj(t)*Cup  O O O O O;
    -t*Cdn' O O O O O;
    conj(t)*Cdn  O O O O O;
    V Cup*F Cup'*F Cdn*F Cdn'*F Id]
end
\end{lstlisting}
or a vector can be defined (or similar)
\begin{lstlisting}[numbers=none]
vecH = [H(i) for i = 1:length(psi)]
\end{lstlisting}
and the function to generate the MPO explicitly is
\begin{lstlisting}[numbers=none]
Ns = length(psi)
physind = size(F,1)
mpo = makeMPO(H,physind,Ns)#or vecH
\end{lstlisting}
The resulting {\tt mpo} variable contains the rank-4 tensors of the MPO.  The basic idea of the algorithm is to search each $\sigma\times\sigma$ block in the input MPO and then fill out the resulting MPO tensor.  

If the MPO has different Fock space sizes on each of the tensors, then the {\tt physind} variable can be replaced by a vector of sizes.  The vector can be any size and it will repeat every modulo the length of the vector.
\begin{lstlisting}[numbers=none]
physind = [2] #repeats every 1 site
mpo = makeMPO(H,physind,Ns)#or vecH
#or
physind = [2,5,4] #repeats every 3 sites
mpo = makeMPO(H,physind,Ns)#or vecH
\end{lstlisting}
One should then define an appropriate function that uses at least two sets of operators to make the block form of the MPO. The only point to be minded there is that the MPS should also have a matching physical index size.

\subsubsection{Applying MPOs to MPSs}

So far, both MPOs and MPSs have been discussed separately.  As one example of an operation where the MPO can be applied to the MPS to form (in the case of a Hamiltonian) $|\mathcal{H}\psi\rangle$.  This is generally an inefficient operation and does not fully take advantage of the locality of the MPS for measurements.  So, this operation should be used only when absolutely necessary as there is often a more efficient, local alternative available.

To apply the MPO to the MPS, each site can be treated independently. The basic operation is to contract the MPO into the MPS as
\begin{equation}
\includegraphics[width=\columnwidth]{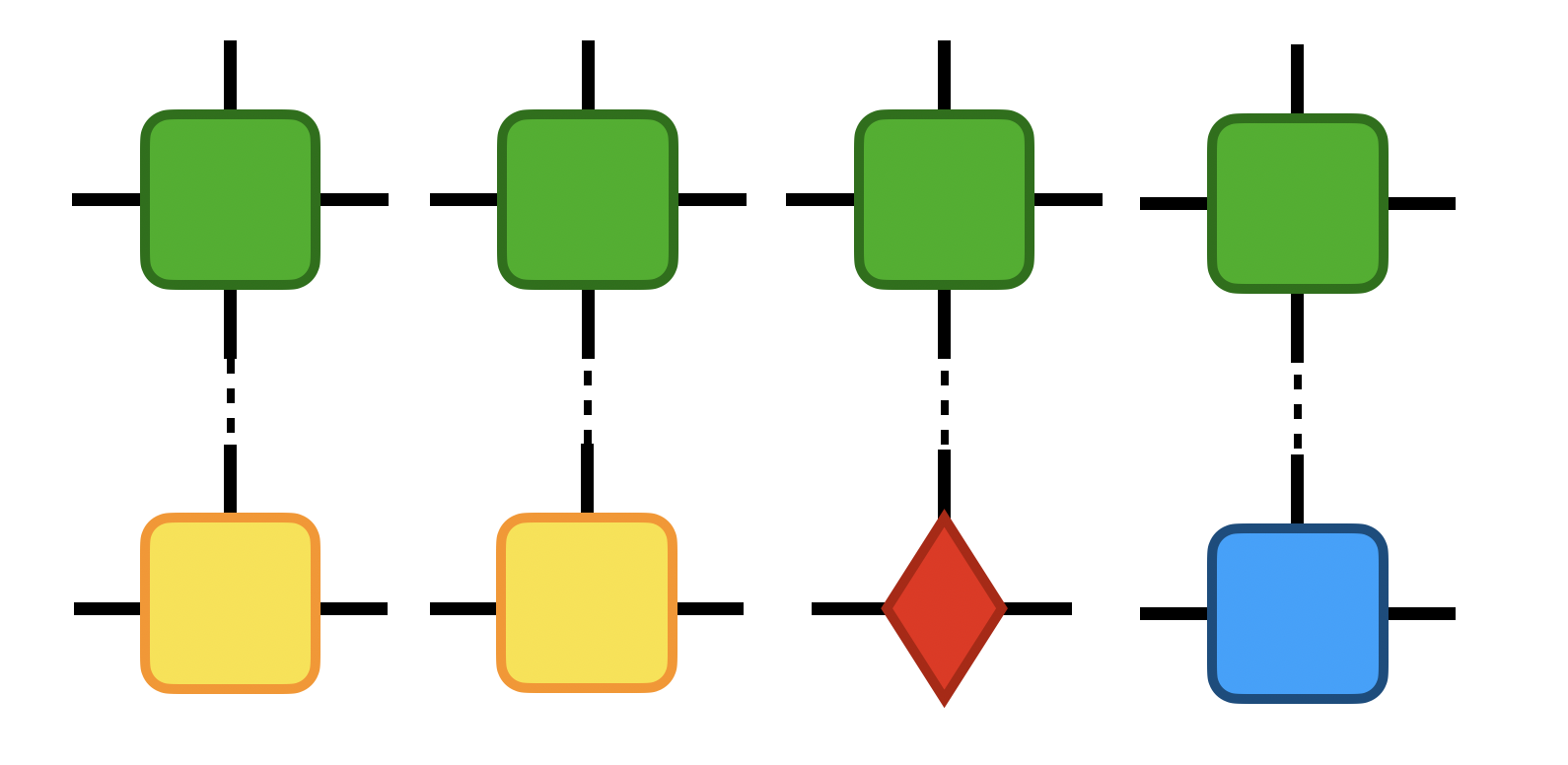}\nonumber
\end{equation}
where it should be noted that the horizontal lines are not contracted.

The process is repeated until the end of the MPS is reached and all MPO tensors have been applied to the MPS tensors. The gauge of the MPS does not matter here as the operation is fully nonlocal. So, the MPS tensor can be replaced by the orthogonality center or left- or right-normalized tensors.

The next step will combine the horizontal indices and return the MPS to uniformly include rank-3 tensors. Note that at this point, the bond dimension size is much larger than it was originally. The size of the bond dimension of the MPS tensor multiplied by the bond dimension of the MPO tensor.  This next step will introduce a mechanism to truncate the bond dimension.

The first step is to perform an SVD on the double tensor according to the diagram
\begin{equation}
\includegraphics[width=\columnwidth]{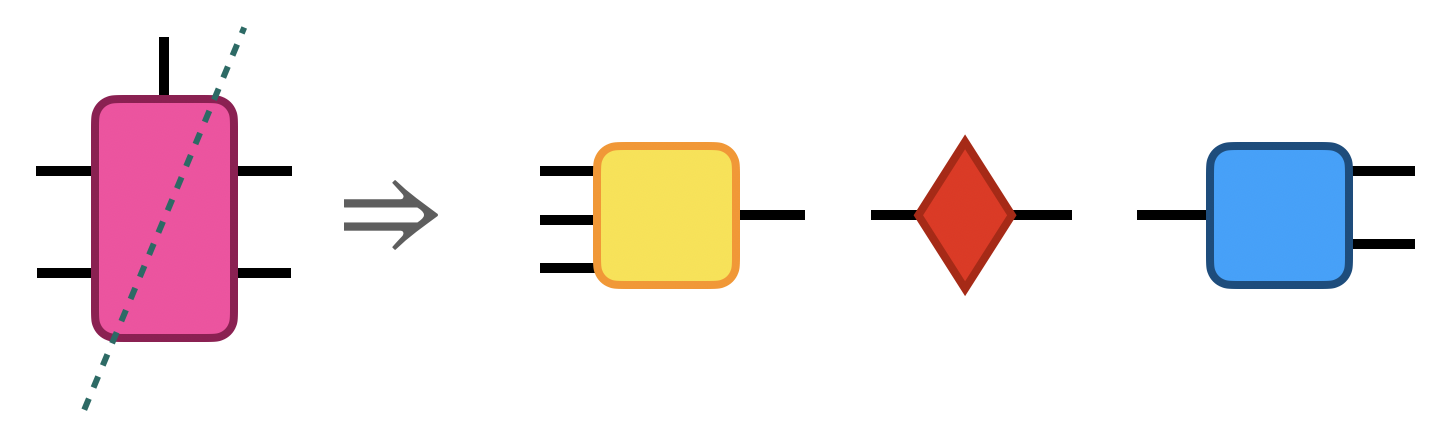}\nonumber
\end{equation}
If this is performed on the first site in a finite MPS, then the two indices on the left are both size one.  These can be trivially reshaped into each other (this is also true for the final two indices on the last tensor). This returns a left-normalized rank-3 tensor for the first site.

\begin{figure}[b]
\includegraphics[width=0.5\columnwidth]{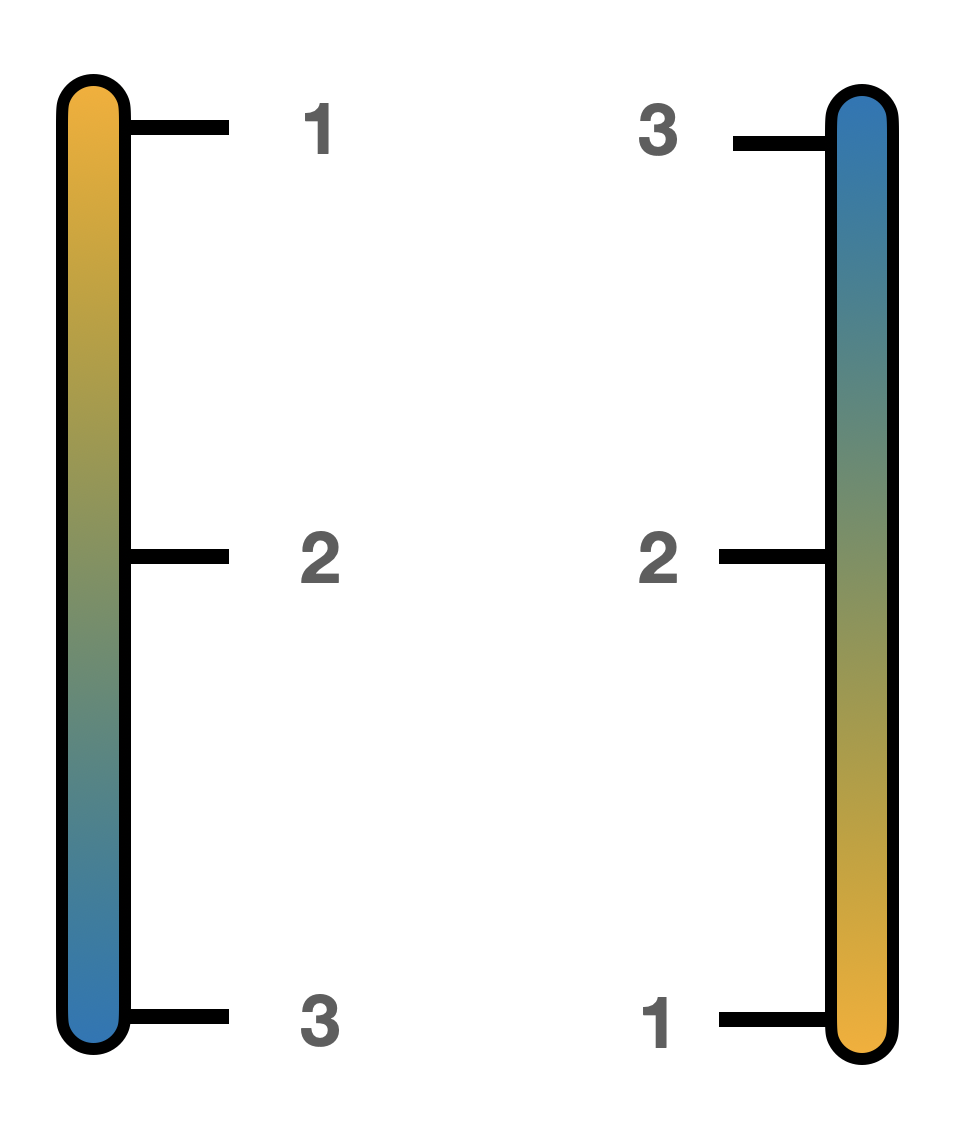}
\caption{The order for the indices of the left and right environments as chosen for DMRjulia. \label{LRenv}}
\end{figure}

The convention used in DMRjulia for the order of indices on the left and right environment tensors is shown in Fig.~\ref{LRenv}.  The left environment has descending order and the right environment has ascending order.

The $\hat D$ matrix is contracted into the $\hat V^\dagger$ tensor.  Note that there are two outer indices that can be contracted into the next tensor to the right
\begin{equation}
\includegraphics[width=\columnwidth]{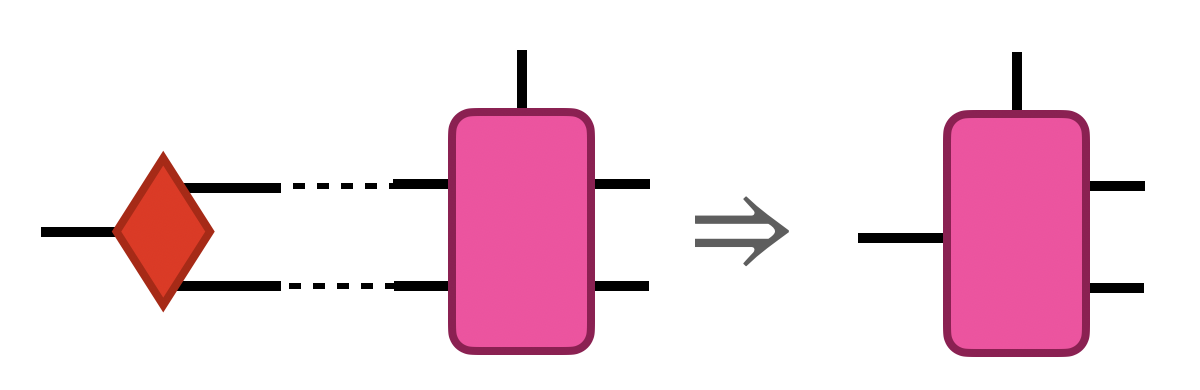}\nonumber
\end{equation}
This operation creates a single index on the left instead of two. It also allows for the repeat of the steps so far to obtain the next tensor to the right.

The steps could be done in reverse, but it is merely taken as convention here to go from left to right.  DMRjulia's function to complete this operation is called as
\begin{lstlisting}[numbers=none]
newpsi = applyMPO(psi,mpo)
\end{lstlisting}
Optional inputs {\tt m=} and {\tt cutoff=} set the truncation of the new MPS on the reverse sweep.  Else, the full, new, larger bond dimension is kept.

\subsubsection{Initial states}

There are a few options to choose an initial state. It is recommended to start with a classical product state (bond dimension one everywhere) unless there is a good reason not to.

For a spin model, the generation of the MPS would look like
\begin{lstlisting}[numbers=none]
#Ns = length of MPS
QS = 2 #quantum states
initTensor = [zeros(1,QS,1) for i=1:Ns]
for i = 1:Ns
  #alternating up and down state
  updn = i%2 == 1 ? 1 : 2
  initTensor[i][1,updn,1] = 1.0
end
psi = MPS(initTensor,1)
#or
psi = MPS(initTensor)
\end{lstlisting}
but a simplified interface for making the MPS is contained in DMRjulia as
\begin{lstlisting}[numbers=none]
psi = MPS(Float64,QS,Ns)
#or
psi = MPS(QS,Ns)
\end{lstlisting}
which generates a MPS of bond dimension 1.  The additional flags {\tt oc=} and {\tt m=} can be added as arguments to the {\tt MPS} function to generate a different maximum bond dimension size, {\tt m}, or a different orthogonality center, {\tt oc}.

There is also an option to generate a random MPS as
\begin{lstlisting}[numbers=none]
psi = randMPS(Float64,QS,Ns,m=50)
#or
psi = randMPS(QS,Ns,m=50)
\end{lstlisting}
which has the same additional arguments as the simplified {\tt MPS} function above.  The difference between {\tt MPS} and {\tt randMPS} is that the former generates a ferromagnetic state and the latter gives a staggered magnetization.

For a Hubbard model, the choice method to generate the wavefunction is to use raising operators.
\begin{lstlisting}[numbers=none]
#Ns = length of MPS
Ne = Ns #half-filling
Ne_up = ceil(Int64,div(Ne,2))
Ne_dn = Ne-Ne_up
QS = 4 #quantum states
out = fermionOps()
Cup,Cdn,Nup,Ndn,Ndens,F,O,Id = out

initTensor = [zeros(1,QS,1) for i=1:Ns]
for i = 1:Ns
   initTensor[i][1,1,1] = 1.0
end

for i = 1:Ne_up
  bb = initTensor[i]
  aa = contract([2,1,3],Cup',2,bb,2)
  initTensor[i] = aa
  for j = 1:i-1
    xx = initTensor[j]
    yy = contract([2,1,3],F,2,xx,2)
    initTensor[j] = yy
  end
end
for i = 1:Ne_dn
  bb = initTensor[i]
  aa = contract([2,1,3],Cdn',2,bb,2)
  initTensor[i] = aa
  for j = 1:i-1
    xx = initTensor[j]
    yy = contract([2,1,3],F,2,xx,2)
    initTensor[j] = yy
  end
end

psi = MPS(initTensor)
\end{lstlisting}
where the application of $\hat F$ operators is justified in Apx.~\ref{jordanwigner}.  However, there is again a simplified interface for doing this in the {\tt applyOps!} function:
\begin{lstlisting}[numbers=none]
#Ns for number of sites
QS = 4 #size of physical index
psi = MPS(QS,Ns)

out = fermionOps()
Cup,Cdn,Nup,Ndn,Ndens,F,O,Id = out

Cupdag = Matrix(Cup')
sites = [2,4]
applyOps!(psi,sites,Cupdag,trail=F)
\end{lstlisting}
which applies up fermions on sites 2 and 4.

The second input in the {\tt MPS} function is for the orthogonality center and it is default set to the first site unless otherwise specified.

\subsubsection{Complex numbers with MPOs to MPSs}

On occaision, there will be the need to introduce complex numbers into the MPS and MPO expressions.  There is a simple interface for this.  For both the MPS and MPO, the leading input can be a {\tt DataType} which in Julia would appear as
\begin{lstlisting}[numbers=none]
psi = MPS(ComplexF64,initTensor)
#some mpo already created
mpo = MPO(ComplexF64,mpo)
\end{lstlisting}
This will keep everything uniform at the outset and make it easier for the DMRG function to handle the inputs.  One physical situation where complex numbers are required is on a breathing lattice \cite{flynn2020two}.

\subsection{Environments}

It can be noted that when updating, for example, the first tensor on the lattice that none of the other right environments are changed.  So, when the computation is moved to the second site, the second right environment tensor will still be correct for the reduced site problem. The environment for the left and right partitions of the wavefunction should be ideally stored for efficient numerical computation.  There is nothing explicitly preventing the re-contraction of the left- and right-normalized tensors starting from the end of the lattice, but this is inefficient and ruins the local nature of the computations if this is done at each step.

The container for the MPO or just a generic vector could be used here for the environment types, but for ease of coding, it was found that defining a new explicit type was useful for reading the code and in some other situations as for large tensors.
\begin{lstlisting}[numbers=none]
mutable struct environment{W} 
  V::Array{W,1}
end
\end{lstlisting}
There is no standard variable for the environment tensors, as was available for the MPS and MPO.  The variable {\tt V} is used here but it is not to be confused with the $\hat V^\dagger$ from the SVD.

Note that by calling
\begin{lstlisting}[numbers=none]
#any number of tensors input
A = randMPS(2,10).A #for example
#...
envA,envB,envC = environment(A,B,C)
\end{lstlisting}
that an environment type will be output.

The environments can be initialized with the function {\tt makeEnv}
\begin{lstlisting}[numbers=none]
Lenv,Renv = makeEnv(psi,mpo)
\end{lstlisting}
which admits any number of MPOs in the last input ({\it e.g.} for two MPOs, {\tt makeEnv(psi,mpo,mpo)}).

\subsection{Large tensors}

Very large computations require a tremendous amount of memory. This is particularly true for large bond dimensions on the order of many thousands. This can be simply initialized before the DMRG function with
\begin{lstlisting}[numbers=none]
bigmps = largeMPS(psi)
bigmpo = largeMPO(mpo)
\end{lstlisting}
The DMRG function will admit these tensor types and automatically create a large environment type inside of the program (the {\tt makeEnv} function will return the large environment types with large input types).  The output files are saved to the disk in the same folder where the computation is run.  For ease of deletion, the files are given a special extension {\tt .dmrjulia} but this file is simply the type exported by Julia's {\tt Serialization} function.

The fields stored in the large MPS and MPO types are to avoid the need to recall the tensors from the disk for simple quantities.
\begin{lstlisting}[numbers=none]
#abstract type: MPS
mutable struct largematrixproductstate
  A::Array{String,1}
  oc::intType
  type::DataType
end
  
#abstract type: MPO
mutable struct largematrixproductoperator
  H::Array{String,1}
  type::DataType
end

#abstract type: Env
mutable struct largeenvironment
  V::Array{String,1}
  type::DataType
end
\end{lstlisting}
For the MPS, this includes the orthogonality center, and for each of the large types, the {\tt type} corresponding to the type of element in each tensor is stored. The strings that are stored are the file names on disk where the computer must load the tensor from.

To load an MPS saved to the disk, the elements can be called as they are in any other context with {\tt psi[2]} or similar.  The square brackets will call a function {\tt getindex!} which will then load the file from the disk.  There are also several other helper functions defined in the Supplemental Material \cite{tensor_recipes}.

If the computation has finished and the wavefunction should be reloaded from the disk in full, the following lines of code will be sufficient
\begin{lstlisting}[numbers=none]
Ns = 100 #number of lattice points
psi = loadMPS(Ns)
mpo = loadMPO(Ns)
Lenv = loadLenv(Ns)
Renv = loadRenv(Ns)
\end{lstlisting}

\section{The density matrix renormalization group}\label{DMRG}

In light of the above operations, there is a very straightforward way to implement the DMRG algorithm to find the ground state.  For a very basic, starter code, an example is provided in Appendix~\ref{DMRGcode}.  It may be useful to follow along with that short code while seeing the concepts here.

From the MPS network, a variational algorithm can be formulated by removing a few of the tensors in the MPS.  For that operation, a partial derivative $\partial/\partial A^{*\sigma_i}_{a_{i-1}a_i}$ would remove the dual tensor $A^{*\sigma_i}_{a_{i-1}a_i}$ from the network so that it can be variationally optimized.

The ground state energy is expressed as
\begin{equation}\label{energyeigen}
E=\frac{\langle\Psi|\mathcal{H}|\Psi\rangle}{\langle\Psi|\Psi\rangle}
\end{equation}
or $\langle\Psi|\mathcal{H}|\Psi\rangle - E\langle\Psi|\Psi\rangle=0$ and in diagrammatic form as
\begin{equation}
\includegraphics[width=\columnwidth]{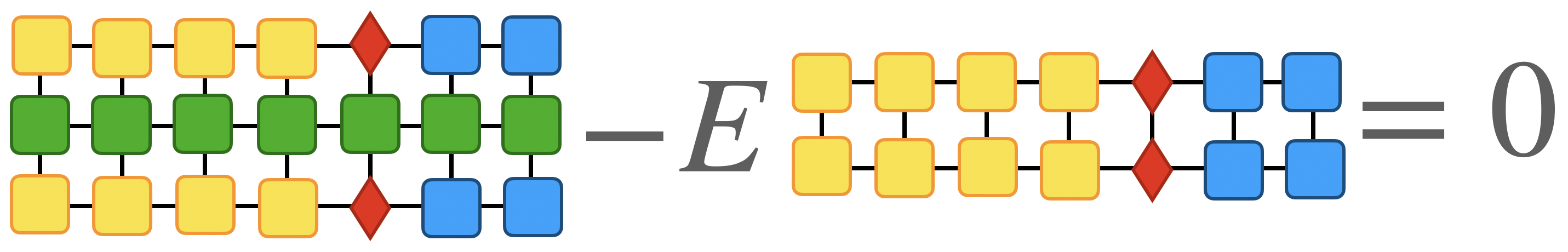}\nonumber
\end{equation}

For the problem reduced to two-sites, the variational principle reads
\begin{equation}\label{variationalMPS}
\frac{\partial^2}{\partial A^{*\sigma_i}_{a_{i-1}a_i}\partial A^{*\sigma_{i+1}}_{a_ia_{i+1}}}\Big(\langle\Psi|\mathcal{H}|\Psi\rangle - E\langle\Psi|\Psi\rangle\Big)=0
\end{equation}
and it is straightforward to generalize the problem to $N_r$-site reduced cells where $N_r\in \mathbb{Z}^+$ ({\it i.e.}, $N_r>0$). In diagrammatic notation, this appears as
\begin{equation}
\includegraphics[width=\columnwidth]{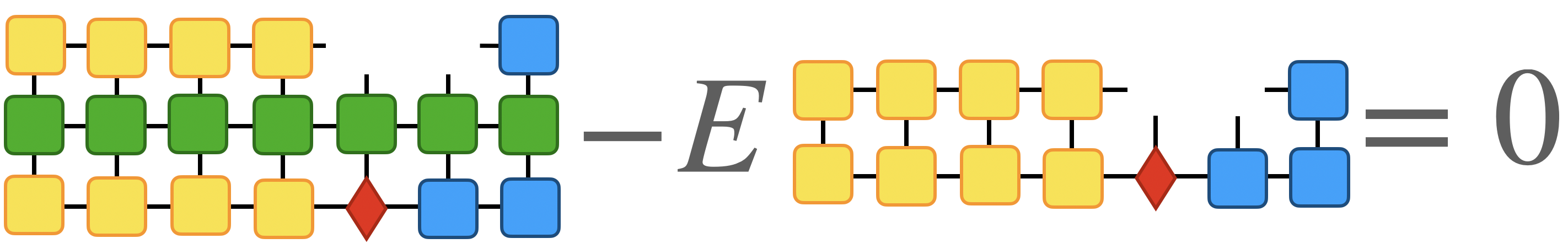}\nonumber
\end{equation}
All but two sites are present in the system, and all other tensors can be contracted up into a single rank-3 environment tensor.  This form can be recognized as an eigenvalue problem of the form $\mathcal{H}\Psi=E\Psi$.  This is the full eigenvalue problem if the full MPS was input into the problem, not some other eigenvalue problem native to only two sites--it is for the whole lattice.  When truncated, the smaller tensor will be a good approximation of the full problem.

Evaluating the eigenvalue problem directly here is expensive (see Apx.~\ref{complexity}) and that can be replaced with a more efficient algorithm.  The appearance of the wavefunction in the second term of Eq.~\eqref{variationalMPS} is always enforced to give a norm of 1, although this can be extended when necessary (see Sec.~\ref{quantchem}).  With this knowledge, an algorithm diagonalizing the problem in a Krylov subspace can be used to obtain the updated tensors.  Here, a Lanczos method will be used.

Using the development above, the steps involved in a DMRG calculation are:
\begin{enumerate}
\item For a given MPS gauged to some site representing the orthogonality center, select two sites to form the reduced site problem.

\item Contract the tensors together.

\begin{center}
\includegraphics[width=0.5\columnwidth]{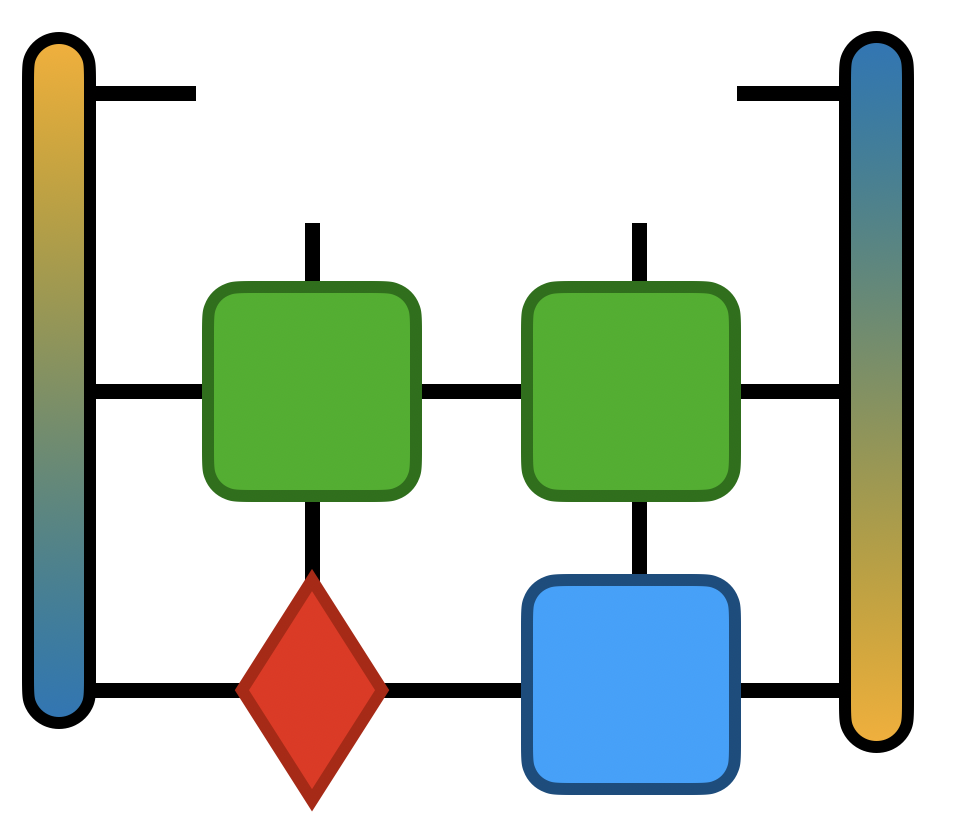}
\end{center}
where the tensors on the left and right of the diagram represent the rest of the environments found by contracting all other tensors up to a single rank-3 tensor.

\item Perform an operation to update the tensors.

This can be a random perturbation, Lanczos \cite{lanczos1950iteration}, Davidson \cite{davidson1975iterative}, power or any other method that would modify the tensors.  The most efficient choice is the Lanczos method and that will be used here.  The input tensor is normalized so that the input tensor is of norm 1. Then a Lanczos recursion of the form
\begin{equation}\label{Lanczos}
|\psi_{n+1}\rangle=\mathcal{H}|\psi_n\rangle-\alpha_n|\psi_n\rangle-\beta_{n}|\psi_{n-1}\rangle
\end{equation}
where $\psi$ is the two-site input tensor and $n\in\{1,\ldots,N_\mathrm{Lanczos}\}$ is the iteration of the Lanczos algorithm.  The coefficients are defined as
\begin{equation}
\alpha_n=\langle\psi_n|\mathcal{H}|\psi_n\rangle\quad\mathrm{and}\quad\beta_n^2=\langle\psi_{n-1}|\psi_{n-1}\rangle
\end{equation}
with $\beta_1=0$.  To evaluate Eq.~\eqref{Lanczos}, the tensors from step 1 must be contracted.  This produces the first term, $\mathcal{H}|\psi_n\rangle$ given by the diagram in step 2.

The second term in Eq.~\eqref{Lanczos} requires $\alpha_n$ which is the contraction of the above diagram onto the dual MPS tensors.  For the first step of the Lanczos routine, the first two terms are added to each other.  The first non-zero $\beta$ coefficient is then obtained by contracting the resulting tensor onto itself (finding the norm).

Diagonalization of the Hamiltonian in the Krylov subspace, taking the tridiagonal form
\begin{equation}
\mathcal{H}=\left(\begin{array}{cccc}
\alpha_1 & \beta_1 & 0 &\cdots \\
\beta_1 & \alpha_2 & \beta_2 & \cdots \\
0 & \beta_2 & \alpha_3 & \ddots \\
\vdots & \vdots &\ddots  &\ddots 
\end{array}\right),
\end{equation}
will give the energy of the system and also the coefficients to form a linear combination of Krylov vectors and form the new ground state.

Too much optimization of only two of the tensors in the network is not ideal.  Instead, this algorithm should only be run for a few iterations (typically, $N_\mathrm{Lanczos}=2$) before moving on to the next step.  This allows some update of the tensors, but not over correcting them in light of the environment probably not being correct.

Note that this Lanczos algorithm functions on the approximate ground-state MPS, as opposed to other formulations for the generic case \cite{senechal2008introduction,baker2021lanczos}.

\item Perform the SVD to break apart the combined tensor and move the orthogonality center as

\begin{center}
\includegraphics[width=\columnwidth]{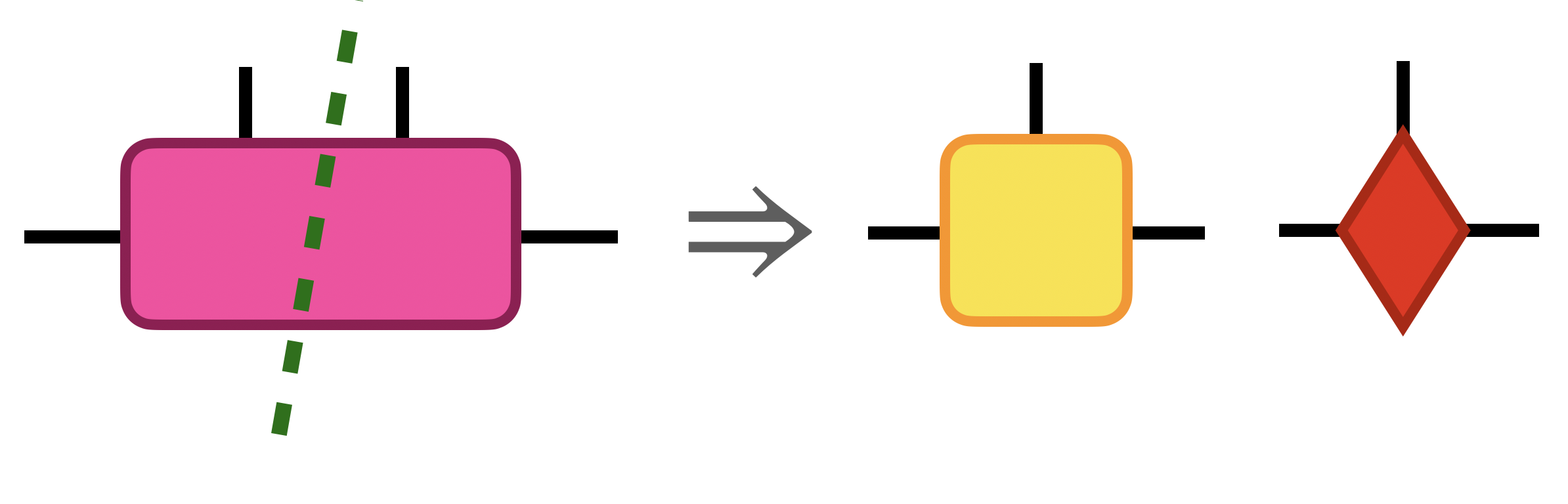}
\end{center}

where the orthogonality center was contracted onto the next site in the lattice.  This step involves a truncation of low-weight states in the density matrix. The long-range entangled states are not physically relevant in more cases, so they are discarded. This will impose a cutoff on the range of entanglement captured by the resulting system.

\item Update the environment tensor for the previous site with the newly discovered left-normalized tensor.
\begin{center}
\includegraphics[width=0.35\columnwidth]{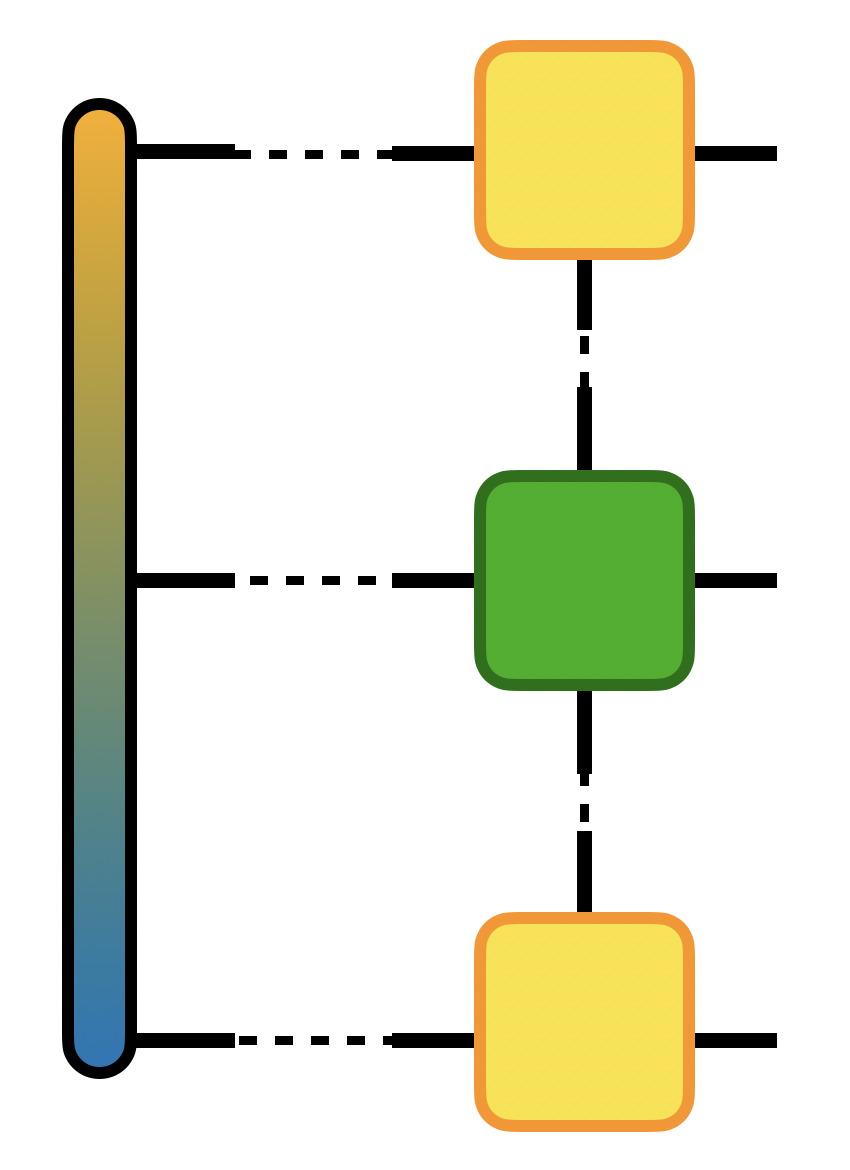}
\end{center}
\item Repeat the list, scanning back and forth across the lattice until a set number of iterations (called sweeps) or a tolerance is reached for the energy, entanglement, or some other quantity.
\end{enumerate}
This is known as the two-site algorithm and was the original formulation of DMRG.  Once the end of the lattice is reached, it is declared that one "sweep" is completed.  In this library, one sweep will mean the return of the orthogonality center to the original site (so one sweep traverses the lattice twice, back and forth).  When performing the algorithm in the reverse direction, all the steps are the same but instead of passing the orthogonality center to the right (as is shown in the above diagrams), it is passed to the left. Likewise, the right environment is updated with the right-normalized tensors in this case.

There are many other ways to write the reduced site operator (beyond the two site reduced problem given here), but this will be saved for a future discussion.  The original implementation of the two-site algorithm here is very efficient in many cases and remains unbiased.  It should produce reliable, fast results in a lot of situations.

\section{Using the density matrix renormalization group}\label{useDMRG}

Fast code and efficient algorithms can only take a user so far. The real means by which a tensor network computation becomes fast is by an informed user.  While every possible system or case can not be accounted for, it is important to understand how best to find a ground state and use measurement operations to achieve fast results.

Some operators defined in this section are defined in the appendices and Supplemental Material \cite{tensor_recipes}.  To aid in the understanding, the spin matrices are given by $\mathbf{S}=(\hat S^x,\hat S^y,\hat S^z)$ with $\hat S^\pm=S^x\pm i\hat S^y$ with $i\equiv\sqrt{-1}$.

\subsection{Finding ground states}

Even though it has been commented in several places here that DMRG finds the ground state quickly, efficiently, and makes well informed choices about how to compress the wavefunction, there is still a lot of information to cover in regards to how to obtain the correct ground state.  The method does not automatically produce the result as happens with exact diagonalization.  The method converges to the best representation of the MPS for the parameters chosen. The need to survey how well the resulting answer is comes from practice with the method.  The discussion here should be used as a starting point for future computations.

\subsubsection{Convergence criteria}

There are three essential choices that can be made when converging results.  One is to converge in the energy.  This turns out to be a rapidly convergence quantity for the reasons surrounding Eq.~\eqref{expectVal}.  This is typically the first quantity to converge in a given computation.

The next quantity to converge, typically, is the entanglement. This is because some of the small singular values take a longer time to converge and that the entropy is computed with a logarithm, so it is a longer time to make sure all digits are sufficiently converged.

Typically, the last quantity to converge is the wavefunction itself.  It may be that spin up and spin down wavefunctions in a spin symmetric system are slightly asymmetric but converge with more sweeps.  Rerunning DMRG once performing a measurement on the wavefunction is possible, and this is the recommended route for ensuring total convergence of the wavefunction.  Note that the {\tt dmrg} function here can admit an option to save the left and right environments. 

 The recommendation is to master the "secret menu" of options for at least the {\tt dmrg} function in order to see how to use it efficiently, which includes which criteria to pick for convergence.  Or the algorithm can be run inefficiently, which might be an option if the user's time is best spent on some other aspect of a given problem.  In the full {\tt dmrg} function, the extra variables to set are {\tt goal} for the target value and the boolean value {\tt cvgE} which converges in energy (default true, less strict criterion) or the entanglement on the bond set with {\tt SvNbond} ({\tt cvgE=false}, more strict convergence criterion).

Choosing the correct convergence criterion largely depends on the problem of study and what properties need to be converged.  It is cautioned that running the program for only as long as necessary is a good idea if it is also ensured that the method has actually converged.  One strategy is to run a large computation with a large bond dimension.  Then, the computation can be rerun for a smaller bond dimension.  Typically if the extrapolated quantity in terms of the bond dimensions leads to the largest computation, then this is a promising sign that the computation has converged.

\subsubsection{Quantifying uncertainty: truncation error}

The truncation error from Sec.~\ref{truncerrdiscuss} is the error uncertainty for a given wavefunction.  Based on the overlap of the wavefunctions surrounding the arguments near Eq.~\eqref{truncerr} can be used to provide an estimate on how accurate the final quantities are in DMRG when combined with Eq.~\eqref{expectVal}.  The full trace in Eq.~\eqref{expectVal} would give the proper expectation value for an operator $\mathcal{A}$
\begin{equation}
\left\langle\mathcal{A}\right\rangle\equiv\mathrm{Tr}\left(\hat \rho\mathcal{A}\right)=\sum_k \rho_k\left\langle\Phi_k\left|\mathcal{A}\right|\Phi_k\right\rangle
\end{equation}
but if the summation over $k$ is truncated, then the calculation represented in Eq.~\eqref{truncerr} applies and adds a factor of $(1-\delta)\langle\mathcal{A}\rangle$. which accounts for the uncertainty in the approximate expectation value of $\mathcal{A}$. This could easily be the energy, and this is why the reporting of the truncation error is done in many tensor network studies. Some other discussion (and diagrams) on the truncation error are contained in Ref.~\onlinecite{bakerCJP21,*baker2019m}.

Typical values for the truncation error vary depending on the system of study.  Machine precision ($10^{-10}$--$10^{-12}$ or less) is often obtained for simple models such as one-dimensional models with only near neighbor connectivity. For larger two-dimensional systems, a truncation error in the range of $10^{-5}$ is typically very good.

Note that it is assumed that the wavefunction is reasonably converged and that the relation of the eigenvalues of the density matrix was done in comparison with the exact wavefunction, so the calculation should be fully converged before using this error metric. 

\subsubsection{Choice of the cutoff}

Choosing a suitable cutoff $\eta$ is typically done by experience. It should be kept in mind that the magnitude of the singular values squared relates to the cutoff. In normal simulations, $10^{-9}$ is often used; however, a much larger $\eta$ value typically gives good energies. 

Choosing the cutoff to be too low may cause the solution to be found to not be close to the ground state. Keep in mind that the cutoff will impose an effective length scale for the entanglement of a given system, and this will limit the range of correlations allowed in the system.  So, keeping enough many body states to accurately describe the ground state is necessary.  

\subsubsection{Comments on annealing parameters for the ground state}

There are other types of DMRG that converge better to the ground-state and will be discussed later. These often include a noise term that helps the system avoid convergence to local minima.  One strategy that has been popular in tensor network codes for some time is the use of a noise parameter. It is hoped that by arming the user with tools to prepare good initial states ({\it i.e.}, by keeping a low bond dimension or 2-10 in a DMRG run with a few number of sweeps) that the need to define a scheduler will not be necessary.  Still, if one is needed for a given application, then DMRjulia can accommodate this.  It is the experience of the authors that annealing parameters in DMRG is not useful beyond setting the original wavefunction, which it is recommended here to do instead of the more elaborate scheduler.

If one did wish to use a schedule of parameters in the code, then the following example should serve as a good starting point.

\begin{lstlisting}[language=Julia]
nsweeps = 20
m = [2^i for i = 1:nsweeps]
cutoffs = [1/i for i = 1:nsweeps]
noise = [1/i for i = 1:nsweeps]
for n = 1:nsweeps
  dmrg(psi,mpo,maxm=m[n],
      cutoff=cutoffs[n],noise=noise[n],
      Lenv=storeLenv,Renv=storeRenv)
end
\end{lstlisting}

If the schedule is used, the common wisdom is to keep parameters constant for two sweeps before increasing them (not shown above). This allows for the second sweep to fully update all tensors.  Then, if necessary, the results can be extrapolated from the second run of the two.   Again, this sort of annealing of parameters in a schedule is not necessary in many situations in the experience of the authors. Rather, the suggestion is that finding a suitable starting state and allowing the algorithm to choose the best parameters unchecked is often best.

Annealing in another parameter of the system (for example, generating a series of MPOs to increase the magnitude of an interaction on the lattice) can be useful to converging results for a specific Fock space size. However, slowing increasing the bond dimension often adds to the computational time without helping converge the results.

\subsection{Measurements}

There are many possible measurements that could be applied to the MPS.  The advantage of using the tensor network here is that only a subset of tensors must be considered to obtain an expectation value, no matter the size of the lattice!  This is accomplished as a direct extension of the gauge condition from Sec.~\ref{gauge} and is explained diagrammatically in Ref.~\onlinecite{bakerCJP21,*baker2019m}.  The functions necessary to implement those features in the library are given here.

A few simple examples of measurements and how to use the measurement functions in the library is presented here for the entanglement, local measurements for one site, two sites, and $r$-site correlation functions.  An efficient algorithm for the expectation values is also given.  For one final measurement tool, the determination of the correlation length on the lattice is determined from the transfer matrix.

\subsubsection{Entanglement}

Note that the determination of the entanglement is readily available from the SVD, so this makes tensor network methods highly useful from the perspective of studying topological physics.  
\begin{lstlisting}[numbers=none]
site = cld(length(psi),2) #or other
move!(psi,site)
U,D,V = svd(psi[site],[[1,2],[3]])
rho = [D[k,k] for k = 1:size(D,1)]
\end{lstlisting}

Some code to find the entropy for each bond is
\begin{lstlisting}[numbers=none]
SvN = zeros(eltype(psi),length(psi))
for i = 1:length(psi)
  move!(psi,i)
  U,D,V = svd(psi[i],[[1,2],[3]])
  for k = 1:size(D,1)
    rhoval = D[k,k]^2
    SvN[i] -= rhoval * log(rhoval)
  end
end
\end{lstlisting}
which takes the SVD according to the diagram
\begin{equation}
\includegraphics[width=0.25\columnwidth]{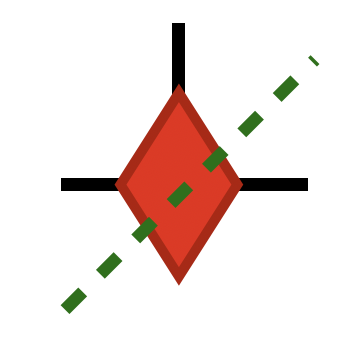}\nonumber
\end{equation}
This will generate all of the entropies for all the bonds in the system in the vector {\tt SvN}.

\subsubsection{Single site measurements}

In the appendix on exact diagonalization (Apx.~\ref{ED}), it is commented that local operators for the full system appear as a tensor product of operators. This results in a very large operator to be multiplied onto a wavefunction. The tensor network has the capability to keep the operators local while performing measurements.  The simplest is the single site measurement which can be efficiently implemented by a careful re-gauging of the tensors.  It was shown graphically in Ref.~\onlinecite{bakerCJP21,baker2019m} that by gauging the MPS to the site of interest that the rest of the tensors contract to the identity.  

When measuring $\langle \hat S^z_i\hat S^z_j\rangle$, the following diagram must be computed:
\begin{equation}
\includegraphics[width=0.25\columnwidth]{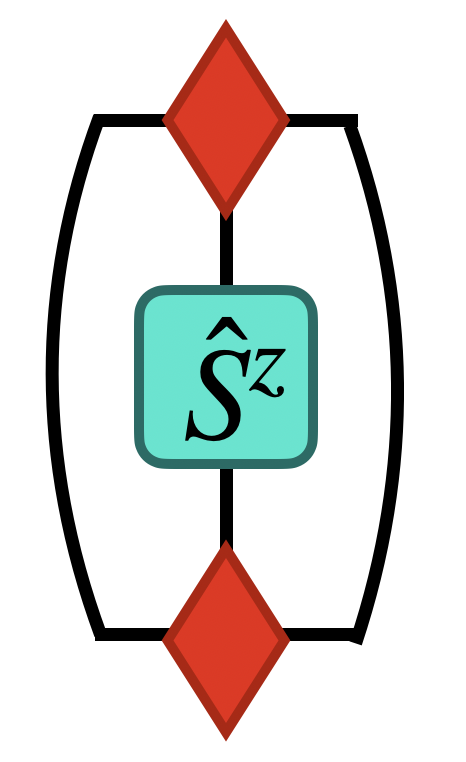}\nonumber
\end{equation}

Therefore, the only operation that must be performed is the movement of the gauge and then contraction of a particular operator onto the physical index of the MPS.

\begin{lstlisting}[numbers=none]
#definition of psi
#optional solution of psi
#definition of Sz
g = eltype(psi)
meas = Array{g,1}(undef,length(psi))
for i = 1:length(psi)
  move!(psi,i)
  C = contract([2,1,3],Sz,2,psi[i],2)
  meas[i] = ccontract(psi[i],C)
end
\end{lstlisting}

\subsubsection{Two site measurements}

Starting from either edge of the MPS, the left- or right-normalized tensors will contract to the identity so long as there are no operators applied.  When attempting to determine the correlation function for a 2-point function, orthogonality is broken between the two sites where operators are applied.   The tensors in between do not contract to the identity because the operator was applied on one of the sides between a tensor in the middle and the edge. Once the second operator is reached, orthogonality is present for the remaining tensors because they are right-normalized.

\begin{equation}
\includegraphics[width=0.75\columnwidth]{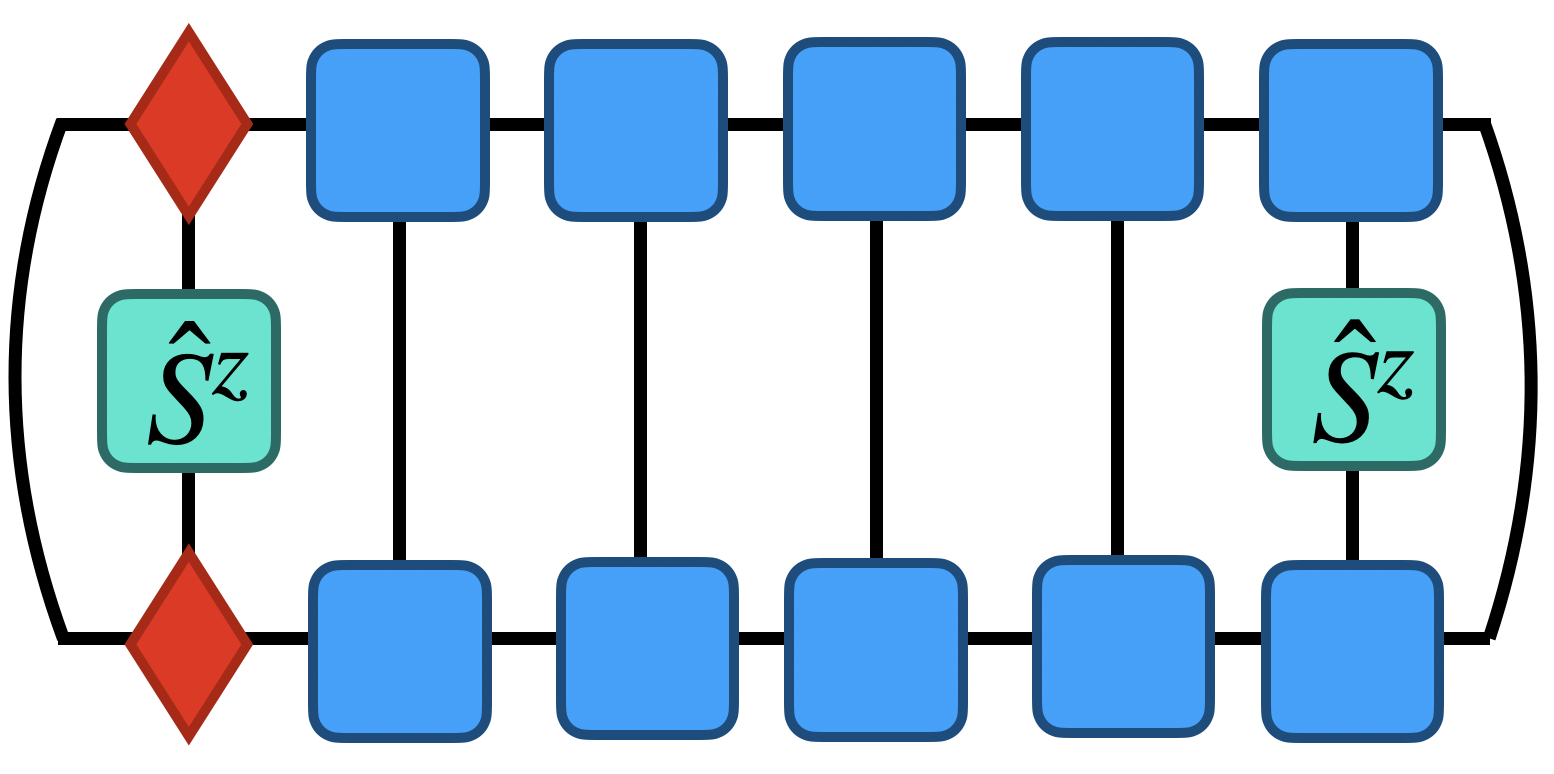}\nonumber
\end{equation}

The strategy should then to be to gauge the MPS to the site $i$ representing the first operator and perform all contractions out to the site $j$.  After the site $j$, the tensors contract to the identity, so they do not contribute to the correlation function. To make this operation efficient, the environments should be saved before moving the second site over.

DMRjulia's syntax for this operation is (the first example is $\langle\hat c^\dagger_{i\uparrow}\hat c_{j\downarrow}\rangle$ and the second is $\langle\hat S_i^+\hat S_j^-\rangle$)
\begin{lstlisting}[numbers=none]
#with trailing operator
Cupd = Array(Cup')
correlationmatrix(psi,Cupd,Cdn,trail=F)
#without trailing operator
correlationmatrix(psi,Sp,Sm)
\end{lstlisting}
This function should only be used when the resulting correlation matrix is symmetric and for two-point correlation functions (two inputs).  Otherwise, a more general function for $r$-point correlation functions can be used as is explained next.

\subsubsection{Measuring $r$-point correlation functions}

If the $r$-point correlation function \cite{economou1983green},
\begin{equation}
\langle\Psi|\varphi(x_1)\varphi(x_2)\ldots\varphi(x_r)|\Psi\rangle
\end{equation}
must be measured, then a uniform function that can handle all $r$-point correlations functions can be used.

The basic idea is to list the operators as they are applied in order.   The operators are not allowed to cross, instead the following will be repeated for every permutation of the operator order.  For example, the 4-point correlation function
\begin{equation}
\langle\Psi|\hat c^\dagger_i\hat c^\dagger_j\hat c_k\hat c_\ell|\Psi\rangle
\end{equation}
can be evaluated where $i\leq j\leq k\leq\ell$ before then permuting the indices.

For this computation, orthogonality is not the most important quantity to keep track of.  Technically, this was not necessary in the {\tt correlationmatrix} function, but it is useful to think of evaluating the problem with the gauge condition enforced in case an isolated function must be taken.  Instead, the 4-point correlation function is simply contracted up from the right, saving the boundary each time.  The right most operator (in order $\ell,k,j,i$) is incremented only when the end of the chain is reached. The right environments do not need to be recomputed and can instead be generated at the beginning with {\tt makeEnv}.  The left environments are updated each time any operator is incremented.

To use this function, DMRjulia uses (for the correlation function $\langle\hat S^+_i\hat S^+_j\hat S^-_k\hat S^-_\ell\rangle$)
\begin{lstlisting}[numbers=none]
correlation(psi,Sp,Sp,Sm,Sm)
\end{lstlisting}
or any number of operators and optionally with the {\tt trail} input from earlier.

This function will also compute the two site measurements from the previous sections, but it is slightly more efficient to use the implementation there. This is because this function does not recognize that the result may be Hermitian and so will compute both the $i\leq j$ and $j<i$ components, marking some overhead over the other function.

\subsubsection{Expectation values}\label{expectvalues}

When contracting an entire MPS and an optional number of MPOs, it is best to use the ideal contraction order presented in Apx.~\ref{complexity}.  When doing this, the notation used for the contraction order delivers a repeating pattern that can be used to define one function for any expectation value that would be taken ({\it i.e.}, any number of MPOs).

The order for two MPOs is shown in Fig.~\ref{MPOorder}.  The generic step is that the first contraction is between the left environment tensor and the first dual MPS tensor. If contracted as

\begin{equation}
\includegraphics[width=0.5\columnwidth]{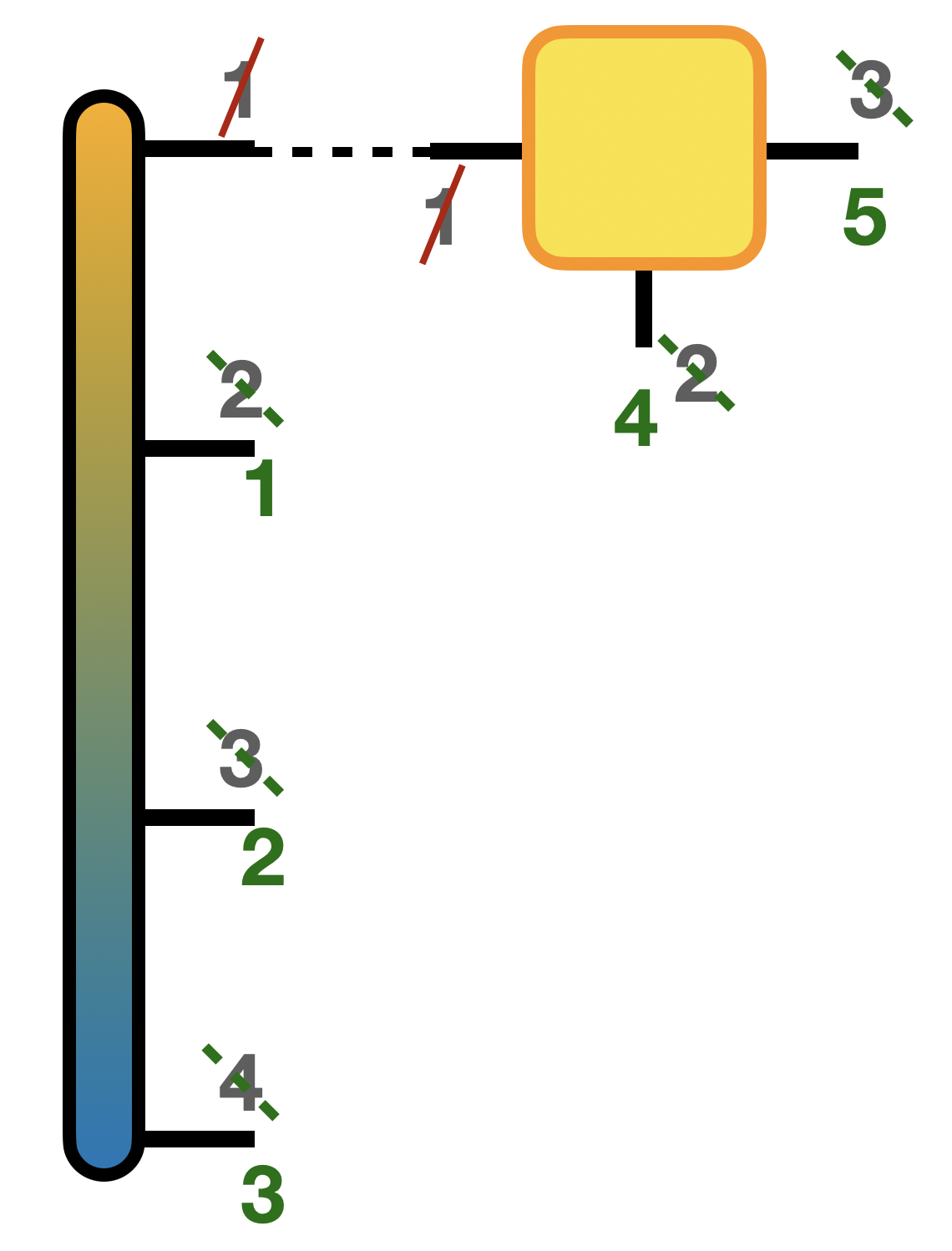}\nonumber
\end{equation}

Then in each step, the indices 1 and $g$ where $g$ is the number of MPOs plus 2 (in this case, 1 and 4 are the indices that are always contracted on the left environment).

\begin{equation}
\includegraphics[width=0.45\columnwidth]{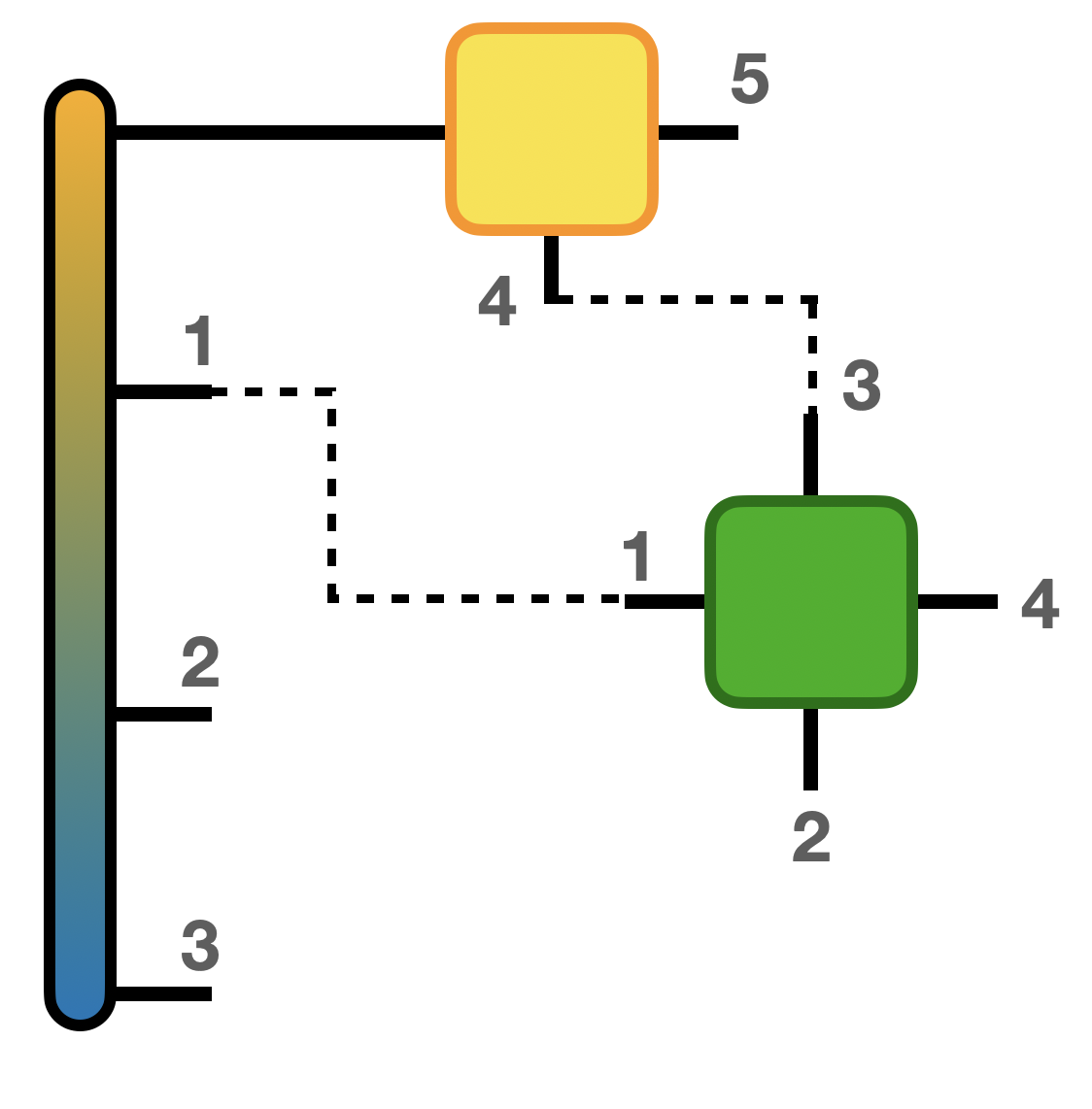}\quad\quad\quad
\includegraphics[width=0.45\columnwidth]{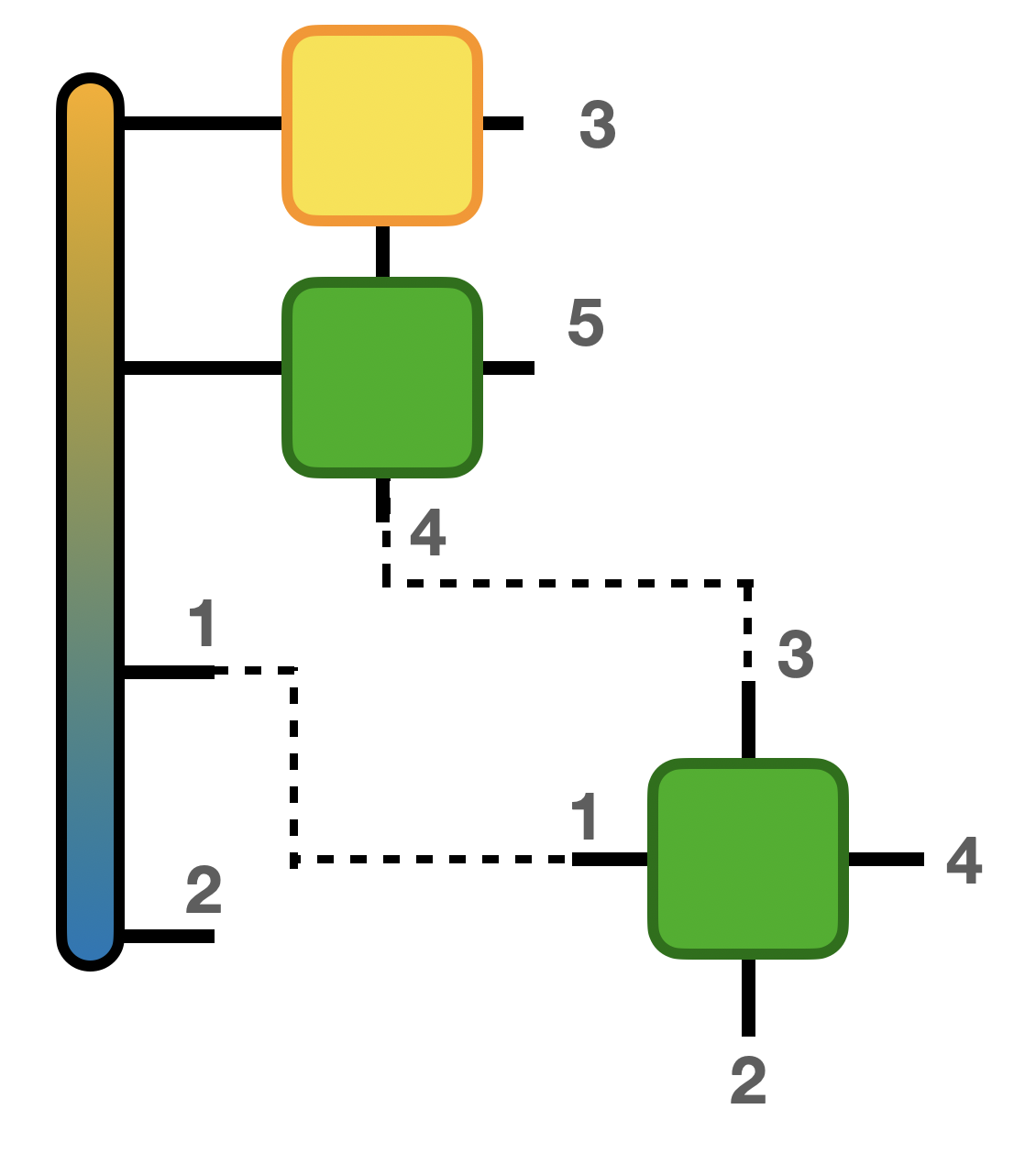}\nonumber
\end{equation}

The tensors are then contracted until the end of the MPS using this strategy.  The result is then contracted to the right environment on the last site and the output is reported.

Note that using {\tt expect} for generic wavefunction measurements will result in all tensors being contracted.  It is far more efficient (by a factor of the lattice size) to use the gauge of the MPS as in the previous three sections to find these quantities.

\begin{lstlisting}[numbers=none]
expect(psi,mpo) #<H>
expect(psi,mpo,mpo) #<H^2>
expect(psi,mpo,mpo,mpo) #<H^3>
#when the dual does not match the MPS
expect(altpsi,psi,mpo)
\end{lstlisting}
And in general any number of MPOs are admissable to the expect function (provided that they are added as the last input).

The resulting code to use this function that handles all of these expectation values is surprisingly short considering how widely applicable this function is.

\subsubsection{Correlation lengths}\label{correlationlengths}

The idea of a transfer matrix extends back to the original solution of an Ising chain \cite{reif2009fundamentals}. The extension here is to identify that a tensor in the MPS and its dual tensor, contracted along the physical index, form the transfer matrix.  Some analysis of the transfer matrix to explain the appearance of exponential and power-law decays was discussed in Ref.~\onlinecite{bakerCJP21,*baker2019m}.

Applying two operators on the lattice such as the example correlation function, $\langle\Psi|\hat S^+_i\hat S^-_j|\Psi\rangle$, contains a set of transfer matrices between sites $i$ and $j$. As was discussed in Ref.~\onlinecite{bakerCJP21,*baker2019m}, the transfer matrix can be thought of in terms of its eigenvalue decomposition. Doing this allows us to track how each eigenvalue develops with more and more transfer matrices.  Namely, the characteristic form of each eigenvalue can be thought of in the form $\exp(-x/\xi)$ where $1/\xi=-\ln\rho_k$, effectively identifying the eigenvalue with a correlation length.  The distance $x$ is the difference between two lattice points.

To apply this to the determination of the transfer matrices, the computation of transfer matrices between two sites with spin operators appears diagrammatically as

\begin{equation}
\includegraphics[width=0.75\columnwidth]{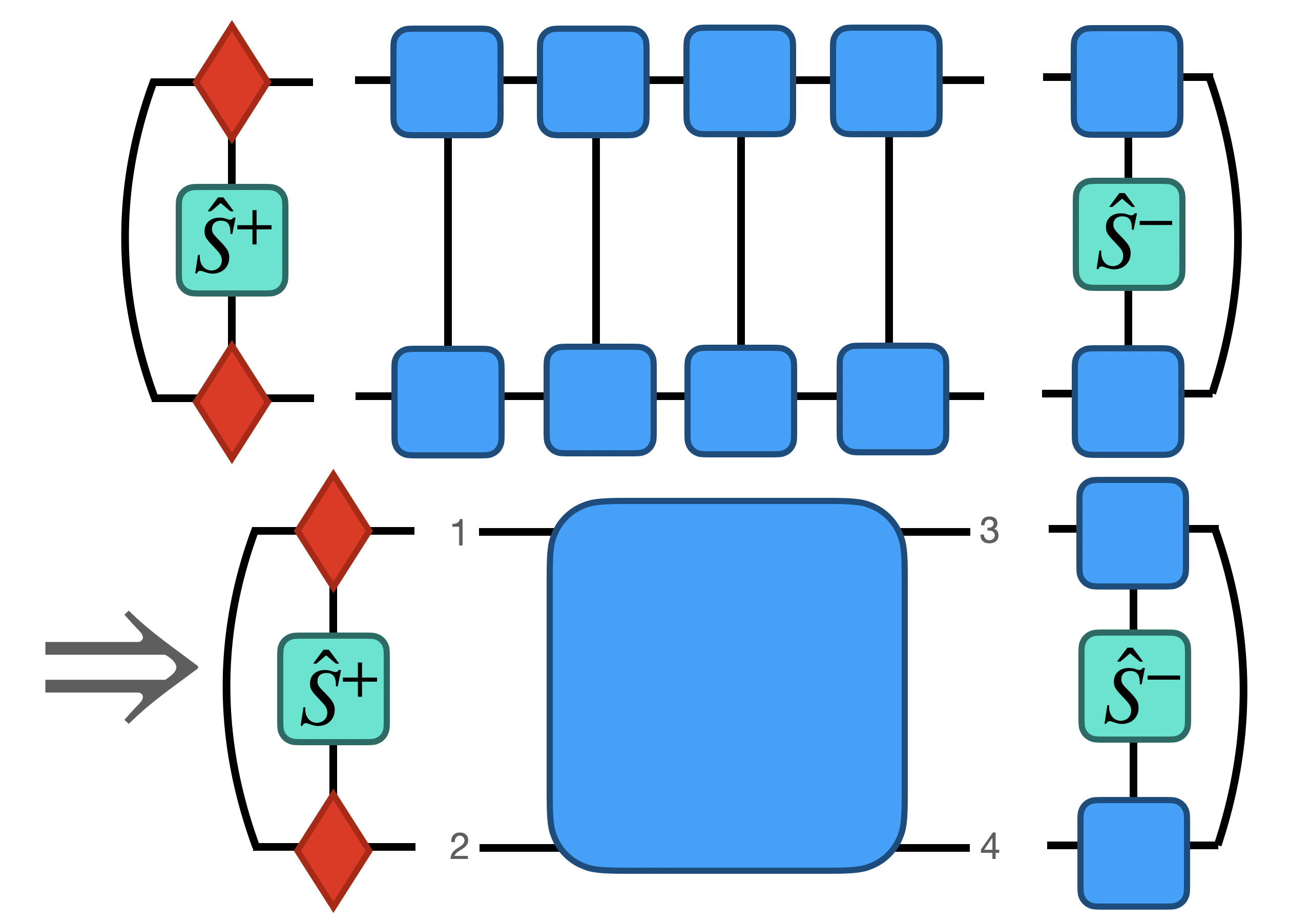}\nonumber
\end{equation}
where the order of indices used in the library are listed on the center tensor: the transfer matrix.  Taking the eigenvalue decomposition of this tensor reveals the eigenvalues which can be used to compute the correlation length.  By manipulating the exponential form above, the correlation length can be computed. Some code to do this in DMRjulia is
\begin{lstlisting}[numbers=none]
tm = transfermatrix(psi,2,4)
D,U = eigen(tm,[[1,3],[2,4]])
#to reload
transfermatrix(tmat,5,8,transfermat=tm)
\end{lstlisting}
where the example shown determines the transfer matrix between sites 1 and 5 where presumably an operator was applied.  From the results of the eigenvalue decomposition, the correlation length can be computed from the eigenvalue decomposition of the above. For one site, this takes the form \cite{schollwock2011density}
\begin{equation}
\mathbb{T}=\sum_k|k\rangle\rho_k\langle k|
\end{equation}
and more details in terms of diagrams are listed in Ref.~\onlinecite{bakerCJP21,*baker2019m}. The index $k$ indexes the eigenvalues of the matrix with associated eigenvalues $|k\rangle$. The full correlation function is then straightforwardly \cite{schollwock2011density}
\begin{equation}
\frac{\langle\Psi|\hat S^+_i\hat S^-_j|\Psi\rangle}{\langle\Psi|\Psi\rangle}=\sum_{k=1}^{m^2}\mathbb{L}|k\rangle\rho_k^x\langle k|\mathbb{R}
\end{equation}
where the total size $m^2$ assumes that the ends of the transfer matrix are uniform of size $m$.  The operators $\mathbb{L}$ and $\mathbb{R}$ correspond to the sites where the operators were applied (left and right sites in the above diagram) and the appropriate environments.

Note that even in models with only near-neighbor interactions that the correlation length can diverge.  This illustrates that the range of interactions in the MPO often has little to do with the resulting correlation length.  Other analysis tools, such as this transfer matrix approach, should be used instead of analysis of the Hamiltonian only.

The path taken on a two-dimensional lattice can inform which direction the correlation length applies to.

\subsection{Some cases of interest}

Some of these systems will require more computational resources (memory and time) to evaluate.  The inclusion of these limitations here should not be a substitute for running these systems and gaining first-hand knowledge.  Instead, they should be attempted to try and solve larger problems.  Using as much knowledge as possible to ensure that computations has converged should always be applied to all problems ({\it i.e.}, checking against exact limits and calculations, other publications, etc.).

\subsubsection{Periodic boundary conditions}\label{pbcs}

\begin{figure}
\includegraphics[width=0.5\columnwidth]{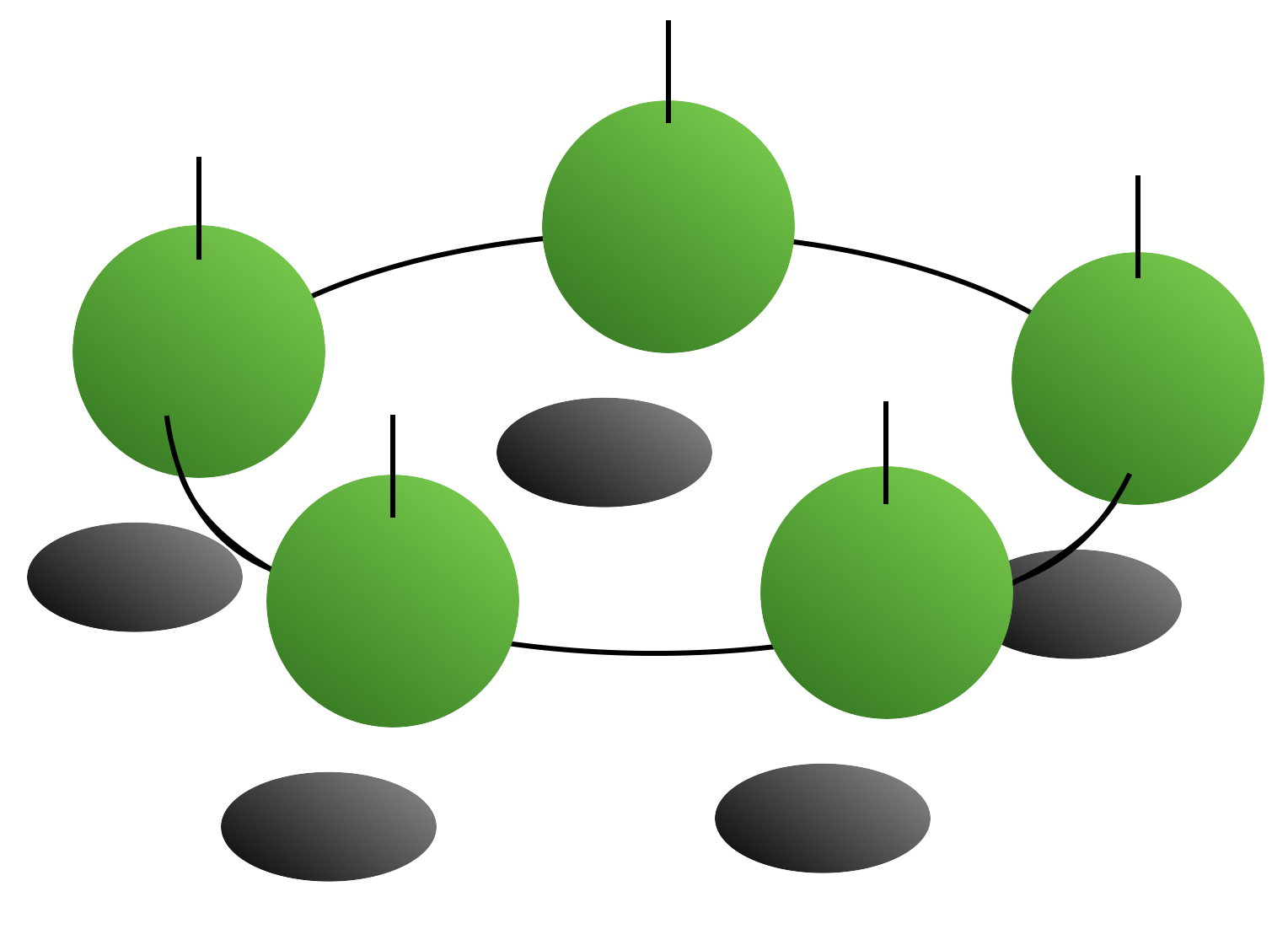}
\caption{Five MPS tensors on a ring. Periodic boundary conditions in tensor networks can converge poorly. A choice of which bond to fix while the others are optimized must be done.  Then, another bond is selected as the fixed bond.\label{pbc}}
\end{figure}

Periodic boundary conditions are notoriously hard for a tensor network to capture.  In the most basic sense, it is not clear where to start contracting the network.  Considering a ring such as demonstrated in Fig.~\ref{pbc}, in order to start the computation from a finite MPS, one of the bonds will need to be chosen to be opened initially.  This means that the bond will not be optimized in the DMRG step, and that the environments for the bond are fixed.

Once the other bonds are converged, then another bond is chosen to be fixed (the environment tensors are updated to now describe this bond). Then the ring is then split another this bond, with the others optimized.  This repeated action of splitting a bond and optimizing the rest produces a convergence pattern that involves jumps in the energy.  This is because re-setting the problem to the new bond often means that the previous sites were optimized in the wrong environment.  This illustrates the difficulty in using periodic boundary conditions: not all environments can be optimized at the same time. 

Many of these computations can be suitably handled by an infinite method, and this will be covered in a subsequent paper.

Note that one additional option could be used to obtain a periodic boundary condition.  If a "long-bond" is added to the MPO ({\it e.g.}, one interaction starting on site 1 and going to the last site), then this will also generate a suitable periodic boundary condition while maintaining a fixed environment.  This strategy can be hard to converge and increases the size of the MPO. Nevertheless, it might be useful as a fast means to obtaining results for small periodic systems.

\subsubsection{Quantum chemistry systems}\label{quantchem}

Quantum chemistry Hamiltonians involve terms beyond quadratic terms. The quartic term that appears is due to the Coulomb interaction and causes some difficulty due to its long range \cite{fetter2012quantum}.  The many-body Hamiltonian in second quantization is
\begin{equation}
\mathcal{H}=\sum_{ij}\left(t_{ij}\hat c^\dagger_i\hat c_i+\sum_{k\ell}V_{ijk\ell}\hat c^\dagger_i\hat c^\dagger_j\hat c_\ell\hat c_k\right)
\end{equation}
where the one ($t_{ij}$) and two ($V_{ijk\ell}$) electron integrals are taken from some other source which can efficiently compute them for a given basis set. Note that the basis sets for these problems are often non-orthogonal and that this is often the most efficient representation of the problem. When using a non-orthogonal basis set, the eigenvalue problem surrounding Eq.~\eqref{energyeigen} must now include an overlap matrix so that the resulting generalized eigenvalue problem can be solved.

When solving a many-body problem with a quartic term in the second quantized Hamiltonian.

The MPO for this term is known analytically from Ref.~\onlinecite{white1999ab}.  However, the recent development of automatic constructors for matrix product operators is a much more straight-forward way to construct any Hamiltonian which will be covered in a subsequent paper.  It is worth understanding how to prepare MPOs by hand with the techniques in this paper before moving on to the automatic MPO constructor.

The work of Garnett Chan prominently brought the tools of DMRG to the forefront in quantum chemistry \cite{chan2002highly}, and it is recommended to at least understand the additional features input into the BLOCK code before attempting quantum chemistry simulations with DMRG \cite{chan2004algorithm,ghosh2008orbital,sharma2012spin,olivares2015ab}. Many chemistry extensions of this method have been taken in several directions (see, for example, Ref.~\onlinecite{parker2014communication}). There are also tricks that are employed dealing with the spin symmetry of a system, and these may be useful for finding ground state energies.

\subsubsection{Higher dimensional systems}

Not only is the rank of a tensor a fluid concept, but the dimension of a particular system is also a fluid concept that can be flattened into another dimension for implementation in a tensor network.  There is no specific prescription for how to write the one-dimensional path through the higher dimensional lattice, as was seen in Sec.~\ref{2DspinMPO}, but some choice can keep the average interaction length shorter.

Models in higher dimensions can be more demanding computationally. The area law makes the scaling of these algorithms much worse as we go to higher dimension. The fact that higher dimensional systems are harder to evaluate should not prevent good-faith attempts to try and solve them.  Cases in three dimensions might be out of reach, but two-dimensional systems are often suitably treated by efficient code \cite{stoudenmire2012studying}.

Some alternative types of tensor networks can handle two-dimensional calculations more efficiently \cite{verstraete2004renormalization}, but these will be covered in a subsequent paper.

\subsubsection{Symmetry sectors from pinning fields}

Solving DMRG may not automatically allow for a broken symmetry solution.  Examples of this might be situations where a spin-density wave should appear on a lattice or if some modulation of the spins should appear in a given system.  For example, it can be noticed that a staggered magnetic texture should appear for some lattices.  For example, if on a two-dimensional lattice, each column along the lattice should switch from an average value $\langle \hat S^y_i\rangle=\pm1$, then it is natural to expect that DMRG will find some linear combination of both of those states, showing a spin texture ({\it i.e.}, measuring $\langle \hat S^y_i\rangle$ on each site) that is $\langle \hat S^y_i\rangle\approx0$.

The question then becomes how to bias the state into one of the symmetry sectors.  This can be done by applying a pinning field on the lattice for a few sweeps.  Then, the system is relaxed into the ground state with the regular Hamiltonian (no pinning field).

For a practical example, see the situation in Ref.~\onlinecite{flynn2020two}.  There, a pinning field of a large magnitude (B$\sim1000$) was applied with $B\hat S^y$ on the left edge of the two-dimensional lattice and $-B\hat S^y$ on the right edge for two sweeps only.  This biases the $y$ moments towards the left or right side.  Then, the pinning field was removed and the solution of the ground state will naturally bias towards the alternating spin texture.

This strategy has been used in many situations on many different lattices and problems.  It could be remarked that this requires analytic insight into a given model to even know that the broken symmetry solution should appear.  In general, it is highly recommended that all numerical studies seek some theoretical justification, but when in doubt, the staggered field can be applied anyway.  If no staggered spin texture is contained in the ground state then when the magnetic field is removed, then the regularly ordered spin texture should return.  This would rule out a staggering.

Implementing the necessary onsite terms into the code is straightforward if using the MPO constructor.  The onsite terms are simply entered into the lower-left entry of the MPO.  However, it is also easy to add these terms to the MPO when using the automatic MPO constructor introduced in a subsequent paper in this series.

There is also a possibility to start from a random initial state in this case.

\subsubsection{Momentum space}

It is technically possible to introduce a momentum space representation into DMRG.  This has been investigated when solving two-dimensional systems that are placed onto a cylinder, where one of the directions is periodic \cite{motruk2016density,ehlers2017hybrid}.  Since there is one edge free, there is no issue with periodic boundary conditions like was discussed in Sec.~\ref{pbcs}.

However, in general the application of a discrete Fourier transformation onto a lattice will introduce long ranged correlations for interactions that appear local in real space \cite{kittel1987quantum}, and this will introduce a larger MPO representation and potentially make computing with DMRG longer.  So, it is not always useful to evaluate the model in momentum space, although it is possible.

\section{Conclusion}

A historical background on why DMRG is named as it is and what basic principles inform it were discussed. Then, four basic operations of reshaping, permuting indices, contraction, and decomposition was introduced with a diagrammatic notation.  Basic aspects of the code including matrix product states, matrix product operators, and their definitions were introduced. The algorithmic steps were then reviewed, followed by a discussion on how to measure observable quantities.

This article focuses on the implementation with dense tensors, which do not take advantage of quantum number symmetries that can simultaneously prevent numerical instabilities from forcing the wavefunction into a new quantum number sector ({\it i.e.}, changing the particle number in a Hubbard model) and also represents a vast time increase when used.  The next article in the series will focus on this implementation in an efficient way.  

\section{Acknowledgements}

Many people contributed meaningfully to the discussions that led to this code. The code was continued from some assignments in 2016 from a course by Steven R.~White. Useful discussions were also held with Michael O.~Flynn, Matthew Fishman, E.~Miles Stoudenmire, Alexandre Foley, Wangwei Lan, Rex Godby, Steven R.~White, Glen Evenbly, Shane Parker, Jodi Christiansen, and David Sénéchal.

T.E.B.~thanks funding provided by the postdoctoral fellowship from Institut quantique and Institut Transdisciplinaire d'Information Quantique (INTRIQ). This research was undertaken thanks in part to funding from the Canada First Research Excellence Fund (CFREF). 

T.E.B.~is grateful to the US-UK Fulbright Commission for financial support under the Fulbright U.S. Scholarship programme as hosted by the University of York.  This research was undertaken in part thanks to funding from the Bureau of Education and Cultural Affairs from the United States Department of State.

\bibliography{TEB_papers,refs,TEB_books}

\begin{thebibliography}{88}%
\makeatletter
\providecommand \@ifxundefined [1]{%
 \@ifx{#1\undefined}
}%
\providecommand \@ifnum [1]{%
 \ifnum #1\expandafter \@firstoftwo
 \else \expandafter \@secondoftwo
 \fi
}%
\providecommand \@ifx [1]{%
 \ifx #1\expandafter \@firstoftwo
 \else \expandafter \@secondoftwo
 \fi
}%
\providecommand \natexlab [1]{#1}%
\providecommand \enquote  [1]{``#1''}%
\providecommand \bibnamefont  [1]{#1}%
\providecommand \bibfnamefont [1]{#1}%
\providecommand \citenamefont [1]{#1}%
\providecommand \href@noop [0]{\@secondoftwo}%
\providecommand \href [0]{\begingroup \@sanitize@url \@href}%
\providecommand \@href[1]{\@@startlink{#1}\@@href}%
\providecommand \@@href[1]{\endgroup#1\@@endlink}%
\providecommand \@sanitize@url [0]{\catcode `\\12\catcode `\$12\catcode
  `\&12\catcode `\#12\catcode `\^12\catcode `\_12\catcode `\%12\relax}%
\providecommand \@@startlink[1]{}%
\providecommand \@@endlink[0]{}%
\providecommand \url  [0]{\begingroup\@sanitize@url \@url }%
\providecommand \@url [1]{\endgroup\@href {#1}{\urlprefix }}%
\providecommand \urlprefix  [0]{URL }%
\providecommand \Eprint [0]{\href }%
\providecommand \doibase [0]{http://dx.doi.org/}%
\providecommand \selectlanguage [0]{\@gobble}%
\providecommand \bibinfo  [0]{\@secondoftwo}%
\providecommand \bibfield  [0]{\@secondoftwo}%
\providecommand \translation [1]{[#1]}%
\providecommand \BibitemOpen [0]{}%
\providecommand \bibitemStop [0]{}%
\providecommand \bibitemNoStop [0]{.\EOS\space}%
\providecommand \EOS [0]{\spacefactor3000\relax}%
\providecommand \BibitemShut  [1]{\csname bibitem#1\endcsname}%
\let\auto@bib@innerbib\@empty
\bibitem [{\citenamefont {Schollw{\"o}ck}(2005)}]{schollwock2005density}%
  \BibitemOpen
  \bibfield  {author} {\bibinfo {author} {\bibfnamefont {Ulrich}\ \bibnamefont
  {Schollw{\"o}ck}},\ }\bibfield  {title} {\enquote {\bibinfo {title} {The
  density-matrix renormalization group},}\ }\href@noop {} {\bibfield  {journal}
  {\bibinfo  {journal} {Reviews of modern physics}\ }\textbf {\bibinfo {volume}
  {77}},\ \bibinfo {pages} {259} (\bibinfo {year} {2005})}\BibitemShut
  {NoStop}%
\bibitem [{\citenamefont {Schollw{\"o}ck}(2011)}]{schollwock2011density}%
  \BibitemOpen
  \bibfield  {author} {\bibinfo {author} {\bibfnamefont {Ulrich}\ \bibnamefont
  {Schollw{\"o}ck}},\ }\bibfield  {title} {\enquote {\bibinfo {title} {The
  density-matrix renormalization group in the age of matrix product states},}\
  }\href@noop {} {\bibfield  {journal} {\bibinfo  {journal} {Annals of
  physics}\ }\textbf {\bibinfo {volume} {326}},\ \bibinfo {pages} {96--192}
  (\bibinfo {year} {2011})}\BibitemShut {NoStop}%
\bibitem [{\citenamefont {Hastings}(2004)}]{hastings2004locality}%
  \BibitemOpen
  \bibfield  {author} {\bibinfo {author} {\bibfnamefont {Matthew~B}\
  \bibnamefont {Hastings}},\ }\bibfield  {title} {\enquote {\bibinfo {title}
  {Locality in quantum and markov dynamics on lattices and networks},}\
  }\href@noop {} {\bibfield  {journal} {\bibinfo  {journal} {Physical review
  letters}\ }\textbf {\bibinfo {volume} {93}},\ \bibinfo {pages} {140402}
  (\bibinfo {year} {2004})}\BibitemShut {NoStop}%
\bibitem [{\citenamefont {White}(1992)}]{white1992density}%
  \BibitemOpen
  \bibfield  {author} {\bibinfo {author} {\bibfnamefont {Steven~R}\
  \bibnamefont {White}},\ }\bibfield  {title} {\enquote {\bibinfo {title}
  {Density matrix formulation for quantum renormalization groups},}\
  }\href@noop {} {\bibfield  {journal} {\bibinfo  {journal} {Physical review
  letters}\ }\textbf {\bibinfo {volume} {69}},\ \bibinfo {pages} {2863}
  (\bibinfo {year} {1992})}\BibitemShut {NoStop}%
\bibitem [{\citenamefont {Krishna-Murthy}\ \emph {et~al.}(1980)\citenamefont
  {Krishna-Murthy}, \citenamefont {Wilkins},\ and\ \citenamefont
  {Wilson}}]{krishna1980renormalization}%
  \BibitemOpen
  \bibfield  {author} {\bibinfo {author} {\bibfnamefont {HR}~\bibnamefont
  {Krishna-Murthy}}, \bibinfo {author} {\bibfnamefont {JW}~\bibnamefont
  {Wilkins}}, \ and\ \bibinfo {author} {\bibfnamefont {KG}~\bibnamefont
  {Wilson}},\ }\bibfield  {title} {\enquote {\bibinfo {title}
  {Renormalization-group approach to the anderson model of dilute magnetic
  alloys. i. static properties for the symmetric case},}\ }\href@noop {}
  {\bibfield  {journal} {\bibinfo  {journal} {Physical Review B}\ }\textbf
  {\bibinfo {volume} {21}},\ \bibinfo {pages} {1003} (\bibinfo {year}
  {1980})}\BibitemShut {NoStop}%
\bibitem [{\citenamefont {Affleck}\ \emph {et~al.}(1988)\citenamefont
  {Affleck}, \citenamefont {Kennedy}, \citenamefont {Lieb},\ and\ \citenamefont
  {Tasaki}}]{affleck1988valence}%
  \BibitemOpen
  \bibfield  {author} {\bibinfo {author} {\bibfnamefont {Ian}\ \bibnamefont
  {Affleck}}, \bibinfo {author} {\bibfnamefont {Tom}\ \bibnamefont {Kennedy}},
  \bibinfo {author} {\bibfnamefont {Elliott~H}\ \bibnamefont {Lieb}}, \ and\
  \bibinfo {author} {\bibfnamefont {Hal}\ \bibnamefont {Tasaki}},\ }\bibfield
  {title} {\enquote {\bibinfo {title} {Valence bond ground states in isotropic
  quantum antiferromagnets},}\ }in\ \href@noop {} {\emph {\bibinfo {booktitle}
  {Condensed matter physics and exactly soluble models}}}\ (\bibinfo
  {publisher} {Springer},\ \bibinfo {year} {1988})\ pp.\ \bibinfo {pages}
  {253--304}\BibitemShut {NoStop}%
\bibitem [{\citenamefont {{\"O}stlund}\ and\ \citenamefont
  {Rommer}(1995)}]{ostlund1995thermodynamic}%
  \BibitemOpen
  \bibfield  {author} {\bibinfo {author} {\bibfnamefont {Stellan}\ \bibnamefont
  {{\"O}stlund}}\ and\ \bibinfo {author} {\bibfnamefont {Stefan}\ \bibnamefont
  {Rommer}},\ }\bibfield  {title} {\enquote {\bibinfo {title} {Thermodynamic
  limit of density matrix renormalization},}\ }\href@noop {} {\bibfield
  {journal} {\bibinfo  {journal} {Physical review letters}\ }\textbf {\bibinfo
  {volume} {75}},\ \bibinfo {pages} {3537} (\bibinfo {year}
  {1995})}\BibitemShut {NoStop}%
\bibitem [{\citenamefont {Baker}()}]{dmrjulia}%
  \BibitemOpen
  \bibfield  {author} {\bibinfo {author} {\bibfnamefont {Thomas~E.}\
  \bibnamefont {Baker}},\ }\href@noop {} {\enquote {\bibinfo {title}
  {{DMRjulia}},}\ }\bibinfo {howpublished}
  {\url{https://github.com/bakerte/DMRjulia.jl}}\BibitemShut {NoStop}%
\bibitem [{\citenamefont {Baker}\ \emph
  {et~al.}(2021{\natexlab{a}})\citenamefont {Baker}, \citenamefont
  {Desrosiers}, \citenamefont {Tremblay},\ and\ \citenamefont
  {Thompson}}]{bakerCJP21}%
  \BibitemOpen
  \bibfield  {author} {\bibinfo {author} {\bibfnamefont {Thomas~E}\
  \bibnamefont {Baker}}, \bibinfo {author} {\bibfnamefont {Samuel}\
  \bibnamefont {Desrosiers}}, \bibinfo {author} {\bibfnamefont {Maxime}\
  \bibnamefont {Tremblay}}, \ and\ \bibinfo {author} {\bibfnamefont {Martin~P}\
  \bibnamefont {Thompson}},\ }\bibfield  {title} {\enquote {\bibinfo {title}
  {{M{\'e}thodes de calcul avec r{\'e}seaux de tenseurs en physique}},}\ }\href
  {\doibase https://doi.org/10.1139/cjp-2019-0611} {\bibfield  {journal}
  {\bibinfo  {journal} {Canadian Journal of Physics}\ }\textbf {\bibinfo
  {volume} {99}},\ \bibinfo {pages} {4} (\bibinfo {year}
  {2021}{\natexlab{a}})}\BibitemShut {NoStop}%
\bibitem [{\citenamefont {Baker}\ \emph {et~al.}(2019)\citenamefont {Baker},
  \citenamefont {Desrosiers}, \citenamefont {Tremblay},\ and\ \citenamefont
  {Thompson}}]{baker2019m}%
  \BibitemOpen
  \bibfield  {author} {\bibinfo {author} {\bibfnamefont {Thomas~E}\
  \bibnamefont {Baker}}, \bibinfo {author} {\bibfnamefont {Samuel}\
  \bibnamefont {Desrosiers}}, \bibinfo {author} {\bibfnamefont {Maxime}\
  \bibnamefont {Tremblay}}, \ and\ \bibinfo {author} {\bibfnamefont {Martin~P}\
  \bibnamefont {Thompson}},\ }\bibfield  {title} {\enquote {\bibinfo {title}
  {Basic tensor network computations in physics},}\ }\href@noop {} {\bibfield
  {journal} {\bibinfo  {journal} {arXiv preprint arXiv:1911.11566, p.~19}\ }
  (\bibinfo {year} {2019})}\BibitemShut {NoStop}%
\bibitem [{\citenamefont {Hallberg}(2006)}]{hallberg2006new}%
  \BibitemOpen
  \bibfield  {author} {\bibinfo {author} {\bibfnamefont {Karen~A}\ \bibnamefont
  {Hallberg}},\ }\bibfield  {title} {\enquote {\bibinfo {title} {New trends in
  density matrix renormalization},}\ }\href@noop {} {\bibfield  {journal}
  {\bibinfo  {journal} {Advances in Physics}\ }\textbf {\bibinfo {volume}
  {55}},\ \bibinfo {pages} {477--526} (\bibinfo {year} {2006})}\BibitemShut
  {NoStop}%
\bibitem [{\citenamefont {Or{\'u}s}(2014)}]{orus2014practical}%
  \BibitemOpen
  \bibfield  {author} {\bibinfo {author} {\bibfnamefont {Rom{\'a}n}\
  \bibnamefont {Or{\'u}s}},\ }\bibfield  {title} {\enquote {\bibinfo {title} {A
  practical introduction to tensor networks: Matrix product states and
  projected entangled pair states},}\ }\href@noop {} {\bibfield  {journal}
  {\bibinfo  {journal} {Annals of Physics}\ }\textbf {\bibinfo {volume}
  {349}},\ \bibinfo {pages} {117--158} (\bibinfo {year} {2014})}\BibitemShut
  {NoStop}%
\bibitem [{\citenamefont {Bridgeman}\ and\ \citenamefont
  {Chubb}(2017)}]{bridgeman2017hand}%
  \BibitemOpen
  \bibfield  {author} {\bibinfo {author} {\bibfnamefont {Jacob~C}\ \bibnamefont
  {Bridgeman}}\ and\ \bibinfo {author} {\bibfnamefont {Christopher~T}\
  \bibnamefont {Chubb}},\ }\bibfield  {title} {\enquote {\bibinfo {title}
  {Hand-waving and interpretive dance: an introductory course on tensor
  networks},}\ }\href@noop {} {\bibfield  {journal} {\bibinfo  {journal}
  {Journal of Physics A: Mathematical and Theoretical}\ }\textbf {\bibinfo
  {volume} {50}},\ \bibinfo {pages} {223001} (\bibinfo {year}
  {2017})}\BibitemShut {NoStop}%
\bibitem [{\citenamefont {Townsend}(2000)}]{townsend2000modern}%
  \BibitemOpen
  \bibfield  {author} {\bibinfo {author} {\bibfnamefont {John~S}\ \bibnamefont
  {Townsend}},\ }\href@noop {} {\emph {\bibinfo {title} {A modern approach to
  quantum mechanics}}}\ (\bibinfo  {publisher} {University Science Books},\
  \bibinfo {year} {2000})\BibitemShut {NoStop}%
\bibitem [{\citenamefont {Desai}(2010)}]{desai2010quantum}%
  \BibitemOpen
  \bibfield  {author} {\bibinfo {author} {\bibfnamefont {Bipin~R}\ \bibnamefont
  {Desai}},\ }\href@noop {} {\emph {\bibinfo {title} {Quantum mechanics with
  basic field theory}}}\ (\bibinfo  {publisher} {Cambridge University Press},\
  \bibinfo {year} {2010})\BibitemShut {NoStop}%
\bibitem [{\citenamefont {Shankar}(2012)}]{shankar2012principles}%
  \BibitemOpen
  \bibfield  {author} {\bibinfo {author} {\bibfnamefont {Ramamurti}\
  \bibnamefont {Shankar}},\ }\href@noop {} {\emph {\bibinfo {title} {Principles
  of quantum mechanics}}}\ (\bibinfo  {publisher} {Springer Science \& Business
  Media},\ \bibinfo {year} {2012})\BibitemShut {NoStop}%
\bibitem [{\citenamefont {Baker}\ \emph
  {et~al.}(2021{\natexlab{b}})\citenamefont {Baker} \emph
  {et~al.}}]{tensor_recipes}%
  \BibitemOpen
  \bibfield  {author} {\bibinfo {author} {\bibfnamefont {Thomas~E.}\
  \bibnamefont {Baker}} \emph {et~al.},\ }\bibfield  {title} {\enquote
  {\bibinfo {title} {Tensor recipes for tensor network computations in
  {DMRjulia}},}\ }\href@noop {} {\bibfield  {journal} {\bibinfo  {journal}
  {arXiv preprint arXiv:2109.XXXX}\ } (\bibinfo {year}
  {2021}{\natexlab{b}})}\BibitemShut {NoStop}%
\bibitem [{\citenamefont {Fraser}(2021)}]{fraser2021twin}%
  \BibitemOpen
  \bibfield  {author} {\bibinfo {author} {\bibfnamefont {James~D}\ \bibnamefont
  {Fraser}},\ }\bibfield  {title} {\enquote {\bibinfo {title} {The twin origins
  of renormalization group concepts},}\ }\href@noop {} {\bibfield  {journal}
  {\bibinfo  {journal} {Studies in History and Philosophy of Science Part A}\
  }\textbf {\bibinfo {volume} {89}},\ \bibinfo {pages} {114--128} (\bibinfo
  {year} {2021})}\BibitemShut {NoStop}%
\bibitem [{\citenamefont {Stueckelberg}\ and\ \citenamefont
  {Petermann}(1953)}]{stueckelberg1953normalization}%
  \BibitemOpen
  \bibfield  {author} {\bibinfo {author} {\bibfnamefont {ECG}\ \bibnamefont
  {Stueckelberg}}\ and\ \bibinfo {author} {\bibfnamefont {A}~\bibnamefont
  {Petermann}},\ }\bibfield  {title} {\enquote {\bibinfo {title} {Normalization
  of constants in the quanta theory},}\ }\href@noop {} {\bibfield  {journal}
  {\bibinfo  {journal} {Helv. Phys. Acta}\ }\textbf {\bibinfo {volume} {26}},\
  \bibinfo {pages} {499--520} (\bibinfo {year} {1953})}\BibitemShut {NoStop}%
\bibitem [{\citenamefont {Gell-Mann}\ and\ \citenamefont
  {Low}(1954)}]{gell1954quantum}%
  \BibitemOpen
  \bibfield  {author} {\bibinfo {author} {\bibfnamefont {Murray}\ \bibnamefont
  {Gell-Mann}}\ and\ \bibinfo {author} {\bibfnamefont {Francis~E}\ \bibnamefont
  {Low}},\ }\bibfield  {title} {\enquote {\bibinfo {title} {Quantum
  electrodynamics at small distances},}\ }\href@noop {} {\bibfield  {journal}
  {\bibinfo  {journal} {Physical Review}\ }\textbf {\bibinfo {volume} {95}},\
  \bibinfo {pages} {1300} (\bibinfo {year} {1954})}\BibitemShut {NoStop}%
\bibitem [{\citenamefont {Wilson}(1975)}]{wilson1975renormalization}%
  \BibitemOpen
  \bibfield  {author} {\bibinfo {author} {\bibfnamefont {Kenneth~G}\
  \bibnamefont {Wilson}},\ }\bibfield  {title} {\enquote {\bibinfo {title} {The
  renormalization group: Critical phenomena and the kondo problem},}\
  }\href@noop {} {\bibfield  {journal} {\bibinfo  {journal} {Reviews of modern
  physics}\ }\textbf {\bibinfo {volume} {47}},\ \bibinfo {pages} {773}
  (\bibinfo {year} {1975})}\BibitemShut {NoStop}%
\bibitem [{\citenamefont {Gardner}(2005)}]{gardner2005phaselock}%
  \BibitemOpen
  \bibfield  {author} {\bibinfo {author} {\bibfnamefont {Floyd~M}\ \bibnamefont
  {Gardner}},\ }\href@noop {} {\emph {\bibinfo {title} {Phaselock
  techniques}}}\ (\bibinfo  {publisher} {John Wiley \& Sons},\ \bibinfo {year}
  {2005})\BibitemShut {NoStop}%
\bibitem [{\citenamefont {Ash}(1965)}]{ash1965information}%
  \BibitemOpen
  \bibfield  {author} {\bibinfo {author} {\bibfnamefont {Robert}\ \bibnamefont
  {Ash}},\ }\href@noop {} {\emph {\bibinfo {title} {Information Theory}}}\
  (\bibinfo  {publisher} {Dover Publications Inc., New York},\ \bibinfo {year}
  {1965})\BibitemShut {NoStop}%
\bibitem [{\citenamefont {Shannon}(1948)}]{shannon1948mathematical}%
  \BibitemOpen
  \bibfield  {author} {\bibinfo {author} {\bibfnamefont {Claude~E}\
  \bibnamefont {Shannon}},\ }\bibfield  {title} {\enquote {\bibinfo {title} {A
  mathematical theory of communication},}\ }\href@noop {} {\bibfield  {journal}
  {\bibinfo  {journal} {The Bell system technical journal}\ }\textbf {\bibinfo
  {volume} {27}},\ \bibinfo {pages} {379--423} (\bibinfo {year}
  {1948})}\BibitemShut {NoStop}%
\bibitem [{\citenamefont {von Neumann}(1955)}]{von1955mathematical}%
  \BibitemOpen
  \bibfield  {author} {\bibinfo {author} {\bibfnamefont {John}\ \bibnamefont
  {von Neumann}},\ }\bibfield  {title} {\enquote {\bibinfo {title}
  {Mathematical foundations of quantum mechanics},}\ }\href@noop {} {\bibfield
  {journal} {\bibinfo  {journal} {Investigations in Physics}\ } (\bibinfo
  {year} {1955})}\BibitemShut {NoStop}%
\bibitem [{\citenamefont {Reif}(2009)}]{reif2009fundamentals}%
  \BibitemOpen
  \bibfield  {author} {\bibinfo {author} {\bibfnamefont {Frederick}\
  \bibnamefont {Reif}},\ }\href@noop {} {\emph {\bibinfo {title} {Fundamentals
  of statistical and thermal physics}}}\ (\bibinfo  {publisher} {Waveland
  Press},\ \bibinfo {year} {2009})\BibitemShut {NoStop}%
\bibitem [{\citenamefont {Simon}()}]{simonknots}%
  \BibitemOpen
  \bibfield  {author} {\bibinfo {author} {\bibfnamefont {Steven~H.}\
  \bibnamefont {Simon}},\ }\href@noop {} {\enquote {\bibinfo {title}
  {Topological quantum: Lecture notes and proto-book},}\ }\bibinfo {note}
  {\url{http://www-thphys.physics.ox.ac.uk/people/SteveSimon/topological2019/Topobook-Nov1-2019.pdff}}\BibitemShut
  {NoStop}%
\bibitem [{\citenamefont {Kitaev}\ and\ \citenamefont
  {Preskill}(2006)}]{kitaev2006topological}%
  \BibitemOpen
  \bibfield  {author} {\bibinfo {author} {\bibfnamefont {Alexei}\ \bibnamefont
  {Kitaev}}\ and\ \bibinfo {author} {\bibfnamefont {John}\ \bibnamefont
  {Preskill}},\ }\bibfield  {title} {\enquote {\bibinfo {title} {Topological
  entanglement entropy},}\ }\href@noop {} {\bibfield  {journal} {\bibinfo
  {journal} {Physical review letters}\ }\textbf {\bibinfo {volume} {96}},\
  \bibinfo {pages} {110404} (\bibinfo {year} {2006})}\BibitemShut {NoStop}%
\bibitem [{\citenamefont {Jiang}\ \emph {et~al.}(2012)\citenamefont {Jiang},
  \citenamefont {Wang},\ and\ \citenamefont {Balents}}]{jiang2012identifying}%
  \BibitemOpen
  \bibfield  {author} {\bibinfo {author} {\bibfnamefont {Hong-Chen}\
  \bibnamefont {Jiang}}, \bibinfo {author} {\bibfnamefont {Zhenghan}\
  \bibnamefont {Wang}}, \ and\ \bibinfo {author} {\bibfnamefont {Leon}\
  \bibnamefont {Balents}},\ }\bibfield  {title} {\enquote {\bibinfo {title}
  {Identifying topological order by entanglement entropy},}\ }\href@noop {}
  {\bibfield  {journal} {\bibinfo  {journal} {Nature Physics}\ }\textbf
  {\bibinfo {volume} {8}},\ \bibinfo {pages} {902--905} (\bibinfo {year}
  {2012})}\BibitemShut {NoStop}%
\bibitem [{\citenamefont {Schwartz}(2014)}]{schwartz2014quantum}%
  \BibitemOpen
  \bibfield  {author} {\bibinfo {author} {\bibfnamefont {Matthew~D}\
  \bibnamefont {Schwartz}},\ }\href@noop {} {\emph {\bibinfo {title} {Quantum
  field theory and the standard model}}}\ (\bibinfo  {publisher} {Cambridge
  University Press},\ \bibinfo {year} {2014})\BibitemShut {NoStop}%
\bibitem [{\citenamefont {Mandelbrot}(1967)}]{mandelbrot1967long}%
  \BibitemOpen
  \bibfield  {author} {\bibinfo {author} {\bibfnamefont {Benoit}\ \bibnamefont
  {Mandelbrot}},\ }\bibfield  {title} {\enquote {\bibinfo {title} {How long is
  the coast of britain? statistical self-similarity and fractional
  dimension},}\ }\href@noop {} {\bibfield  {journal} {\bibinfo  {journal}
  {science}\ }\textbf {\bibinfo {volume} {156}},\ \bibinfo {pages} {636--638}
  (\bibinfo {year} {1967})}\BibitemShut {NoStop}%
\bibitem [{\citenamefont {Cardy}(1996)}]{cardy1996scaling}%
  \BibitemOpen
  \bibfield  {author} {\bibinfo {author} {\bibfnamefont {John}\ \bibnamefont
  {Cardy}},\ }\href@noop {} {\emph {\bibinfo {title} {Scaling and
  renormalization in statistical physics}}},\ Vol.~\bibinfo {volume} {5}\
  (\bibinfo  {publisher} {Cambridge university press},\ \bibinfo {year}
  {1996})\BibitemShut {NoStop}%
\bibitem [{\citenamefont {Francesco}\ \emph {et~al.}(2012)\citenamefont
  {Francesco}, \citenamefont {Mathieu},\ and\ \citenamefont
  {S{\'e}n{\'e}chal}}]{francesco2012conformal}%
  \BibitemOpen
  \bibfield  {author} {\bibinfo {author} {\bibfnamefont {Philippe}\
  \bibnamefont {Francesco}}, \bibinfo {author} {\bibfnamefont {Pierre}\
  \bibnamefont {Mathieu}}, \ and\ \bibinfo {author} {\bibfnamefont {David}\
  \bibnamefont {S{\'e}n{\'e}chal}},\ }\href@noop {} {\emph {\bibinfo {title}
  {Conformal field theory}}}\ (\bibinfo  {publisher} {Springer Science \&
  Business Media},\ \bibinfo {year} {2012})\BibitemShut {NoStop}%
\bibitem [{\citenamefont {Ehrenfest}(1933)}]{ehrenfest1933phasenumwandlungen}%
  \BibitemOpen
  \bibfield  {author} {\bibinfo {author} {\bibfnamefont {Paul}\ \bibnamefont
  {Ehrenfest}},\ }\href@noop {} {\emph {\bibinfo {title} {Phasenumwandlungen im
  ueblichen und erweiterten Sinn, classifiziert nach den entsprechenden
  Singularitaeten des thermodynamischen Potentiales}}}\ (\bibinfo  {publisher}
  {NV Noord-Hollandsche Uitgevers Maatschappij},\ \bibinfo {year}
  {1933})\BibitemShut {NoStop}%
\bibitem [{\citenamefont {Jaeger}(1998)}]{jaeger1998ehrenfest}%
  \BibitemOpen
  \bibfield  {author} {\bibinfo {author} {\bibfnamefont {Gregg}\ \bibnamefont
  {Jaeger}},\ }\bibfield  {title} {\enquote {\bibinfo {title} {The ehrenfest
  classification of phase transitions: introduction and evolution},}\
  }\href@noop {} {\bibfield  {journal} {\bibinfo  {journal} {Archive for
  history of exact sciences}\ }\textbf {\bibinfo {volume} {53}},\ \bibinfo
  {pages} {51--81} (\bibinfo {year} {1998})}\BibitemShut {NoStop}%
\bibitem [{\citenamefont {Bethe}(1935)}]{bethe1935statistical}%
  \BibitemOpen
  \bibfield  {author} {\bibinfo {author} {\bibfnamefont {Hans~A}\ \bibnamefont
  {Bethe}},\ }\bibfield  {title} {\enquote {\bibinfo {title} {Statistical
  theory of superlattices},}\ }\href@noop {} {\bibfield  {journal} {\bibinfo
  {journal} {Proceedings of the Royal Society of London. Series A-Mathematical
  and Physical Sciences}\ }\textbf {\bibinfo {volume} {150}},\ \bibinfo {pages}
  {552--575} (\bibinfo {year} {1935})}\BibitemShut {NoStop}%
\bibitem [{\citenamefont {Kadanoff}(1966)}]{kadanoff1966scaling}%
  \BibitemOpen
  \bibfield  {author} {\bibinfo {author} {\bibfnamefont {Leo~P}\ \bibnamefont
  {Kadanoff}},\ }\bibfield  {title} {\enquote {\bibinfo {title} {Scaling laws
  for ising models near t c},}\ }\href@noop {} {\bibfield  {journal} {\bibinfo
  {journal} {Physics Physique Fizika}\ }\textbf {\bibinfo {volume} {2}},\
  \bibinfo {pages} {263} (\bibinfo {year} {1966})}\BibitemShut {NoStop}%
\bibitem [{\citenamefont {Peskin}\ and\ \citenamefont
  {Schroeder}(1995)}]{peskin1995quantum}%
  \BibitemOpen
  \bibfield  {author} {\bibinfo {author} {\bibfnamefont {Michael~E}\
  \bibnamefont {Peskin}}\ and\ \bibinfo {author} {\bibfnamefont {Daniel~V}\
  \bibnamefont {Schroeder}},\ }\bibfield  {title} {\enquote {\bibinfo {title}
  {Quantum field theory},}\ }\href@noop {} {\bibfield  {journal} {\bibinfo
  {journal} {The Advanced Book Program, Perseus Books Reading, Massachusetts}\
  } (\bibinfo {year} {1995})}\BibitemShut {NoStop}%
\bibitem [{\citenamefont {De~Bruijn}(1981)}]{de1981asymptotic}%
  \BibitemOpen
  \bibfield  {author} {\bibinfo {author} {\bibfnamefont {Nicolaas~Govert}\
  \bibnamefont {De~Bruijn}},\ }\href@noop {} {\emph {\bibinfo {title}
  {Asymptotic methods in analysis}}},\ Vol.~\bibinfo {volume} {4}\ (\bibinfo
  {publisher} {Courier Corporation},\ \bibinfo {year} {1981})\BibitemShut
  {NoStop}%
\bibitem [{\citenamefont {Horowitz}\ \emph {et~al.}(1980)\citenamefont
  {Horowitz}, \citenamefont {Hill},\ and\ \citenamefont
  {Robinson}}]{horowitz1980art}%
  \BibitemOpen
  \bibfield  {author} {\bibinfo {author} {\bibfnamefont {Paul}\ \bibnamefont
  {Horowitz}}, \bibinfo {author} {\bibfnamefont {Winfield}\ \bibnamefont
  {Hill}}, \ and\ \bibinfo {author} {\bibfnamefont {Ian}\ \bibnamefont
  {Robinson}},\ }\href@noop {} {\emph {\bibinfo {title} {The art of
  electronics}}},\ Vol.~\bibinfo {volume} {2}\ (\bibinfo  {publisher}
  {Cambridge university press Cambridge},\ \bibinfo {year} {1980})\BibitemShut
  {NoStop}%
\bibitem [{\citenamefont {Zinn-Justin}(2007)}]{zinn2007phase}%
  \BibitemOpen
  \bibfield  {author} {\bibinfo {author} {\bibfnamefont {Jean}\ \bibnamefont
  {Zinn-Justin}},\ }\href@noop {} {\emph {\bibinfo {title} {Phase transitions
  and renormalization group}}}\ (\bibinfo  {publisher} {Oxford University Press
  on Demand},\ \bibinfo {year} {2007})\BibitemShut {NoStop}%
\bibitem [{\citenamefont {Pathria}\ and\ \citenamefont
  {Beale}(2011)}]{pathria2011statistical}%
  \BibitemOpen
  \bibfield  {author} {\bibinfo {author} {\bibfnamefont {R}~\bibnamefont
  {Pathria}}\ and\ \bibinfo {author} {\bibfnamefont {PD}~\bibnamefont
  {Beale}},\ }\href@noop {} {\emph {\bibinfo {title} {Statistical Mechanics 3rd
  ed.}}}\ (\bibinfo  {publisher} {Academic Press, Boston},\ \bibinfo {year}
  {2011})\BibitemShut {NoStop}%
\bibitem [{\citenamefont {Baker}\ \emph {et~al.}(2018)\citenamefont {Baker},
  \citenamefont {Burke},\ and\ \citenamefont {White}}]{bakerPRB18}%
  \BibitemOpen
  \bibfield  {author} {\bibinfo {author} {\bibfnamefont {Thomas~E.}\
  \bibnamefont {Baker}}, \bibinfo {author} {\bibfnamefont {Kieron}\
  \bibnamefont {Burke}}, \ and\ \bibinfo {author} {\bibfnamefont {Steven~R.}\
  \bibnamefont {White}},\ }\bibfield  {title} {\enquote {\bibinfo {title}
  {Accurate correlation energies in one-dimensional systems from small
  system-adapted basis functions},}\ }\href {\doibase
  10.1103/PhysRevB.97.085139} {\bibfield  {journal} {\bibinfo  {journal}
  {Phys.~Rev.~B}\ }\textbf {\bibinfo {volume} {97}},\ \bibinfo {pages} {085139}
  (\bibinfo {year} {2018})}\BibitemShut {NoStop}%
\bibitem [{\citenamefont {L{\"o}wdin}\ and\ \citenamefont
  {Shull}(1956)}]{lowdin1956natural}%
  \BibitemOpen
  \bibfield  {author} {\bibinfo {author} {\bibfnamefont {Per-Olov}\
  \bibnamefont {L{\"o}wdin}}\ and\ \bibinfo {author} {\bibfnamefont {Harrison}\
  \bibnamefont {Shull}},\ }\bibfield  {title} {\enquote {\bibinfo {title}
  {Natural orbitals in the quantum theory of two-electron systems},}\
  }\href@noop {} {\bibfield  {journal} {\bibinfo  {journal} {Physical Review}\
  }\textbf {\bibinfo {volume} {101}},\ \bibinfo {pages} {1730} (\bibinfo {year}
  {1956})}\BibitemShut {NoStop}%
\bibitem [{\citenamefont {L{\"o}wdin}(1955)}]{lowdin1955quantum}%
  \BibitemOpen
  \bibfield  {author} {\bibinfo {author} {\bibfnamefont {Per-Olov}\
  \bibnamefont {L{\"o}wdin}},\ }\bibfield  {title} {\enquote {\bibinfo {title}
  {Quantum theory of many-particle systems. i. physical interpretations by
  means of density matrices, natural spin-orbitals, and convergence problems in
  the method of configurational interaction},}\ }\href@noop {} {\bibfield
  {journal} {\bibinfo  {journal} {Physical Review}\ }\textbf {\bibinfo {volume}
  {97}},\ \bibinfo {pages} {1474} (\bibinfo {year} {1955})}\BibitemShut
  {NoStop}%
\bibitem [{\citenamefont {Noack}\ and\ \citenamefont
  {White}(1993)}]{noack1993real}%
  \BibitemOpen
  \bibfield  {author} {\bibinfo {author} {\bibfnamefont {RM}~\bibnamefont
  {Noack}}\ and\ \bibinfo {author} {\bibfnamefont {SR}~\bibnamefont {White}},\
  }\bibfield  {title} {\enquote {\bibinfo {title} {Real-space quantum
  renormalization group and anderson localization},}\ }\href@noop {} {\bibfield
   {journal} {\bibinfo  {journal} {Physical Review B}\ }\textbf {\bibinfo
  {volume} {47}},\ \bibinfo {pages} {9243} (\bibinfo {year}
  {1993})}\BibitemShut {NoStop}%
\bibitem [{\citenamefont {Eisert}\ \emph {et~al.}(2010)\citenamefont {Eisert},
  \citenamefont {Cramer},\ and\ \citenamefont {Plenio}}]{eisert2010colloquium}%
  \BibitemOpen
  \bibfield  {author} {\bibinfo {author} {\bibfnamefont {Jens}\ \bibnamefont
  {Eisert}}, \bibinfo {author} {\bibfnamefont {Marcus}\ \bibnamefont {Cramer}},
  \ and\ \bibinfo {author} {\bibfnamefont {Martin~B}\ \bibnamefont {Plenio}},\
  }\bibfield  {title} {\enquote {\bibinfo {title} {Colloquium: Area laws for
  the entanglement entropy},}\ }\href@noop {} {\bibfield  {journal} {\bibinfo
  {journal} {Reviews of Modern Physics}\ }\textbf {\bibinfo {volume} {82}},\
  \bibinfo {pages} {277} (\bibinfo {year} {2010})}\BibitemShut {NoStop}%
\bibitem [{\citenamefont {Kuwahara}\ and\ \citenamefont
  {Saito}(2020)}]{kuwahara2020area}%
  \BibitemOpen
  \bibfield  {author} {\bibinfo {author} {\bibfnamefont {Tomotaka}\
  \bibnamefont {Kuwahara}}\ and\ \bibinfo {author} {\bibfnamefont {Keiji}\
  \bibnamefont {Saito}},\ }\bibfield  {title} {\enquote {\bibinfo {title} {Area
  law of noncritical ground states in 1d long-range interacting systems},}\
  }\href@noop {} {\bibfield  {journal} {\bibinfo  {journal} {Nature
  communications}\ }\textbf {\bibinfo {volume} {11}},\ \bibinfo {pages} {1--7}
  (\bibinfo {year} {2020})}\BibitemShut {NoStop}%
\bibitem [{\citenamefont {Hawking}(1976)}]{hawking1976particle}%
  \BibitemOpen
  \bibfield  {author} {\bibinfo {author} {\bibfnamefont {SW}~\bibnamefont
  {Hawking}},\ }\bibfield  {title} {\enquote {\bibinfo {title} {Particle
  creation by black holes},}\ }\href@noop {} {\bibfield  {journal} {\bibinfo
  {journal} {Communications in Mathematical Physics}\ }\textbf {\bibinfo
  {volume} {46}},\ \bibinfo {pages} {206--206} (\bibinfo {year}
  {1976})}\BibitemShut {NoStop}%
\bibitem [{\citenamefont {Kohn}(1996)}]{kohn1996density}%
  \BibitemOpen
  \bibfield  {author} {\bibinfo {author} {\bibfnamefont {Walter}\ \bibnamefont
  {Kohn}},\ }\bibfield  {title} {\enquote {\bibinfo {title} {Density functional
  and density matrix method scaling linearly with the number of atoms},}\
  }\href@noop {} {\bibfield  {journal} {\bibinfo  {journal} {Physical Review
  Letters}\ }\textbf {\bibinfo {volume} {76}},\ \bibinfo {pages} {3168}
  (\bibinfo {year} {1996})}\BibitemShut {NoStop}%
\bibitem [{\citenamefont {Prodan}\ and\ \citenamefont
  {Kohn}(2005)}]{prodan2005nearsightedness}%
  \BibitemOpen
  \bibfield  {author} {\bibinfo {author} {\bibfnamefont {Emil}\ \bibnamefont
  {Prodan}}\ and\ \bibinfo {author} {\bibfnamefont {Walter}\ \bibnamefont
  {Kohn}},\ }\bibfield  {title} {\enquote {\bibinfo {title} {Nearsightedness of
  electronic matter},}\ }\href@noop {} {\bibfield  {journal} {\bibinfo
  {journal} {Proceedings of the National Academy of Sciences}\ }\textbf
  {\bibinfo {volume} {102}},\ \bibinfo {pages} {11635--11638} (\bibinfo {year}
  {2005})}\BibitemShut {NoStop}%
\bibitem [{\citenamefont {Movassagh}\ and\ \citenamefont
  {Shor}(2016)}]{movassagh2016supercritical}%
  \BibitemOpen
  \bibfield  {author} {\bibinfo {author} {\bibfnamefont {Ramis}\ \bibnamefont
  {Movassagh}}\ and\ \bibinfo {author} {\bibfnamefont {Peter~W}\ \bibnamefont
  {Shor}},\ }\bibfield  {title} {\enquote {\bibinfo {title} {Supercritical
  entanglement in local systems: Counterexample to the area law for quantum
  matter},}\ }\href@noop {} {\bibfield  {journal} {\bibinfo  {journal}
  {Proceedings of the National Academy of Sciences}\ }\textbf {\bibinfo
  {volume} {113}},\ \bibinfo {pages} {13278--13282} (\bibinfo {year}
  {2016})}\BibitemShut {NoStop}%
\bibitem [{\citenamefont {Movassagh}\ and\ \citenamefont
  {Shor}(2014)}]{movassagh2014power}%
  \BibitemOpen
  \bibfield  {author} {\bibinfo {author} {\bibfnamefont {Ramis}\ \bibnamefont
  {Movassagh}}\ and\ \bibinfo {author} {\bibfnamefont {Peter~W}\ \bibnamefont
  {Shor}},\ }\bibfield  {title} {\enquote {\bibinfo {title} {Power law
  violation of the area law in quantum spin chains},}\ }\href@noop {}
  {\bibfield  {journal} {\bibinfo  {journal} {arXiv preprint arXiv:1408.1657}\
  } (\bibinfo {year} {2014})}\BibitemShut {NoStop}%
\bibitem [{\citenamefont {Abanin}\ \emph {et~al.}(2019)\citenamefont {Abanin},
  \citenamefont {Altman}, \citenamefont {Bloch},\ and\ \citenamefont
  {Serbyn}}]{abanin2019colloquium}%
  \BibitemOpen
  \bibfield  {author} {\bibinfo {author} {\bibfnamefont {Dmitry~A}\
  \bibnamefont {Abanin}}, \bibinfo {author} {\bibfnamefont {Ehud}\ \bibnamefont
  {Altman}}, \bibinfo {author} {\bibfnamefont {Immanuel}\ \bibnamefont
  {Bloch}}, \ and\ \bibinfo {author} {\bibfnamefont {Maksym}\ \bibnamefont
  {Serbyn}},\ }\bibfield  {title} {\enquote {\bibinfo {title} {Colloquium:
  Many-body localization, thermalization, and entanglement},}\ }\href@noop {}
  {\bibfield  {journal} {\bibinfo  {journal} {Reviews of Modern Physics}\
  }\textbf {\bibinfo {volume} {91}},\ \bibinfo {pages} {021001} (\bibinfo
  {year} {2019})}\BibitemShut {NoStop}%
\bibitem [{\citenamefont {White}(1999)}]{white1999all}%
  \BibitemOpen
  \bibfield  {author} {\bibinfo {author} {\bibfnamefont {S}~\bibnamefont
  {White}},\ }\bibfield  {title} {\enquote {\bibinfo {title} {How it all began:
  A personal account},}\ }\href@noop {} {\bibfield  {journal} {\bibinfo
  {journal} {LECTURE NOTES IN PHYSICS-NEW YORK THEN BERLIN-}\ ,\ \bibinfo
  {pages} {vii--xvi}} (\bibinfo {year} {1999})}\BibitemShut {NoStop}%
\bibitem [{\citenamefont {Feynman}\ \emph {et~al.}(1965)\citenamefont
  {Feynman}, \citenamefont {Leighton},\ and\ \citenamefont
  {Sands}}]{feynman1965feynman}%
  \BibitemOpen
  \bibfield  {author} {\bibinfo {author} {\bibfnamefont {Richard~P}\
  \bibnamefont {Feynman}}, \bibinfo {author} {\bibfnamefont {Robert~B}\
  \bibnamefont {Leighton}}, \ and\ \bibinfo {author} {\bibfnamefont {Matthew}\
  \bibnamefont {Sands}},\ }\bibfield  {title} {\enquote {\bibinfo {title} {The
  feynman lectures on physics; vol. i},}\ }\href@noop {} {\bibfield  {journal}
  {\bibinfo  {journal} {American Journal of Physics}\ }\textbf {\bibinfo
  {volume} {33}},\ \bibinfo {pages} {750--752} (\bibinfo {year}
  {1965})}\BibitemShut {NoStop}%
\bibitem [{\citenamefont {White}(1993)}]{white1993density}%
  \BibitemOpen
  \bibfield  {author} {\bibinfo {author} {\bibfnamefont {Steven~R}\
  \bibnamefont {White}},\ }\bibfield  {title} {\enquote {\bibinfo {title}
  {Density-matrix algorithms for quantum renormalization groups},}\ }\href@noop
  {} {\bibfield  {journal} {\bibinfo  {journal} {Physical Review B}\ }\textbf
  {\bibinfo {volume} {48}},\ \bibinfo {pages} {10345} (\bibinfo {year}
  {1993})}\BibitemShut {NoStop}%
\bibitem [{\citenamefont {Rommer}\ and\ \citenamefont
  {{\"O}stlund}(1997)}]{rommer1997class}%
  \BibitemOpen
  \bibfield  {author} {\bibinfo {author} {\bibfnamefont {Stefan}\ \bibnamefont
  {Rommer}}\ and\ \bibinfo {author} {\bibfnamefont {Stellan}\ \bibnamefont
  {{\"O}stlund}},\ }\bibfield  {title} {\enquote {\bibinfo {title} {Class of
  ansatz wave functions for one-dimensional spin systems and their relation to
  the density matrix renormalization group},}\ }\href@noop {} {\bibfield
  {journal} {\bibinfo  {journal} {Physical review b}\ }\textbf {\bibinfo
  {volume} {55}},\ \bibinfo {pages} {2164} (\bibinfo {year}
  {1997})}\BibitemShut {NoStop}%
\bibitem [{\citenamefont {Vidal}(2008)}]{vidal2008class}%
  \BibitemOpen
  \bibfield  {author} {\bibinfo {author} {\bibfnamefont {Guifr{\'e}}\
  \bibnamefont {Vidal}},\ }\bibfield  {title} {\enquote {\bibinfo {title}
  {Class of quantum many-body states that can be efficiently simulated},}\
  }\href@noop {} {\bibfield  {journal} {\bibinfo  {journal} {Physical review
  letters}\ }\textbf {\bibinfo {volume} {101}},\ \bibinfo {pages} {110501}
  (\bibinfo {year} {2008})}\BibitemShut {NoStop}%
\bibitem [{\citenamefont {Boas}(2006)}]{boas2006mathematical}%
  \BibitemOpen
  \bibfield  {author} {\bibinfo {author} {\bibfnamefont {Mary~L}\ \bibnamefont
  {Boas}},\ }\href@noop {} {\emph {\bibinfo {title} {Mathematical methods in
  the physical sciences}}}\ (\bibinfo  {publisher} {Wiley},\ \bibinfo {year}
  {2006})\BibitemShut {NoStop}%
\bibitem [{\citenamefont {Schutz}(2009)}]{schutz2009first}%
  \BibitemOpen
  \bibfield  {author} {\bibinfo {author} {\bibfnamefont {Bernard}\ \bibnamefont
  {Schutz}},\ }\href@noop {} {\emph {\bibinfo {title} {A first course in
  general relativity}}}\ (\bibinfo  {publisher} {Cambridge university press},\
  \bibinfo {year} {2009})\BibitemShut {NoStop}%
\bibitem [{\citenamefont {Penrose}(1971)}]{penrose1971applications}%
  \BibitemOpen
  \bibfield  {author} {\bibinfo {author} {\bibfnamefont {Roger}\ \bibnamefont
  {Penrose}},\ }\bibfield  {title} {\enquote {\bibinfo {title} {Applications of
  negative dimensional tensors},}\ }\href@noop {} {\bibfield  {journal}
  {\bibinfo  {journal} {Combinatorial mathematics and its applications}\
  }\textbf {\bibinfo {volume} {1}},\ \bibinfo {pages} {221--244} (\bibinfo
  {year} {1971})}\BibitemShut {NoStop}%
\bibitem [{\citenamefont {Press}\ \emph {et~al.}(1992)\citenamefont {Press},
  \citenamefont {Teukolsky}, \citenamefont {Vetterling},\ and\ \citenamefont
  {Flannery}}]{press1992numerical}%
  \BibitemOpen
  \bibfield  {author} {\bibinfo {author} {\bibfnamefont {William~H}\
  \bibnamefont {Press}}, \bibinfo {author} {\bibfnamefont {Saul~A}\
  \bibnamefont {Teukolsky}}, \bibinfo {author} {\bibfnamefont {William~T}\
  \bibnamefont {Vetterling}}, \ and\ \bibinfo {author} {\bibfnamefont
  {Brian~P}\ \bibnamefont {Flannery}},\ }\bibfield  {title} {\enquote {\bibinfo
  {title} {Numerical recipes in c++},}\ }\href@noop {} {\bibfield  {journal}
  {\bibinfo  {journal} {The art of scientific computing}\ } (\bibinfo {year}
  {1992})}\BibitemShut {NoStop}%
\bibitem [{\citenamefont {Stoudenmire}\ and\ \citenamefont
  {White}(2013)}]{stoudenmire2013real}%
  \BibitemOpen
  \bibfield  {author} {\bibinfo {author} {\bibfnamefont {EM}~\bibnamefont
  {Stoudenmire}}\ and\ \bibinfo {author} {\bibfnamefont {Steven~R}\
  \bibnamefont {White}},\ }\bibfield  {title} {\enquote {\bibinfo {title}
  {Real-space parallel density matrix renormalization group},}\ }\href@noop {}
  {\bibfield  {journal} {\bibinfo  {journal} {Physical review B}\ }\textbf
  {\bibinfo {volume} {87}},\ \bibinfo {pages} {155137} (\bibinfo {year}
  {2013})}\BibitemShut {NoStop}%
\bibitem [{\citenamefont {Baker}(2017)}]{baker2017methods}%
  \BibitemOpen
  \bibfield  {author} {\bibinfo {author} {\bibfnamefont {Thomas~Edward}\
  \bibnamefont {Baker}},\ }\emph {\bibinfo {title} {{Methods of Calculation
  with the Exact Density Functional using the Renormalization Group}}},\
  \href@noop {} {Ph.D. thesis},\ \bibinfo  {school} {University of California,
  Irvine} (\bibinfo {year} {2017})\BibitemShut {NoStop}%
\bibitem [{\citenamefont {Cervera}\ and\ \citenamefont
  {Baker}(2017)}]{cervera2017IPAM}%
  \BibitemOpen
  \bibfield  {author} {\bibinfo {author} {\bibfnamefont {Carlos~Garcia}\
  \bibnamefont {Cervera}}\ and\ \bibinfo {author} {\bibfnamefont {Thomas~E}\
  \bibnamefont {Baker}},\ }\enquote {\bibinfo {title} {{IPAM Book of DFT:
  Mathematical Foundations of Quantum Mechanics}},}\ \ (\bibinfo {year}
  {2017})\ Chap.~\bibinfo {chapter} {2}\BibitemShut {NoStop}%
\bibitem [{\citenamefont {Nielsen}\ and\ \citenamefont
  {Chuang}(2010)}]{nielsen2010quantum}%
  \BibitemOpen
  \bibfield  {author} {\bibinfo {author} {\bibfnamefont {Michael~A}\
  \bibnamefont {Nielsen}}\ and\ \bibinfo {author} {\bibfnamefont {Isaac~L}\
  \bibnamefont {Chuang}},\ }\href@noop {} {\emph {\bibinfo {title} {Quantum
  Computation and Quantum Information}}}\ (\bibinfo  {publisher} {Cambridge
  University Press},\ \bibinfo {year} {2010})\BibitemShut {NoStop}%
\bibitem [{\citenamefont {Li}\ \emph {et~al.}(2016)\citenamefont {Li},
  \citenamefont {Baker}, \citenamefont {White},\ and\ \citenamefont
  {Burke}}]{liPRB16}%
  \BibitemOpen
  \bibfield  {author} {\bibinfo {author} {\bibfnamefont {Li}~\bibnamefont
  {Li}}, \bibinfo {author} {\bibfnamefont {Thomas~E}\ \bibnamefont {Baker}},
  \bibinfo {author} {\bibfnamefont {Steven~R}\ \bibnamefont {White}}, \ and\
  \bibinfo {author} {\bibfnamefont {Kieron}\ \bibnamefont {Burke}},\ }\bibfield
   {title} {\enquote {\bibinfo {title} {{Pure density functional for strong
  correlation and the thermodynamic limit from machine learning}},}\ }\href
  {\doibase https://doi.org/10.1103/PhysRevB.94.245129} {\bibfield  {journal}
  {\bibinfo  {journal} {Phys.~Rev.~B}\ }\textbf {\bibinfo {volume} {94}},\
  \bibinfo {pages} {245129} (\bibinfo {year} {2016})}\BibitemShut {NoStop}%
\bibitem [{\citenamefont {Hubbard}(1963)}]{hubbard1963electron}%
  \BibitemOpen
  \bibfield  {author} {\bibinfo {author} {\bibfnamefont {John}\ \bibnamefont
  {Hubbard}},\ }\bibfield  {title} {\enquote {\bibinfo {title} {Electron
  correlations in narrow energy bands},}\ }\href@noop {} {\bibfield  {journal}
  {\bibinfo  {journal} {Proceedings of the Royal Society of London. Series A.
  Mathematical and Physical Sciences}\ }\textbf {\bibinfo {volume} {276}},\
  \bibinfo {pages} {238--257} (\bibinfo {year} {1963})}\BibitemShut {NoStop}%
\bibitem [{\citenamefont {Fetter}\ and\ \citenamefont
  {Walecka}(2012)}]{fetter2012quantum}%
  \BibitemOpen
  \bibfield  {author} {\bibinfo {author} {\bibfnamefont {Alexander~L}\
  \bibnamefont {Fetter}}\ and\ \bibinfo {author} {\bibfnamefont {John~Dirk}\
  \bibnamefont {Walecka}},\ }\href@noop {} {\emph {\bibinfo {title} {Quantum
  theory of many-particle systems}}}\ (\bibinfo  {publisher} {Courier
  Corporation},\ \bibinfo {year} {2012})\BibitemShut {NoStop}%
\bibitem [{\citenamefont {Flynn}\ \emph {et~al.}(2021)\citenamefont {Flynn},
  \citenamefont {Baker}, \citenamefont {Jindal},\ and\ \citenamefont
  {Singh}}]{flynn2020two}%
  \BibitemOpen
  \bibfield  {author} {\bibinfo {author} {\bibfnamefont {Michael~O.}\
  \bibnamefont {Flynn}}, \bibinfo {author} {\bibfnamefont {Thomas~E.}\
  \bibnamefont {Baker}}, \bibinfo {author} {\bibfnamefont {Siddharth}\
  \bibnamefont {Jindal}}, \ and\ \bibinfo {author} {\bibfnamefont {Rajiv
  R.~P.}\ \bibnamefont {Singh}},\ }\bibfield  {title} {\enquote {\bibinfo
  {title} {{Two Phases Inside the Bose Condensation Dome of
  ${\mathrm{Yb}}_{2}{\mathrm{Si}}_{2}{\mathrm{O}}_{7}$}},}\ }\href {\doibase
  10.1103/PhysRevLett.126.067201} {\bibfield  {journal} {\bibinfo  {journal}
  {Phys. Rev. Lett.}\ }\textbf {\bibinfo {volume} {126}},\ \bibinfo {pages}
  {067201} (\bibinfo {year} {2021})}\BibitemShut {NoStop}%
\bibitem [{\citenamefont {Lanczos}(1950)}]{lanczos1950iteration}%
  \BibitemOpen
  \bibfield  {author} {\bibinfo {author} {\bibfnamefont {Cornelius}\
  \bibnamefont {Lanczos}},\ }\href@noop {} {\emph {\bibinfo {title} {An
  iteration method for the solution of the eigenvalue problem of linear
  differential and integral operators}}}\ (\bibinfo  {publisher} {United States
  Governm. Press Office Los Angeles, CA},\ \bibinfo {year} {1950})\BibitemShut
  {NoStop}%
\bibitem [{\citenamefont {Davidson}(1975)}]{davidson1975iterative}%
  \BibitemOpen
  \bibfield  {author} {\bibinfo {author} {\bibfnamefont {Ernest~R}\
  \bibnamefont {Davidson}},\ }\bibfield  {title} {\enquote {\bibinfo {title}
  {The iterative calculation of a few of the lowest eigenvalues and
  corresponding eigenvectors of large real-symmetric matrices},}\ }\href@noop
  {} {\bibfield  {journal} {\bibinfo  {journal} {Journal of Computational
  Physics}\ }\textbf {\bibinfo {volume} {17}},\ \bibinfo {pages} {87--94}
  (\bibinfo {year} {1975})}\BibitemShut {NoStop}%
\bibitem [{\citenamefont {S{\'e}n{\'e}chal}(2008)}]{senechal2008introduction}%
  \BibitemOpen
  \bibfield  {author} {\bibinfo {author} {\bibfnamefont {David}\ \bibnamefont
  {S{\'e}n{\'e}chal}},\ }\bibfield  {title} {\enquote {\bibinfo {title} {An
  introduction to quantum cluster methods},}\ }\href@noop {} {\bibfield
  {journal} {\bibinfo  {journal} {arXiv preprint arXiv:0806.2690}\ } (\bibinfo
  {year} {2008})}\BibitemShut {NoStop}%
\bibitem [{\citenamefont {Baker}(2021)}]{baker2021lanczos}%
  \BibitemOpen
  \bibfield  {author} {\bibinfo {author} {\bibfnamefont {Thomas~E}\
  \bibnamefont {Baker}},\ }\bibfield  {title} {\enquote {\bibinfo {title}
  {{Lanczos recursion on a quantum computer for the Green's function and ground
  state}},}\ }\href {\doibase https://doi.org/10.1103/PhysRevA.103.032404}
  {\bibfield  {journal} {\bibinfo  {journal} {Physical Review A}\ }\textbf
  {\bibinfo {volume} {103}},\ \bibinfo {pages} {032404} (\bibinfo {year}
  {2021})}\BibitemShut {NoStop}%
\bibitem [{\citenamefont {Economou}(1983)}]{economou1983green}%
  \BibitemOpen
  \bibfield  {author} {\bibinfo {author} {\bibfnamefont {Eleftherios~N}\
  \bibnamefont {Economou}},\ }\href@noop {} {\emph {\bibinfo {title} {Green's
  functions in quantum physics}}},\ Vol.~\bibinfo {volume} {3}\ (\bibinfo
  {publisher} {Springer},\ \bibinfo {year} {1983})\BibitemShut {NoStop}%
\bibitem [{\citenamefont {White}\ and\ \citenamefont
  {Martin}(1999)}]{white1999ab}%
  \BibitemOpen
  \bibfield  {author} {\bibinfo {author} {\bibfnamefont {Steven~R}\
  \bibnamefont {White}}\ and\ \bibinfo {author} {\bibfnamefont {Richard~L}\
  \bibnamefont {Martin}},\ }\bibfield  {title} {\enquote {\bibinfo {title} {Ab
  initio quantum chemistry using the density matrix renormalization group},}\
  }\href@noop {} {\bibfield  {journal} {\bibinfo  {journal} {The Journal of
  chemical physics}\ }\textbf {\bibinfo {volume} {110}},\ \bibinfo {pages}
  {4127--4130} (\bibinfo {year} {1999})}\BibitemShut {NoStop}%
\bibitem [{\citenamefont {Chan}\ and\ \citenamefont
  {Head-Gordon}(2002)}]{chan2002highly}%
  \BibitemOpen
  \bibfield  {author} {\bibinfo {author} {\bibfnamefont {Garnet Kin-Lic}\
  \bibnamefont {Chan}}\ and\ \bibinfo {author} {\bibfnamefont {Martin}\
  \bibnamefont {Head-Gordon}},\ }\bibfield  {title} {\enquote {\bibinfo {title}
  {Highly correlated calculations with a polynomial cost algorithm: A study of
  the density matrix renormalization group},}\ }\href@noop {} {\bibfield
  {journal} {\bibinfo  {journal} {The Journal of chemical physics}\ }\textbf
  {\bibinfo {volume} {116}},\ \bibinfo {pages} {4462--4476} (\bibinfo {year}
  {2002})}\BibitemShut {NoStop}%
\bibitem [{\citenamefont {Chan}(2004)}]{chan2004algorithm}%
  \BibitemOpen
  \bibfield  {author} {\bibinfo {author} {\bibfnamefont {Garnet Kin-Lic}\
  \bibnamefont {Chan}},\ }\bibfield  {title} {\enquote {\bibinfo {title} {An
  algorithm for large scale density matrix renormalization group
  calculations},}\ }\href@noop {} {\bibfield  {journal} {\bibinfo  {journal}
  {The Journal of chemical physics}\ }\textbf {\bibinfo {volume} {120}},\
  \bibinfo {pages} {3172--3178} (\bibinfo {year} {2004})}\BibitemShut {NoStop}%
\bibitem [{\citenamefont {Ghosh}\ \emph {et~al.}(2008)\citenamefont {Ghosh},
  \citenamefont {Hachmann}, \citenamefont {Yanai},\ and\ \citenamefont
  {Chan}}]{ghosh2008orbital}%
  \BibitemOpen
  \bibfield  {author} {\bibinfo {author} {\bibfnamefont {Debashree}\
  \bibnamefont {Ghosh}}, \bibinfo {author} {\bibfnamefont {Johannes}\
  \bibnamefont {Hachmann}}, \bibinfo {author} {\bibfnamefont {Takeshi}\
  \bibnamefont {Yanai}}, \ and\ \bibinfo {author} {\bibfnamefont {Garnet
  Kin-Lic}\ \bibnamefont {Chan}},\ }\bibfield  {title} {\enquote {\bibinfo
  {title} {Orbital optimization in the density matrix renormalization group,
  with applications to polyenes and $\beta$-carotene},}\ }\href@noop {}
  {\bibfield  {journal} {\bibinfo  {journal} {The Journal of chemical physics}\
  }\textbf {\bibinfo {volume} {128}},\ \bibinfo {pages} {144117} (\bibinfo
  {year} {2008})}\BibitemShut {NoStop}%
\bibitem [{\citenamefont {Sharma}\ and\ \citenamefont
  {Chan}(2012)}]{sharma2012spin}%
  \BibitemOpen
  \bibfield  {author} {\bibinfo {author} {\bibfnamefont {Sandeep}\ \bibnamefont
  {Sharma}}\ and\ \bibinfo {author} {\bibfnamefont {Garnet Kin-Lic}\
  \bibnamefont {Chan}},\ }\bibfield  {title} {\enquote {\bibinfo {title}
  {Spin-adapted density matrix renormalization group algorithms for quantum
  chemistry},}\ }\href@noop {} {\bibfield  {journal} {\bibinfo  {journal} {The
  Journal of chemical physics}\ }\textbf {\bibinfo {volume} {136}},\ \bibinfo
  {pages} {124121} (\bibinfo {year} {2012})}\BibitemShut {NoStop}%
\bibitem [{\citenamefont {Olivares-Amaya}\ \emph {et~al.}(2015)\citenamefont
  {Olivares-Amaya}, \citenamefont {Hu}, \citenamefont {Nakatani}, \citenamefont
  {Sharma}, \citenamefont {Yang},\ and\ \citenamefont {Chan}}]{olivares2015ab}%
  \BibitemOpen
  \bibfield  {author} {\bibinfo {author} {\bibfnamefont {Roberto}\ \bibnamefont
  {Olivares-Amaya}}, \bibinfo {author} {\bibfnamefont {Weifeng}\ \bibnamefont
  {Hu}}, \bibinfo {author} {\bibfnamefont {Naoki}\ \bibnamefont {Nakatani}},
  \bibinfo {author} {\bibfnamefont {Sandeep}\ \bibnamefont {Sharma}}, \bibinfo
  {author} {\bibfnamefont {Jun}\ \bibnamefont {Yang}}, \ and\ \bibinfo {author}
  {\bibfnamefont {Garnet Kin-Lic}\ \bibnamefont {Chan}},\ }\bibfield  {title}
  {\enquote {\bibinfo {title} {The ab-initio density matrix renormalization
  group in practice},}\ }\href@noop {} {\bibfield  {journal} {\bibinfo
  {journal} {The Journal of chemical physics}\ }\textbf {\bibinfo {volume}
  {142}},\ \bibinfo {pages} {034102} (\bibinfo {year} {2015})}\BibitemShut
  {NoStop}%
\bibitem [{\citenamefont {Parker}\ and\ \citenamefont
  {Shiozaki}(2014)}]{parker2014communication}%
  \BibitemOpen
  \bibfield  {author} {\bibinfo {author} {\bibfnamefont {Shane~M}\ \bibnamefont
  {Parker}}\ and\ \bibinfo {author} {\bibfnamefont {Toru}\ \bibnamefont
  {Shiozaki}},\ }\bibfield  {title} {\enquote {\bibinfo {title} {Communication:
  Active space decomposition with multiple sites: Density matrix
  renormalization group algorithm},}\ }\href@noop {} {\bibfield  {journal}
  {\bibinfo  {journal} {Journal of Chemical Physics}\ }\textbf {\bibinfo
  {volume} {141}} (\bibinfo {year} {2014})}\BibitemShut {NoStop}%
\bibitem [{\citenamefont {Stoudenmire}\ and\ \citenamefont
  {White}(2012)}]{stoudenmire2012studying}%
  \BibitemOpen
  \bibfield  {author} {\bibinfo {author} {\bibfnamefont {Edwin~M}\ \bibnamefont
  {Stoudenmire}}\ and\ \bibinfo {author} {\bibfnamefont {Steven~R}\
  \bibnamefont {White}},\ }\bibfield  {title} {\enquote {\bibinfo {title}
  {Studying two-dimensional systems with the density matrix renormalization
  group},}\ }\href@noop {} {\bibfield  {journal} {\bibinfo  {journal} {Annu.
  Rev. Condens. Matter Phys.}\ }\textbf {\bibinfo {volume} {3}},\ \bibinfo
  {pages} {111--128} (\bibinfo {year} {2012})}\BibitemShut {NoStop}%
\bibitem [{\citenamefont {Verstraete}\ and\ \citenamefont
  {Cirac}(2004)}]{verstraete2004renormalization}%
  \BibitemOpen
  \bibfield  {author} {\bibinfo {author} {\bibfnamefont {Frank}\ \bibnamefont
  {Verstraete}}\ and\ \bibinfo {author} {\bibfnamefont {J~Ignacio}\
  \bibnamefont {Cirac}},\ }\bibfield  {title} {\enquote {\bibinfo {title}
  {Renormalization algorithms for quantum-many body systems in two and higher
  dimensions},}\ }\href@noop {} {\bibfield  {journal} {\bibinfo  {journal}
  {arXiv preprint cond-mat/0407066}\ } (\bibinfo {year} {2004})}\BibitemShut
  {NoStop}%
\bibitem [{\citenamefont {Motruk}\ \emph {et~al.}(2016)\citenamefont {Motruk},
  \citenamefont {Zaletel}, \citenamefont {Mong},\ and\ \citenamefont
  {Pollmann}}]{motruk2016density}%
  \BibitemOpen
  \bibfield  {author} {\bibinfo {author} {\bibfnamefont {Johannes}\
  \bibnamefont {Motruk}}, \bibinfo {author} {\bibfnamefont {Michael~P}\
  \bibnamefont {Zaletel}}, \bibinfo {author} {\bibfnamefont {Roger~SK}\
  \bibnamefont {Mong}}, \ and\ \bibinfo {author} {\bibfnamefont {Frank}\
  \bibnamefont {Pollmann}},\ }\bibfield  {title} {\enquote {\bibinfo {title}
  {Density matrix renormalization group on a cylinder in mixed real and
  momentum space},}\ }\href@noop {} {\bibfield  {journal} {\bibinfo  {journal}
  {Physical Review B}\ }\textbf {\bibinfo {volume} {93}},\ \bibinfo {pages}
  {155139} (\bibinfo {year} {2016})}\BibitemShut {NoStop}%
\bibitem [{\citenamefont {Ehlers}\ \emph {et~al.}(2017)\citenamefont {Ehlers},
  \citenamefont {White},\ and\ \citenamefont {Noack}}]{ehlers2017hybrid}%
  \BibitemOpen
  \bibfield  {author} {\bibinfo {author} {\bibfnamefont {G}~\bibnamefont
  {Ehlers}}, \bibinfo {author} {\bibfnamefont {SR}~\bibnamefont {White}}, \
  and\ \bibinfo {author} {\bibfnamefont {RM}~\bibnamefont {Noack}},\ }\bibfield
   {title} {\enquote {\bibinfo {title} {Hybrid-space density matrix
  renormalization group study of the doped two-dimensional hubbard model},}\
  }\href@noop {} {\bibfield  {journal} {\bibinfo  {journal} {Physical Review
  B}\ }\textbf {\bibinfo {volume} {95}},\ \bibinfo {pages} {125125} (\bibinfo
  {year} {2017})}\BibitemShut {NoStop}%
\bibitem [{\citenamefont {Kittel}(1987)}]{kittel1987quantum}%
  \BibitemOpen
  \bibfield  {author} {\bibinfo {author} {\bibfnamefont {Charles}\ \bibnamefont
  {Kittel}},\ }\href@noop {} {\emph {\bibinfo {title} {Quantum theory of
  solids}}}\ (\bibinfo  {publisher} {Wiley},\ \bibinfo {year}
  {1987})\BibitemShut {NoStop}%
\end{thebibliography}%

\begin{appendix}

\section{Computational complexity in big-$O$ notation}\label{complexity}

The idea behind computational complexity estimates is to find the algorithmic scaling for the worst-case scenario. Once this is accomplished by assuming that all indices have a size $\chi$ and then continuing the number of operations required to complete a specific task.  The cost will be represented in big-$O$ notation, and the meaning of this notation will become clear once it is used on an example here.  Algorithms with smaller costs (lower polynomials in $\chi$) will be faster in general.

In the text, the computational efficiency of evaluating the contraction in Eq.~\eqref{contractex} is now determined here.  To do this, the number of loops to be evaluated must be counted.  In this case, the contraction scales as $O(\chi^6)$. There are two indices that are summed over (in this analysis all indices are dimension $\chi$) and four that are not.  When writing a code to do this using the "long" way of writing out all the loops manually, there is required 6 loops in total perform the contraction (two inner loops for the contracted indices and 4 outer loops to scan all elements of the resulting tensor).

Diagrammatically, one can visually see how an algorithm scales by counting the number of uncontracted indices and adding it to the number of contracted indices.  The end result is the same if applied to Eq.~\eqref{contractex}.

For a second example, the contraction order in Sec.~\ref{expectvalues} can be revisited for two MPSs (no MPOs) to find the norm of the wavefunction the long way.  The contraction can be illustrated in two different ways to show how changing the contraction order can change the speed of a code.

In the context of tensor networks, the computational complexity is a good analysis tool to compare two algorithms. However, note that computational complexity does not contain the full story in general.  There can be a prefactor on the scaling that makes a particular algorithm run very poorly, and that prefactor can depend on $\chi$ in some cases.  This will matter more when discussing quantum number tensors.

\subsection{Contracting MPS tensors}

The algorithm presented in Sec.~\ref{expectvalues} is the best scaling method.  The first tensor is contracted to the first tensor, represented as
\begin{equation}
\sum_{\sigma_1}\left(A^\dagger\right)^{\sigma_1}_{a_1'}A^{\sigma_1}_{a_1}=L^E_{a'_1a_1}
\end{equation}
where $L^E$ is the environment from the left side of the system (see step 5 in Sec.~\ref{DMRG}). The cost of this contraction with physical index size $\hat D$ is $O(dm^2)$ where $m$ is the auxiliary link index size on all tensors.

The next tensor to be contracted into the environment tensor is $A^{\sigma_2}_{a_1a_2}$ and gives
\begin{equation}
\sum_{a_1}L^E_{a_1'a_1}A^{\sigma_2}_{a_1a_2}=L^E_{a_1'\sigma_2a_2}
\end{equation}
at a cost of $O(dm^3)$. Then contracting the dual tensor, the new environment obtained is
\begin{equation}
\sum_{a_1',\sigma_2}L^E_{a_1'\sigma_2a_2}\left(A^\dagger\right)^{\sigma_2}_{a_1'a_2'}=L^E_{a_2'a_2}
\end{equation}
where this costs $O(dm^3)$. So, the cost of contracting the tensors in the ideal order is $O(dm^3)$ because this was the largest cost of all the steps.  Formally, one can add all three steps together and only retain the largest cost when evaluating the complexity.

\subsection{Contracting transfer matrices}

Another example of an equivalent operation is to contract all pairs of tensors and their duals along the physical indices.  Then, the tensors can be contracted together along the link indices,
\begin{equation}
\sum_{\sigma_1}\left(A^\dagger\right)^{\sigma_i}_{a_{i-1}'a_i'}A^{\sigma_i}_{a_{i-1}a_i}=T^{a'_{i-1}a_i'}_{a_{i-1}a_i}
\end{equation}
where $T$ is a transfer matrix.  This cost $O(dm^4)$.  To then multiply two transfer matrices together, this will be
\begin{equation}
\sum_{a_i'a_i}T^{a'_{i-1}a_i'}_{a_{i-1}a_i}T^{a'_{i}a_{i+1}'}_{a_{i}a_{i+1}}=T^{a_i'a_{i+1}'}_{a_{i-1}a_{i+1}}
\end{equation}
and cost $O(m^6)$.  This is the cost of computing the correlation length in Sec.~\ref{correlationlengths}.  More discussion can be found in Ref.~\onlinecite{bakerCJP21,*baker2019m}.  

Comparing this estimate from the cost in the last section shows that this is not the ideal way to contract tensors for the MPS.  The difference here is that no environment on the left or right of the transfer matrix is available, so the cost increased.  If the transfer matrix must be found, then this cost will be required to be paid to obtain it.

\subsection{Computational complexity of SVDs}

The time complexity for the SVD is $O(\mathrm{min}(ab^2,a^2b))$ for a rectangular matrix of size $a\times b$.  This is known from existing arguments \cite{press1992numerical}. Depending on how a tensor is reshaped, the cost of the SVD will vary.
  
\subsection{Computational complexity of DMRG}

The overall scaling for DMRG--based on the previous sections--goes as $O(Ndm^3)$ for a number of sites $N$, size of the physical dimension $d$, bond dimension of the MPO of $\chi$, and bond dimension size $m$. The contractions for the MPOs are assumed to have a lower bond dimension, so do not contribute to the worst case scenario of large MPS tensors.  If this was not the case (which would be rare), then the bond dimension would be that of the MPO.

\section{Fermionic operators}\label{jordanwigner}

Fermions can be implemented in MPS computations, although it requires some explanation to demonstrate how to do this. This brief introduction to the Jordan-Wigner transformation will attempt to educate any interested reader in how to use these.  The main point of advice here is to use the examples in the library and make small modifications on the code to arrive at whatever use the code eventually needs to satisfy.  Checking against small examples is also highly encouraged.

The representation of the proper commutation rules for fermions in a tensor network code requires special treatment to ensure that the anti-commutator takes the correct form ($\{\hat c^\dagger_{i\sigma},\hat c_{j\sigma'}\}=\delta_{i,j}\delta_{\sigma,\sigma'}$) for the fermionic operators.  The formal justification of this construction can be found elsewhere \cite{fetter2012quantum}, but how it is used and implemented is the focus here.

The bare operators for raising and lowering operators can be discovered by explicitly evaluating each of the matrix elements.  In the physical Fock space defined by
\begin{equation}\label{HubbardBasis}
\{|0\rangle,|\uparrow\rangle,|\downarrow\rangle,|\uparrow\downarrow\rangle\}
\end{equation}
Then the wavefunctions can be enumerated in this basis.  The following relations must be held by the operators $\hat c_\uparrow^\dagger$ and $\hat c_\downarrow^\dagger$:
\begin{align}
\hat c_\uparrow^\dagger|0\rangle=|\uparrow\rangle\quad\mathrm{and}\quad\hat c_\uparrow^\dagger|\downarrow\rangle=|\uparrow\downarrow\rangle\\
\hat c_\downarrow^\dagger|0\rangle=|\downarrow\rangle\quad\mathrm{and}\quad\hat c_\downarrow^\dagger|\uparrow\rangle=-|\uparrow\downarrow\rangle
\end{align}
with all other elements equal to zero.  Similarly, the relationships are reversed for the lowering operator.  Note that the minus sign is crucial to the definition of the fermions.  The reason that it appears here is because adding the fermions out of order (or more specifically, out of normal ordering \cite{fetter2012quantum,peskin1995quantum}) must satisfy the anti-commutation rules.  It can be verified explicitly that these are satisfied here with matrix multiplication of the tensors.  The negative sign could have been added to the appropriate element of $\hat c_{\uparrow}^\dagger$.  This would define the opposite convention as is used here. So long as the sign convention is obeyed throughout the code, then this is sufficient to define these operators.

The resulting operators in the basis of Eq.~\eqref{HubbardBasis} is
\begin{equation}
\hat c_\uparrow^\dagger=\left(\begin{array}{cccc}
0 & 1 & 0 & 0\\
0 & 0 & 0 & 0\\
0 & 0 & 0 & 1\\
0 & 0 & 0 & 0\\
\end{array}\right)
\quad\mathrm{and}\quad
\hat c_\downarrow^\dagger=\left(\begin{array}{cccc}
0 & 0 & 1 & 0\\
0 & 0 & 0 & -1\\
0 & 0 & 0 & 0\\
0 & 0 & 0 & 0\\
\end{array}\right)
\end{equation}

However, these operators on their own are not sufficient to enforce the anti-commutation relations everywhere. Note that the site index was dropped for the basic definitions of the operators.  If, for example, a particle is added on one site and then another particle is added on another site, it is a reasonable question to ask what overall sign the wavefunction has.  This can be unified by defining a third operator that retains the information of the fermion sign, so it is known as the fermion sign operator, $F$, defined as
\begin{equation}
\hat F=\left(\begin{array}{cccc}
1 & 0 & 0 & 0\\
0 & -1 & 0 & 0\\
0 & 0 & -1 & 0\\
0 & 0 & 0 & 1\\
\end{array}\right)
\end{equation}
where the two middle elements are negative because they have one fermion and those elements multiply together on the double occupancy term.  Note that two $\hat F$ operators multiply to the identity matrix, so pairs of operators will cancel their fermion strings on the left-most sites.

Consider one strategy to enforce the correct fermion signs by adding a trail of $\hat F$ operators to each application of $\hat c^\dagger_{i\sigma}$.  In diagrammatic notation, this becomes

\begin{equation}
\includegraphics[width=0.75\columnwidth]{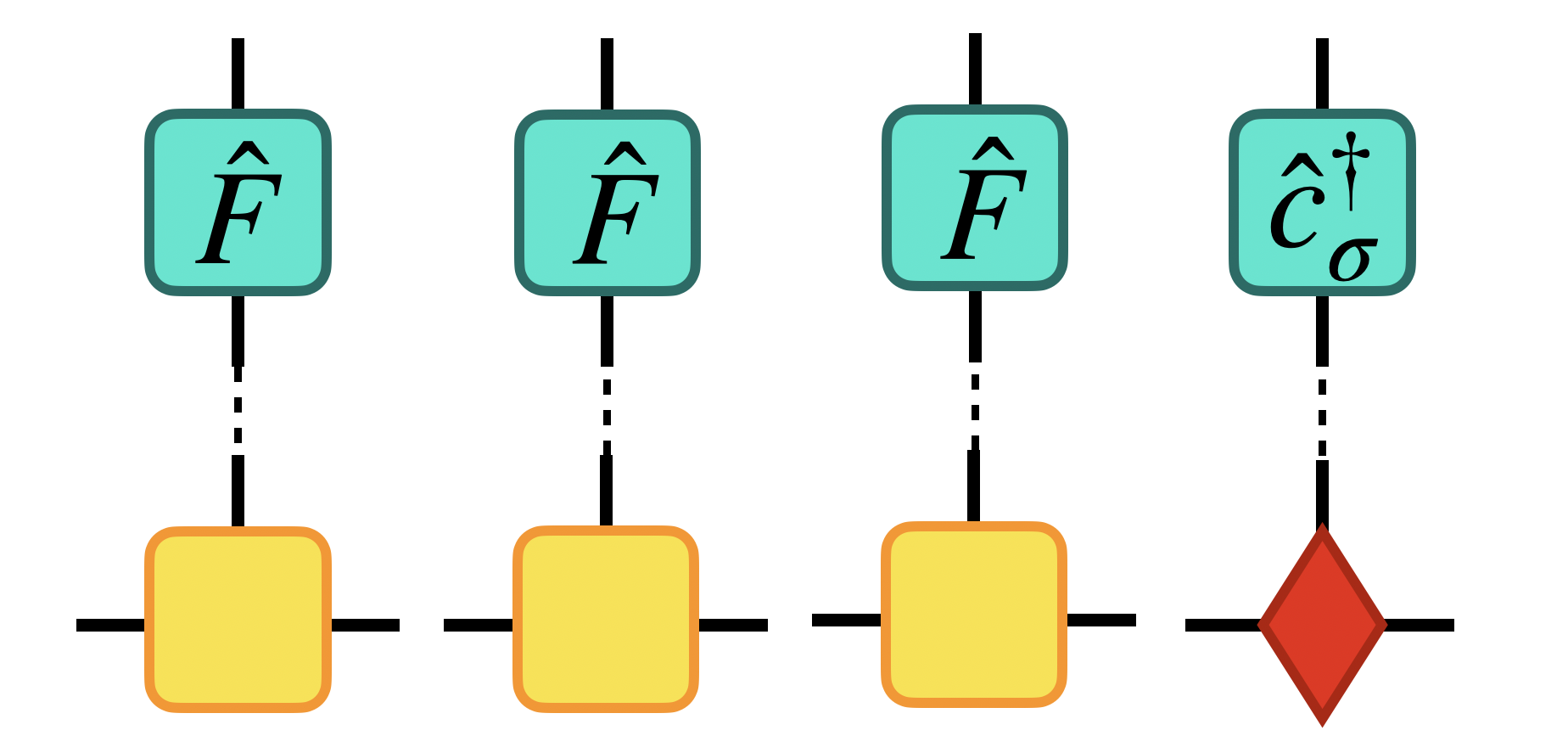}\nonumber
\end{equation}
where the code always assumes the trailing operators will be applied to the left of the given site where $\hat c^\dagger_\sigma$ was applied.  It is no difference to enforce the opposite convention.

To connect with ideas in topology, this enforces a boundary condition on the application of a fermion.  So, braiding two fermions together will cause their fermion string operators to mingle in such a way as to gain the correct sign. So long as this is done consistently, the correct answer will result.

Note that while the fermion string operators must be implemented on each application, and that this might make the use of fermions an inherently non-local problem, the multiplication of two $\hat F$ operators gives the identity.  So, if two fermions are applied, only the fermion strings in between them will matter.  This means that the discussion in the main text on measuring two operators retains the local aspects of that computation.  In other words, only the tensors between the two operators on sites $i$ and $j$ (inclusive) must be considered.

Small numerical round-off errors will allow for the particle number to change if the chemical potential is not set correctly.  One can run the dense tensor network computations for the fermionic operators, but it is must better to implement a quantum number conserving system to do this, and this will be the next article in the series.

\section{Connection to exact diagonalization}\label{ED}

Some users will come from a background in exact diagonalization or wish to implement a small check model with exact diagonalization codes. This section aims to introduce the concepts in exact diagonalization with the functions defined in the library so far.  This will also be useful for debugging small systems.  The functions used here can be made to check the energies and wavefunctions of the DMRG results.

The decomposition of a generic Hamiltonian into a set of operators can be best written as a tensor product of identity operators (one for each site), with some of those operators replaced by another operator.  For example, the Pauli matrix $\sigma_z$ on the second site of many is
\begin{equation}\label{sigmazfull}
\hat\sigma^z_2=\hat I\otimes\hat \sigma^z\otimes\hat I\otimes\ldots\otimes\hat I
\end{equation}
and with a straightforward generalization to other terms. It is then straightforward to construct Hamiltonian operators or any others that are necessary.

Each of the terms then scales exponentially with the number of lattice sites $N$, with a local Fock dimension of size $\hat D$, as $d^N\times d^N$.  This quickly becomes too large for classical memory storage, and this is the point where tensor network methods become more efficient.

In the Julia language, the operator {\tt kron} can be used to contstruct these states.  For an example with a spin system, the following code produces the term from Eq.~\eqref{sigmazfull}
\begin{lstlisting}[numbers=none]
Sx,Sy,Sz,Sp,Sm,Id,O = spinOps()
bigSz = kron(Id,Sz)
\end{lstlisting}
where it should be noted that the {\tt kron} function as of v1.6 is column-majored as opposed to row-majored, meaning that the operators must be input in the reverse here.  However, there is an easy way to convert from the MPO to the full Hamiltonian with the {\tt fullH} function in DMRjulia
\begin{lstlisting}[numbers=none]
ed_Ham = fullH(mpo)
#full diagonalization:
D,U = eigen(ed_Ham)
\end{lstlisting}
This simply connects all of the link indices of the MPO together and chooses the lower left corner (default) where the Hamiltonian accumulates.  The function will be limited in the system size based on the memory constraints of the host computer but it is useful to check the energies for small systems.

There is no function to break apart the full Hamiltonian into the MPO.  Instead, an automatic MPO creator will be implemented in a future work.  However, the basic concept behind breaking up the full wavefunction in the MPS essentially applies to obtain the local MPO form.

The function to convert the wavefunction output from exact diagonalization (or from an analytic form) into a MPS representation.
\begin{lstlisting}[numbers=none]
physind = 2 #size of physical index
#number of sites
Ns = cld(length(exactvec),physind)
psi=convert2MPS(exactvect,physind,Ns)
\end{lstlisting}
and similarly to {\tt fullH}, there is a function to convert the MPS to the full wavefunction
\begin{lstlisting}[numbers=none]
#define MPS as psi
wvfct = fullpsi(psi)
\end{lstlisting}
A function to do this with on the SVD decompositions was included in Ref.~\onlinecite{bakerCJP21,*baker2019m}. The use of QR decompositions are used when it can be assured that the bond dimension is not truncated.  The use of the QR decomposition avoids the re-contraction of the $\hat D$ matrix into the $V^\dagger$ matrix as was necessary in the example code from Ref.~\onlinecite{bakerCJP21,*baker2019m}.  This potentially creates some speed improvements.  Since it is assumed that the input wavefunction should not be truncated here, this justifies the use of the QR decomposition.  The LQ decomposition could be used here if the wavefunction is decomposed from right to left (resulting in the orthogonality center existing on the first site at the end of the computation), but there is no explicit reason why this would be advantageous here, and the movement of the orthogonality center for the first site after using the QR decomposition does not take too much time to run.

If the full wavefunction is to be kept, then the bond dimension $m$ of the resulting tensors will be untruncated in the SVD and, resultingly, exponentially large.  However, if the value is allowed to be truncated through the SVD, then the most relevant states (natural orbitals) will be kept at each step.  More discussion was already covered by us in Ref.~\onlinecite{bakerCJP21,*baker2019m} and some useful discussion on the cutoffs in bond dimension is there too.

To apply Lanczos onto the large Hamiltonian, simply call
\begin{lstlisting}[numbers=none]
out = krylov(fullH,fullpsi,maxiter=4)
wvfct,En = out
\end{lstlisting}
where it should be noted that other packages also have implementations of the Lanczos algorithm for large matrices. 

Note further that the application of a local operator ({\it i.e.}, $\hat\sigma^z_i$) requires a full $d^N\times d^N$ object for each term in exact diagonalization.  For tensor network methods, it is simply size $d\times d$ since only the single site $\hat\sigma^z$ is applied.

It is important to keep in mind, unlike exact diagonalization, that DMRG will iteratively converge the ground-state solution.  So, there are more parameters to keep in mind when running a DMRG simulation.  Comparatively, exact diagonalization is more automatic and gives solutions more immediately.  However, the reduction in memory allows for larger systems to be solved, and this is well worth the trouble of understanding the tensor network methods. 

\section{Getting started in Julia}

Downloading the Julia library is very simple from the offiicial distribution. The code used in this paper is for {\tt v1.6.0} released as of March 24, 2021), although the code is generic enough to date back to {\tt v1.1.1}.  Note that there was a slowdown in the parallelization for {\tt v1.2.0-v1.4.0}, so this should be minded if using those versions.  The code is general enough that it should function with subsequent versions, but the code will be maintained online \cite{dmrjulia}.

To load the library and start using all of the functions here, the main file must be included.
\begin{lstlisting}[numbers=none]
include("<library path>/DMRjulia.jl")
\end{lstlisting}
After this, all functions should be available.  The examples in {\tt /examples} show how to start several computations of interest.

For most functions, typing {\tt ?} and then the function name will print out documentation that is included in the code.

Many functions have both an in-place version (where the input to the function is modified) and one that is not in-place (where the input is copied and manipulated).  The functions that act in-place are generally denoted with an exclamation point ({\it e.g.}, {\tt reshape!} is in-place and {\tt reshape} copies the input). 

There is also a "secret menu" of options on many high level functions that are worth understanding.  This discussion is reserved for the Supplemental Material \cite{tensor_recipes}.

One operation that might be useful in the Julia code as of now is the writing of data to an output file.  This can be accomplished in the {\tt DelimitedFiles} package.  A code such as
\begin{lstlisting}[numbers=none]
#define X and Y
using DelimitedFiles
open("values.txt", "w") do io
  writedlm(io,[X Y]) #two data outputs
end
\end{lstlisting}
There are also several plotting tools in Julia that can make graphs of data without exporting, including an imported tools from Python.

\section{Simple DMRG code}\label{DMRGcode}

The development of the basic tools in this paper are what was found to be to be necessary for efficient implementation of the DMRG algorithm and other extensions of these techniques in an efficient way.  A short code is provided here using those tools.

\begin{widetext}
\begin{lstlisting}[language=Julia]
path = "<path to DMRjulia.jl>"
include(path*"/DMRjulia.jl")

#create reduced site problem (step 2)
function make2site(Lenv,Renv,psiL,psiR,mpoL,mpoR)
  Lpsi = contract(Lenv,[3],psiL,[1])
  LHpsi = contract(Lpsi,[2,3],mpoL,[1,2])
  LHpsipsi = contract(LHpsi,[2],psiR,[1])
  LHHpsipsi = contract(LHpsipsi,[3,4],mpoR,[1,2])
  HpsiLR = contract(LHHpsipsi,[3,5],Renv,[1,2])
  return HpsiLR
end


function simplelanczos(iL,iR,Lenv,Renv,psi,mpo;betatest = 1E-10)
  Hpsi = make2site(Lenv[iL],Renv[iR],psi[iL],psi[iR],mpo[iL],mpo[iR])
  AA = contract(psi[iL],3,psi[iR],1)
  AA = div!(AA,norm(AA))
  alpha1 = real(ccontract(AA,Hpsi))

  psi2 = sub!(Hpsi, AA, alpha1)
  beta1 = norm(psi2)
  if beta1 > betatest
    psi2 = div!(psi2, beta1)

    Hpsi2 = contract(Lenv[iL],3,psi2,1)
    ops = contract(mpo[iL],4,mpo[iR],1)
    Hpsi2 = contract(Hpsi2,[2,3,4],ops,[1,2,4])
    Hpsi2 = contract(Hpsi2,[2,5],Renv[iR],[1,2])

    alpha2 = real(ccontract(psi2,Hpsi2))
    M = [alpha1 beta1; beta1 alpha2]
    D, U = eigen(M)
    energy = D[1,1]
    outAA = conj(U[1,1])*AA + conj(U[2,1])*psi2
  else
    energy = alpha1
    outAA = AA
  end
  return outAA,energy
end

function simpledmrg(psi,mpo;m=20,sweeps=10,cutoff=1E-9)
  Lenv,Renv = makeEnv(psi,mpo)  #make environments
  j = 1
  for n = 1:sweeps
    for s = 1:2*(length(psi)-1)
      i = psi.oc

     #select sites to form renormalized problem (step 1)
      if j > 0
        iL = i
        iR = i + 1
      else
        iL = i - 1
        iR = i
      end

      outAA,energy = simplelanczos(iL,iR,Lenv,Renv,psi,mpo) #Lanczos (step 3)

      if cld(length(psi),2) == psi.oc
        println("sweep = $n, site = $(psi.oc), energy = ",energy[1])
        println()
      end

      U,D,V = svd(outAA,[[1,2],[3,4]],m=m,cutoff=cutoff) #SVD (step 4)

      if j > 0
        psi[iL] = U
        psi[iR] = contract(D,2,V,1)
        #update environments (step 5)
        Lenv[iR] = Lupdate(Lenv[iL],psi[iL],psi[iL],mpo[iL])
      else
        psi[iL] = contract(U,3,D,1)
        psi[iR] = V
        #update environments (step 5)
        Renv[iL] = Rupdate(Renv[iR],psi[iR],psi[iR],mpo[iR])
      end
      psi.oc += j
        
      if j > 0 && psi.oc == length(psi)
          j *= -1
      elseif j < 0 && psi.oc == 1
          j *= -1
      end
    end
  end
end

Ns = 100
spinmag = 0.5

physindsize = convert(Int64,2*spinmag+1)

psi = randMPS(physindsize,Ns)

Sx,Sy,Sz,Sp,Sm,O,Id = spinOps(s=spinmag)
H = [Id O O O O;
     Sp O O O O;
     Sm O O O O;
     Sz O O O O;
     O Sm/2 Sp/2 Sz Id]

mpo = makeMPO(H,physindsize,Ns)

println("#############")
println("simple DMRG (dense version)")
println("#############")

simpledmrg(psi,mpo,sweeps=20,m=45,cutoff=1E-9)
\end{lstlisting}
\end{widetext}

This code can be compared with the basic steps for DMRG in both diagrammatic and text form in Sec.~\ref{DMRG}.

\end{appendix}

\end{document}